\documentclass[resetfootnote]{aastex701}

\newlength{\plotwidth}
\usepackage{multirow}
\usepackage{textalpha}  
\usepackage[usestackEOL]{stackengine} 
\setstackgap{L}{\normalbaselineskip}
\usepackage{bigdelim}   
\usepackage{enumitem}   
\usepackage{nicefrac}
\usepackage[export]{adjustbox}
\usepackage{calc}
\usepackage{capt-of}
\usepackage{makecell}
\usepackage{tikz}
\usepackage[outline]{contour}
\contourlength{0.4pt}
\usepackage{tipa}

\definecolor{TableauGreen}{HTML}{2CA02C}
\definecolor{TableauOrange}{HTML}{FF7F0E}

\newcommand{\approptoinn}[2]{\mathrel{\vcenter{
			\offinterlineskip\halign{\hfil$##$\cr
				#1\propto\cr\noalign{\kern2pt}#1\sim\cr\noalign{\kern-2pt}}}}}

\usepackage{array}
\newcolumntype{Z}[1]{>{\centering\arraybackslash}m{#1}}  

\received{January 10, 2026}
\revised{February 16, 2026}
\accepted{February 18, 2026}

\submitjournal{PASP}

\shorttitle{The Astronomical Telescope of the University of Stuttgart (ATUS)}
\shortauthors{Schindler et al.}

\begin{document}

\title{The Astronomical Telescope of the University of Stuttgart (ATUS):\\Development, Optimization, and Lessons Learned}

\correspondingauthor{Karsten Schindler}
\email{karsten.schindler@irs.uni-stuttgart.de}

\author[0000-0001-7337-2452,gname=Karsten,sname=Schindler]{Karsten Schindler}
\affiliation{Institute of Space Systems, Universität Stuttgart, Pfaffenwaldring 29, 70569 Stuttgart, Germany}\altaffiliation{The German SOFIA Institute (Deutsches SOFIA Institut, DSI) existed as a branch of the Institute of Space Systems (IRS) until 2025.}
\affiliation{SOFIA Science Center, NASA Ames Research Center, Moffett Field, CA 94035, USA}
\email{karsten.schindler@irs.uni-stuttgart.de}

\author{Jürgen Wolf}
\affiliation{Institute of Space Systems, Universität Stuttgart, Pfaffenwaldring 29, 70569 Stuttgart, Germany}
\affiliation{SOFIA Science Center, NASA Ames Research Center, Moffett Field, CA 94035, USA}
\email{juergen.wolf@irs.uni-stuttgart.de}

\author[0000-0002-8522-7006,gname=Alfred,sname=Krabbe]{Alfred Krabbe}
\affiliation{Institute of Space Systems, Universität Stuttgart, Pfaffenwaldring 29, 70569 Stuttgart, Germany}
\email{alfred.krabbe@irs.uni-stuttgart.de}

\begin{abstract}

ATUS, the {Astronomical Telescope of the University of Stuttgart}, is a fully remote-controlled \mbox{0.6~m $f/8.17$} Ritchey-Chrétien telescope optimized for high-cadence, high-fidelity photometry of transient sources. Observations are time-referenced with very high accuracy and precision, making it an ideal platform for time-domain astronomy and space situational awareness. Initially conceived to support instrument developments and operations of SOFIA, the {Stratospheric Observatory for Infrared Astronomy}, it evolved into a scientific instrument for various use cases in instrument development, astronomical research, and teaching. This paper presents an overview of its development and optimization to achieve diffraction-limited images and highly accurate pointing and tracking, even at high speeds. The findings and lessons learned are universally applicable to other telescopes that are currently at the planning stage, or where similar issues might be encountered.

\end{abstract}



\section{Introduction} \label{sec:intro}

ATUS was primarily conceived as a test platform for evaluating new hardware and software under consideration for SOFIA, the {Stratospheric Observatory for Infrared Astronomy} \citep{Temi2018}. With its main camera and filter set identical to SOFIA's {Focal Plane Imager} \citep[FPI\textsuperscript{+},][]{Pfueller2018} and of comparable focal length, an auxiliary apochromatic refractor mimicking the field of view (FoV) of SOFIA's {Fine Field Imager} (FFI), and a geometrically near-identical {Wide Field Imager} (WFI), ATUS resembled the optical conditions of the three viewfinder and tracking cameras of the SOFIA telescope \citep{Wiedemann2012,Wiedemann2016} almost ideally (see Table~\ref{tab:imagers}). This enabled engineering trials and developments that could not have been conducted during SOFIA's operational phase as test time was unavailable. 

Beyond serving as a SOFIA testbed, a second goal emerged: transforming ATUS into a high-performing scientific instrument in its own right. A number of improvements and configuration changes were necessary to turn it into a reliable instrument that enables rigorous observations of stellar occultations and transits of extrasolar planets. Its sub-microsecond time-referenced, gap-free, high-cadence imaging became critical for excelling in observations of transient sources. This article documents that journey.

\begin{figure*}[htb]
\begin{center}
\resizebox{0.9\linewidth}{!}{%
\includegraphics[height=1cm]{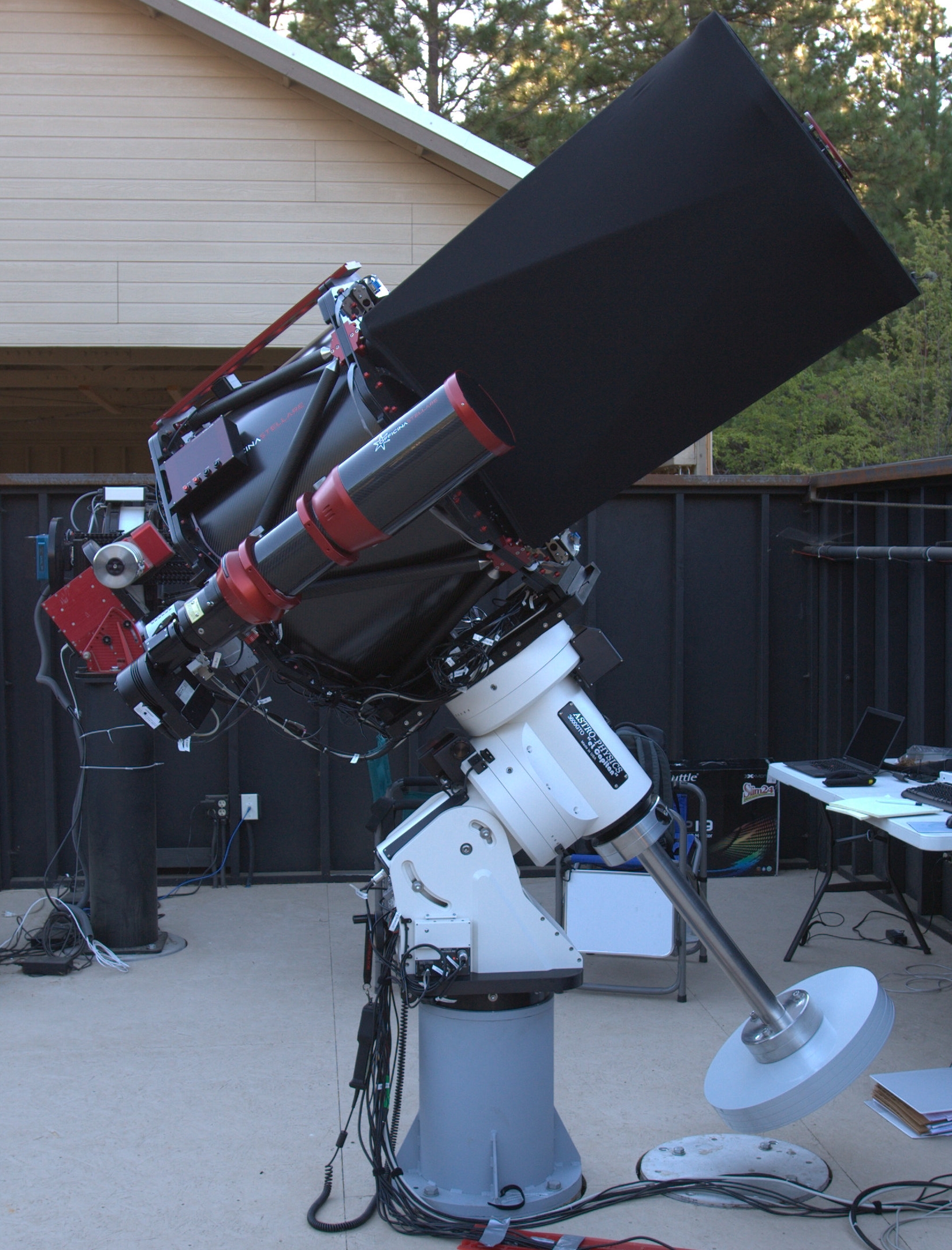} 
\includegraphics[height=1cm]{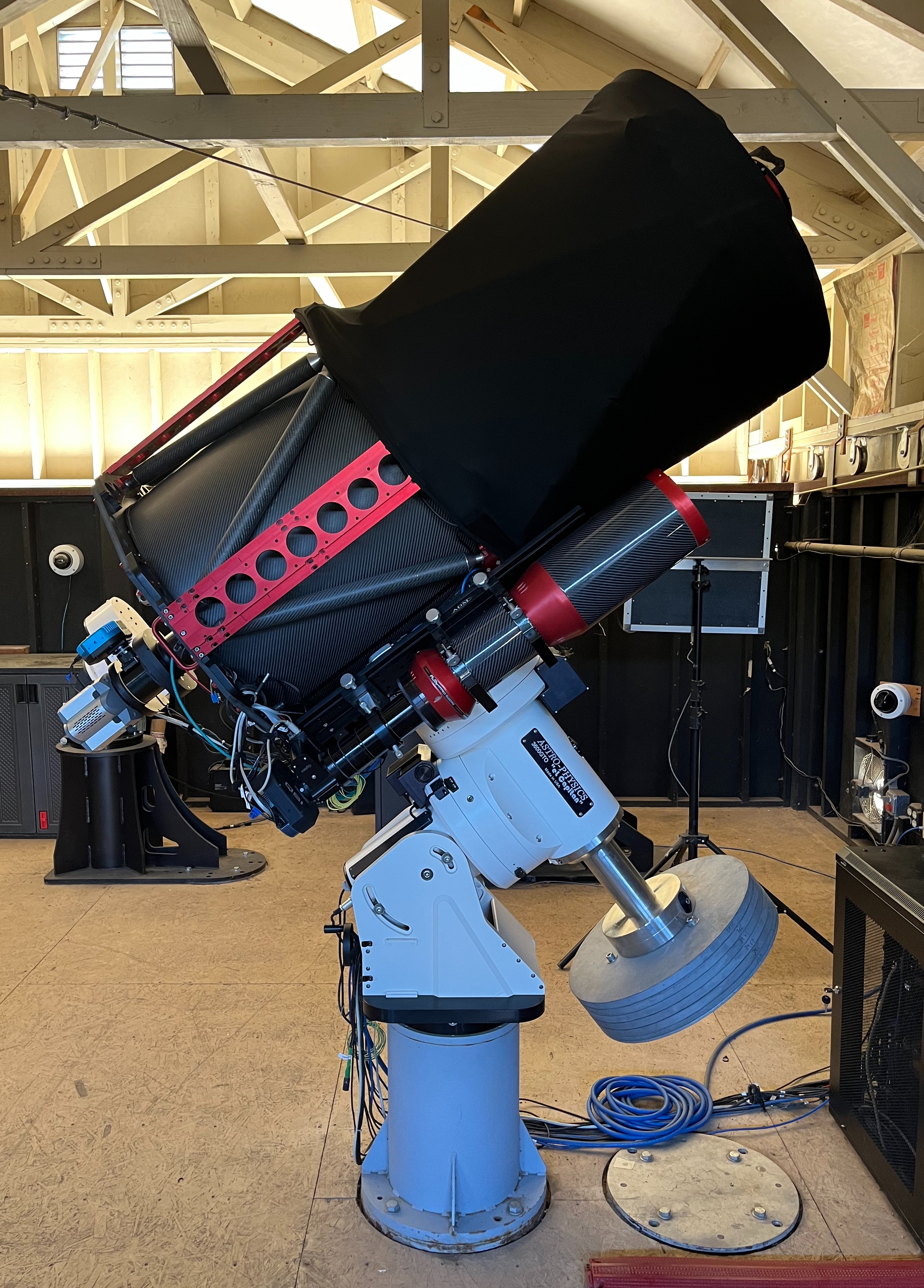} 
}
\end{center}
\caption{ATUS during initial commissioning (left) and in its final configuration (right) with a completely redesigned telescope, a shorter and much stiffer counterweight assembly, an optimized weight distribution to lower the setup's moment of inertia, and a custom off-axis guider (see discussions in Section~\ref{sec:optimization}). The Wide Field Imager (WFI) is mounted on the west side of the telescope tube and thus hidden; it is pictured in Figure~\ref{fig:WFI+PDU}.
\label{fig:ATUS_evolution}}
\end{figure*}

The following sections provide a general system overview, summarize key issues encountered and the remedies that were applied, and add some reflection on operating the instrument in the field for more than eleven years. Figure~\ref{fig:ATUS_evolution} gives an impression of how the setup evolved over time.  


\begin{table}[ht]
\centering
\caption{Field of view and plate scale of the three viewfinder and tracking cameras of the SOFIA telescope, in comparison to the three imaging cameras on the ATUS telescope (see also Table~\ref{tab:instrumentation_overview}).} 
\label{tab:imagers}
\begin{tabular}{cccc}
\tablewidth{0pt}
\hline\hline
                        & FPI\textsuperscript{+}                        & FFI                                           & WFI \\
\hline\hline
\multirow{2}{*}{SOFIA}  & 8.7~\texttimes~8.7~arcmin\textsuperscript{2}  & 67~\texttimes~67~arcmin\textsuperscript{2}    & 6.0~\texttimes~6.0~deg\textsuperscript{2} \\
                            & 0.51~arcsec/pixel                             & 3.9~arcsec/pixel                              & 21.2~arcsec/pixel                         \\
\hline
\multirow{2}{*}{ATUS}   & 9.34~\texttimes~9.34~arcmin\textsuperscript{2}  & 65~\texttimes~44~arcmin\textsuperscript{2}    & 5.7~\texttimes~5.7~deg\textsuperscript{2} \\
                            & 0.547~arcsec/pixel                             & 1.8~arcsec/pixel                              & 19.9~arcsec/pixel                         \\
\hline 
\end{tabular}
\tablecomments{The plate scale of the FPI\textsuperscript{+} varied as SOFIA's secondary mirror was moved to achieve the focal plane position required by each scientific instrument; the FPI\textsuperscript{+} was focused separately via movable tilted mirrors in its optical delay line.}
\end{table} 



\section{System overview} \label{sec:system_overview}

\subsection{Configuration}
ATUS operates in a fully reflective Ritchey-Chrétien (RC) $f/8.17$ configuration. The optical tube assembly (OTA) was designed and manufactured by Officina Stellare, Sarcedo, Italy. A focal reducer (lens triplet) and a field flattener (lens doublet) are available (cf.\ Table~\ref{tab:imagecircle}), but were not put into use, given the site-appropriate sampling of the main camera's detector (see discussion in Section~{\ref{sec:final_IQ}}) and its small size, at which field curvature is not notable. Spectral range is thus not limited by chromatic aberration, e.g.\ as in other popular systems that employ refractive elements. The telescope has an $f/3$ primary mirror (M1) and a slightly oversized secondary mirror (M2), offering a usable image circle radius of 25.5~arcmin, or 36.3~mm at the focal plane. Both mirrors are made of OHARA \mbox{CLEARCERAM-Z HS} ultra-low-expansion glass ceramics, and are coated with aluminum covered by a protective silica layer. Optical design and fabrication of both mirrors, the focal reducer, and the field flattener were conducted by Massimo Riccardi, Occhiobello, Italy. The mechanical envelope of the conical M2 baffle determines the central obscuration ratio of 45.3\% (cf.\ Table~\ref{tab:OTA_comparison}), which translates into an obscured aperture area of 20.6\%. Considerable effort went into the telescope's mechanical design (Section~\ref{sec:OTA}) and optimal mirror alignment (Section~\ref{sec:mirror_spacing}).  


\begin{table*}[t]
\centering
\caption{Relevant optical parameters of the ATUS telescope.} 
\label{tab:imagecircle}
\begin{tabular}{cccc}
\tablewidth{0pt}
\hline\hline
                        & fully reflective                        & with field flattener                                           & with focal reducer \\
\hline\hline
Focal length $f$ & 4899.8~mm	& 5236.2~mm & 3836.1~mm \\
Aperture ratio $f/D$    & 8.17   & 8.73   & 6.39 \\
\multirow{2}{*}{Image circle radius} & 36.35~mm & 24.74~mm & 19.94~mm \\
                 & 25.5~arcmin & 16.2~arcmin & 17.7~arcmin \\
Back focal length (BFL) & 1656.4~mm & 1673.9~mm & 1632.6~mm \\
Flange focal distance (FFD) & 303.5~mm & 228.1~mm & 83.7~mm\\
\hline
Mounted via & & M85 thread & circular dovetail clamp \\
\hline
\end{tabular}
\tablecomments{Back focal length is measured from the M2 vertex. Flange focal distance of the fully reflective configuration is measured from the telescope's backplane, otherwise from the back of the respective lens housing. Only the fully reflective configuration has been used so far.}
\end{table*} 


The instrument is carried by a 3600GTOPE German-equatorial mount made by Astro-Physics, Machesney Park, Illinois (USA), which got subsequently customized (Sections~\ref{sec:mount} and \ref{sec:CW_modification}). It offers slew speeds of up to 3.75\degr/s. A high-resolution Renishaw encoder ring at the polar axis eliminates the periodic error in the worm gear. As on every mount of this type, the telescope needs to change pier sides when its target is crossing the meridian, essentially requiring both axes to rotate by 180\degr. Mount control allows to delay or to advance this ``meridian flip'' up to a defined limit, extending observations past or before the meridian, with the counterweight shaft pointing up. ATUS routinely operated with a meridian delay/advance of 36~min, during which a slew command would bring the telescope earlier on the east side of the pier, or keep tracking on a target on the west side of the pier. Mechanical limit switches on the polar axis, adjustable up to $\approx$ 45~min before and after the meridian, add a layer of safety to prevent pier collisions, should erroneous commands or software limits bring the telescope into an unsafe position.


\subsection{Imaging cameras}

The primary instrument is an Andor iXon DU-888 camera with a back-illuminated, frame-transfer EMCCD sensor and a 10-position filter wheel for $1\nicefrac{1}{4}$-inch (31.75~mm) filter cells, housing a parfocal filter set based on Sloan Digital Sky Survey (SDSS) bandpasses \citep{Fukugita1996}, but optimized for highest transmission and with a larger separation between g' and r' bands to avoid atmospheric sky glow (Astrodon Photometrics Sloan Generation~2). This filter set has been adopted by a range of projects such as Las Cumbres Observatory Global Telescope \citep[LCOGT,][]{Brown2013}, the AAVSO Photometric All Sky Survey \citep[APASS,][]{Henden2009}, and  the MuSCAT instrument series \citep[Multicolor Simultaneous Camera for studying Atmospheres of Transiting exoplanets,][]{Narita2015,Narita2024}, and was also selected for SOFIA's FPI\textsuperscript{+} \citep{Pfueller2018}. Additionally, a 500~nm longpass filter and a 500~--~700~nm bandpass (``VR'') are available, tailored for exoplanet transit observations (reduced sky background during moon-lit nights, \added{as well as reduced} atmospheric extinction at lower elevations) and astrometric position measurements of minor planets (balancing throughput with dispersion due to differential refraction at lower elevations). The camera's EMCCD offers high quantum efficiency ($\eta_{\mathrm{Peak}} > 95\%$, cf.\ Figure~\ref{fig:QE}), high frame rates, and virtually gap-free imaging thanks to the sensor's frame-transfer architecture. Very low dark currents are achieved by the camera's four-stage thermoelectric cooler, capable of cooling the sensor as much as $\approx$~95~K below ambient. This enabled operations with a sensor temperature of $-60\degr\mathrm{C}$ throughout the entire year, where dark current has already reached a negligible level. Before capturing photons with ATUS, the camera head had served as the {Fast Diagnostic Camera} (FDC) for telescope characterization tests during SOFIA's commissioning and early science phase from 2010 to 2012 \citep{Pfueller2012}. The FDC's successful test campaigns motivated the FPI\textsuperscript{+} upgrade of SOFIA's main tracking camera in 2013. The former FDC became the main camera of ATUS, enabling a representative testbed resembling the FPI\textsuperscript{+}. 

A fully custom-designed off-axis guider (OAG) redirects unused light outside the main camera's optical beam into a guiding camera, enabling ultra-deep exposures and steady tracking with sub-pixel accuracy over hours (see Section~\ref{sec:OAG}). Focusing is either achieved by moving M2 with a stepper motor in the direction of the optical axis, or via a dedicated mechanical focuser mounted in front of the OAG and the main camera. M1 can be protected from dust by closing a four-petal cover. Three fans at the telescope backplane aid in cooling M1 to ambient temperature.  

The setup is complemented by a 130~mm $f/6$ apochromatic refractor. Due to differential flexure, its use as a guide scope has been limited (see Section~\ref{sec:tracking_pointing_performance}), but it has proven itself valuable as an imager with an intermediate field size, e.g.\ for acquisition of fast moving targets in Earth's orbit or extended cometary comae. In addition, a commercial 135~mm $f/2.8$ Canon telephoto lens, controlled with an Arduino microcontroller \citep[cf.\ Figure~\ref{fig:WFI+PDU};][]{Doerr2014}, acts as a piggy-back wide-field imager using a former prototype camera. This imager was essential for software tests for SOFIA (see Section~\ref{sec:astrometry.net}), has proven useful for following trajectories of fast objects against background stars (e.g.\ NEAs, geostationary satellites) while tracking at a sidereal rate, and provides valuable context information about sky conditions at the time of critical observations (e.g.\ presence of thin cirrus clouds). 

Table~\ref{tab:instrumentation_overview} in Appendix~\ref{sec:app_imager_configuration} provides a detailed technical summary of all available imagers. Readers who are interested in more details about the setup's ancillary equipment are referred to Appendix~\ref{sec:app_periphery}. A break down of the total instrument mass is provided in Appendix~\ref{sec:app_weight_breakdown}. 

Until October 2024, ATUS was based at Sierra Remote Observatories (SRO), a commercial ``telescope farm'' located in California's Sierra Nevada mountain range at an altitude of about 1400~m near the town of Auberry, which is about an hour's drive north-east of Fresno. At its first light in September 2013, ATUS was only the fourth telescope inside ``SRO 9'' (37\degr 04\arcmin 13.44\arcsec N, 119\degr 24\arcmin 47.088\arcsec W), the first of what would become six large roll-off-roof observatories housing several dozens of setups today. At the time of writing, preparations for a future use under Chilean skies are underway.


\subsection{Time referencing} \label{sec:time_referencing}

A key component of the setup is the Intelligent Reference/TM-4 time reference system \citep{SpectrumInstruments2015} with a GPS-disciplined 10~MHz oven-controlled crystal oscillator (OCXO), which has been used with great success for portable occultation setups \citep[POETS,][and PICO, \citealp{Lockhart2010}]{Souza2006}, by the MORIS instrument on the NASA IRTF \citep{Gulbis2011}, and by the FDC onboard SOFIA. These setups typically used the TM-4 as the initiator that emits precisely timed TTL-level pulses to a target camera, with each pulse either triggering a frame shift of a frame-transfer EMCCD (POETS), or the start of an exposure of a shutter-controlled camera with preset exposure times (PICO). This allows time-synchronized observations, e.g.\ among multiple observing sites, or of several cameras in multi-channel imaging photometers. 

In contrast, ATUS uses the Andor camera as the initiator and the TM-4 as the target. At the beginning of each exposure, the camera emits a TTL-level trigger pulse over its 50~\textOmega~``Fire'' SMB output port. The TM-4 is monitoring its input signal line for a logic transition, and the rising edge of the pulse triggers it to log an event time stamp. To be detected by the TM-4, the pulse width must be greater than 10~ns, pulses must be at least 4~ms apart, and use TTL or CMOS logic levels.  
The decision to make the camera the initiator eliminates additional training for operators and allows camera control through various common software packages and protocols, greatly simplifying operations. Event time stamps are temporarily stored in the TM-4's internal buffer memory before they are broadcasted over a serial connection. 

With a control PC running on Microsoft Windows\footnote{\added{The control PC ran on Windows~7 (09/2013 -- 10/2018), Windows~10 (10/2018 -- 12/2023), and ultimately, Windows~11 (since 12/2023).}}, an ASCOM driver was developed that provides the corresponding event time tag read from the TM-4 serial port with each exposure, allowing it to be written directly into the header of each FITS file. As a result, all data that gets collected via an ASCOM client is properly time-referenced with very high accuracy and precision. Although designed as a stand-alone unit, the TM-4 has been integrated inside the telescope control PC for better thermal protection. Antenna and trigger cables are connected to feed-through connectors at a custom side panel. 

Time tags provided by the TM-4 have a resolution of 100~ns. The Andor camera's ``Fire'' port was designed to precisely trigger short-duration, pulsed light sources, e.g.\ for applications in the biosciences. System characterization and cable length considerations confirmed that the TM-4's time stamp accuracy is better than 1~\textmu s. It was found that the TM-4 supports a maximum continuous rate of 28 triggers per second, which is well above the camera's full-frame readout rates (cf.\ Table~\ref{tab:instrumentation_overview}), and practically above all reasonable sampling rates of photon-starved astronomical sources with a 60~cm aperture. Faster trigger rates cause the buffer memory to overflow, as serial broadcast of logged time stamps cannot keep up. Thus, should a faster sampling rate of a very bright target via a partial EMCCD sensor readout (or by using a different camera) be desired, a TTL frequency divider can be used. 


\section{System development and optimization} \label{sec:optimization}

The following section provides a retrospective on the various improvements during operations at SRO, centered around three major issues that probably any practitioner and instrument builder has to combat: minimizing instrument flexure, optimizing image quality, and achieving reliable pointing and tracking. All three areas are fundamentally connected with each other, making separate discussions sometimes difficult. The order of the following sections loosely reflects the chronological evolution of ATUS, and iterations in all three areas lead to the final instrument configuration with peak performance.

\subsection{Instrument flexure} \label{sec:instrument_flexure}

Anticipating a telescope weight of $\approx$~105~kg, the 300~lbs (136.4~kg) payload capacity rating of the 3600GTOPE German-equatorial mount seemed appropriate. However, upon delivery of the ``Mark I'' OTA, its weight turned out to be significantly larger than specified (cf.\ Table~\ref{tab:OTA_comparison}), and exceeded this rating already without any auxiliary components. Before the instrument could be deployed to the field, the mount had to be modified to be able to carry such load. After setup at SRO, it quickly became clear that the Mk~I OTA had a significant amount of flexure, triggering a full optomechanical redesign and replacement.  


\begin{table*}[htb]
\centering
\caption{Differences of the ``Mark I'' and ``Mark II'' optical tube assembly.} 
\label{tab:OTA_comparison}
\begin{tabular}{p{6cm} c c}
\tablewidth{0pt}
\hline
\strut                                               & {Mk~I}                                                & {Mk~II} \\
\hline\hline
Mirror substrate material                       & LZOS Sitall CO-115M (``Astrositall'')                   & OHARA CLEARCERAM-Z HS \\
Density                                         & 2.46 g/cm\textsuperscript{3} [1]                           & 2.55 g/cm\textsuperscript{3} [2] \\  
\hline
Primary mirror (M1)                             &                                                       & \\
\hspace{5mm}\Centerstack{Shape and weight}      & \Centerstack{7.5:1 aspect ratio, flat back ($\approx 48.3~\textrm{kg}$)}       & \added{\Centerstack{6.4:1 flat near center, tapered\\to 30:1 at edge ($\approx 37.1~\textrm{kg}$)}} \\ 
\hspace{5mm}Mirror cell                         & Whiffle tree (27 contact points)                      & Central hub \\
\hline
Secondary mirror (M2)                           &                                                       & \\
\hspace{5mm}\Centerstack{Focus mechanism}    & \Centerstack{Four shafts (spaced 90 deg),\\guided by linear ball bearings} & \Centerstack{Draw tube, sliding on three pairs of\\preloaded ball bearings (spaced 120 deg)} \\
\hspace{5mm}Stepper motor                       & 200 steps/revolution                                  & 400 steps/revolution \\
\hspace{5mm}Optical encoder                     & none                                                  & incremental, 2000 CPR / 8000 PPR \\
\hspace{5mm}Linear actuator resolution          & 7.9 \textmu m/step                                    & 1.41 \textmu m/step \\ 
\noalign{\vskip 2pt}
\parbox[c]{6cm}{\hangindent=5mm\hangafter=0{Distance from M2 center of mass\newline to front ring median plane}}        & $\approx$ 134~\textrm{mm}                             & $\approx$ 0~mm (a) / $\approx$ 12~\textrm{mm} (b) \\
\noalign{\vskip 2pt}
\hspace{5mm}Central obscuration ratio           & 39.2\%                                                & 45.3\% \\
\hline
Truss                                           &                                                       & \\
\hspace{5mm}Material of truss ball joints       & Stainless steel                                       & Ti-6Al-4V \\
\hspace{5mm}Length ratio front/back truss       & 1.83                                                  & 1.28 \\
\hline
Total weight                                    & 145~kg                                                & 131.2~kg \\
\hline
Flange focal distance (FFD)                     & 244~mm                                                & 207~mm (a) / 305~mm (b) \\
Focal length                                    & 4800~mm                                                & 4740~mm (a)/ 4900~mm (b) \\
\hline
\end{tabular}
\tablerefs{[1] \citet{Abdulkadyrov2012}, [2] \citet{OharaCorporation}}
\tablecomments{(a): as delivered; later also referred to as ``Mk~II.a''; (b): in its final configuration, after several modifications; later also referred to as ``Mk~II.b''.}
\end{table*} 


\subsubsection{Increasing the mount's load capacity} \label{sec:mount}

The mount's modular design is geared towards assembly in the field by no more than two persons without a lifting device. Its weight-optimized structural design thus governed its load capacity, yet higher static and dynamic loads were uncritical for its bearings. As ATUS would become a permanent setup, mass savings were not a priority. 

\begin{figure}[htb]
\centering
\includegraphics[width=\plotwidth]{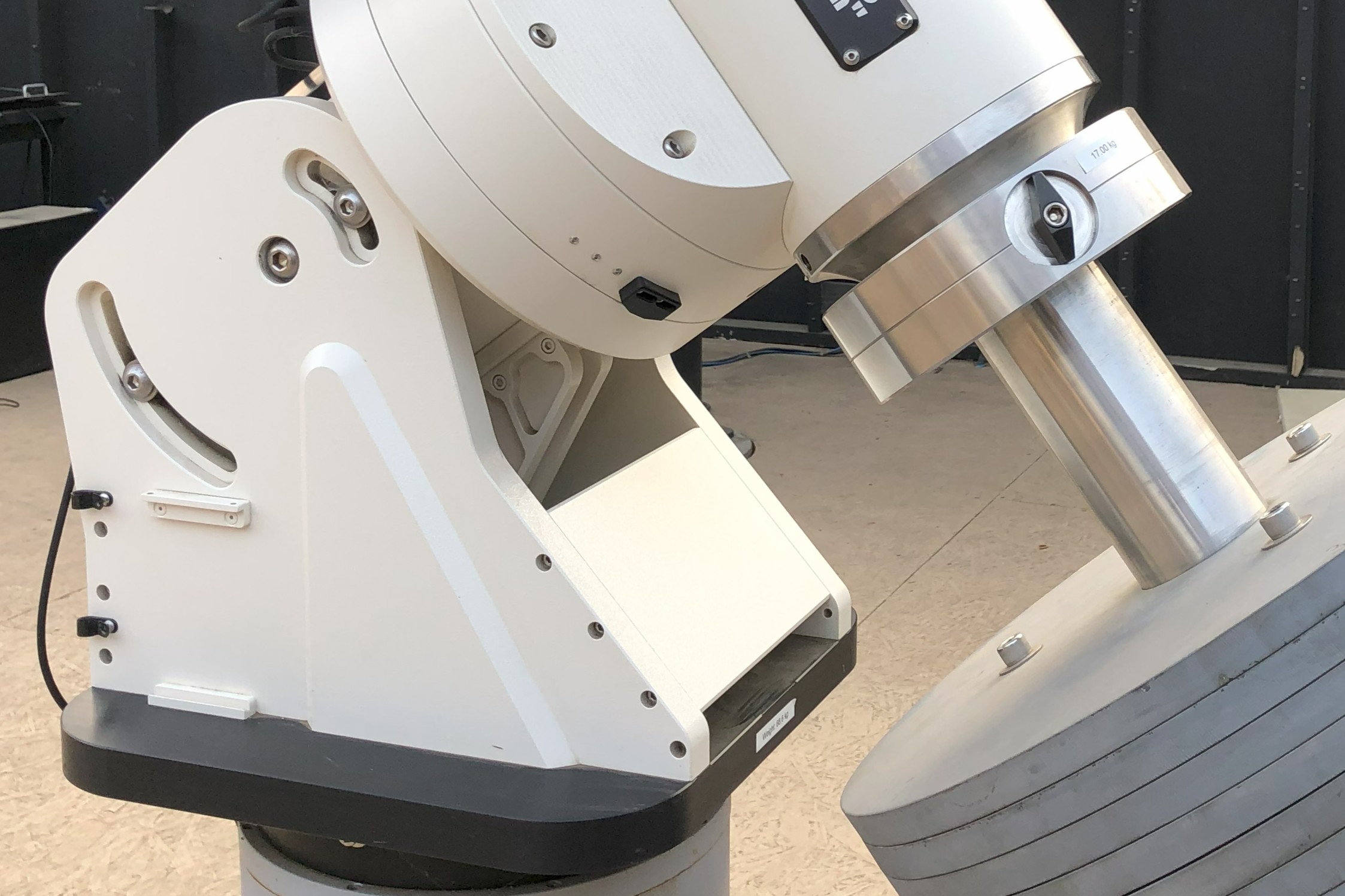} 
\caption{Modified polar fork assembly of the 3600GTOPE mount. Stiffness increased significantly by using thicker, solid side plates, an additional cross brace, and a thicker, solid base.  
\label{fig:mount_polar_fork}}
\end{figure}

To assess mount performance with a payload of roughly 125\% of its specified capacity, the setup was initially tested indoors at the labs of the German SOFIA Institute (Deutsches SOFIA Institut, DSI) at NASA Ames. Slew tests revealed that the stiffness of the weight-optimized polar fork assembly had to be increased, as acceleration-induced torque caused notable vibrations due to torsion. The polar axis thus received thicker side plates and a thicker base, both fabricated without weight-reducing pockets, and an additional, solid cross brace that bolts both side plates together (cf.\ Figure~\ref{fig:mount_polar_fork}). This mechanical modification, done ahead of deployment to SRO, increased the stiffness of the polar fork assembly significantly. 


Balancing such a heavy setup presented another challenge. Stock counterweights (30~lbs / 13.6~kg), again designed to be handled safely by a single person, required an extension of the 2.5-inch (63.5~mm) diameter, 30-inch (762~mm) long counterweight shaft. While such an 11.5-inch (292~mm) extension was readily available as an accessory, it could not be used at SRO due to building height constraints. As a pragmatic and quick remedy, a custom stack of heavier counterweights was made. Steel disks of 20-inch (508~mm) diameter were water-cut from planar steel plates of \nicefrac{3}{4}, \nicefrac{1}{2}, \nicefrac{1}{4} and \nicefrac{1}{8}-inch thickness and powder-coated, forming weights of roughly 30, 20, 10 and 5~kg each. A stack with a total weight of 122~kg was then mounted at the end of the stock counterweight shaft and bolted to a custom end piece; one stock counterweight remained for fine balancing, as it could be moved over the remaining shaft length as required. Figure~\ref{fig:ATUS_evolution} (left) showcases the found solution within tight building constraints; note the small clearance to floor level ($\approx$ 4-inch) and to the 9-feet tall gable ($\approx$ 3-inch) when parked pointing to the celestial pole. The setup operated successfully in this non-ideal configuration until the entire counterweight assembly got replaced in March 2021 (see Section~\ref{sec:CW_modification}). 


\subsubsection{Full redesign of the optical tube assembly} \label{sec:OTA}

At first light, image quality of the initial ``Mark I'' OTA was inconsistent and significantly degraded. Early commissioning results and attempts to improve mirror alignment were hinting at an unstable collimation. Moreover, the M2 mechanism did not allow for a reproducible focus adjustment; aside of a severe hysteresis, it introduced unreproducible changes in image quality, suggesting a variable M2 tilt due to a flawed actuator design. A modification of the M1 cell and the M2 focus mechanism on-site in December 2013 (leading to the iterated ``Mk~I.b'' OTA), in collaboration with the manufacturer, did not resolve the issues. 

\begin{figure*}[htb]
\centering 
\includegraphics[width=0.9\linewidth]{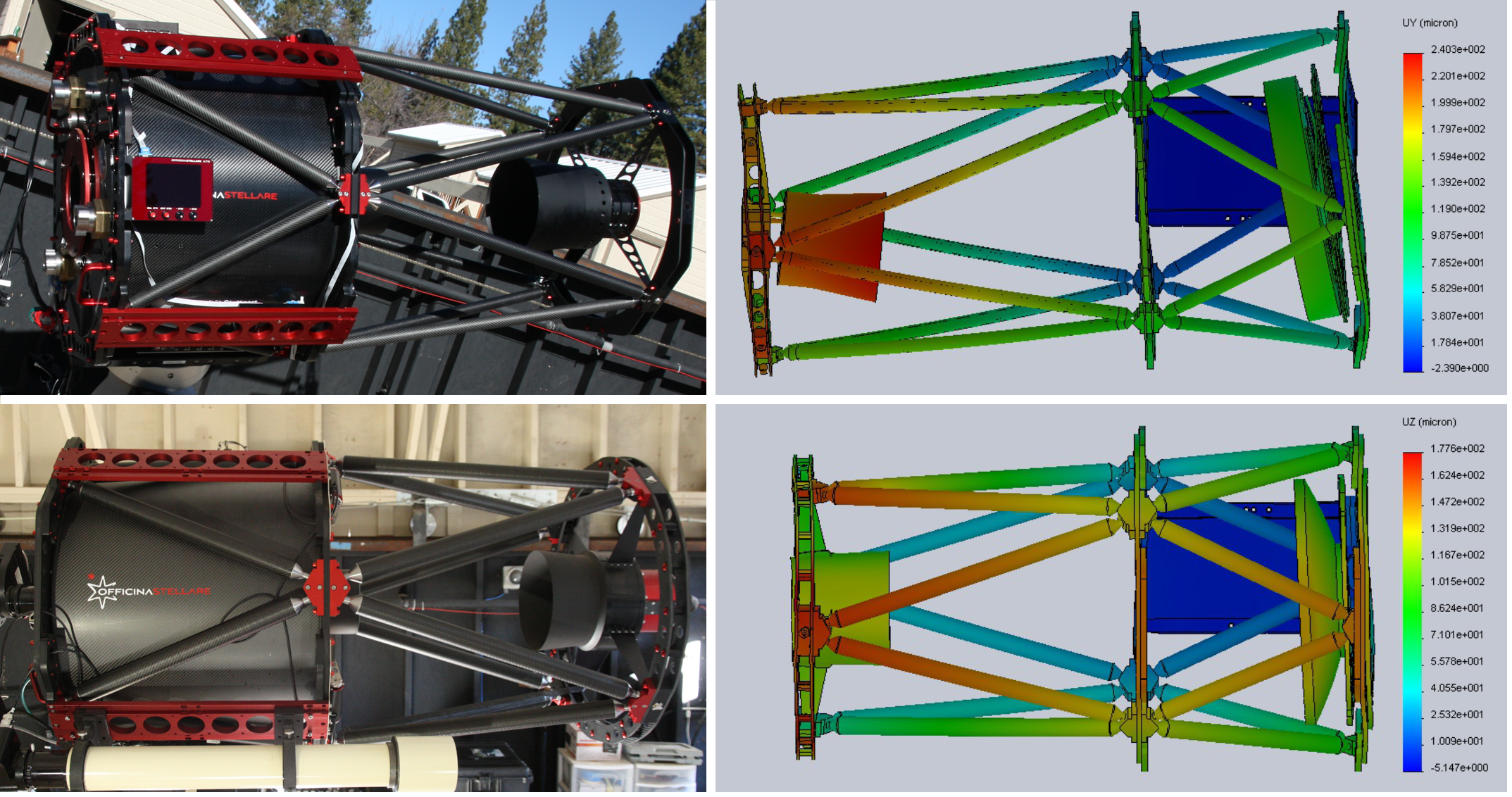} 
\caption 
{Evolution of the ATUS telescope: ``Mark I'' (Mk~I, top, installed at the end of September 2013) and ``Mark II'' (Mk~II, bottom, installed in May 2015). The finite-element analysis (FEA) illustrates the flexure of the telescope structure under its own weight while pointing at the horizon, exaggerated many times over. Lateral misalignment and tilting of the mirrors relative to each other were minimized. FEA plots were included with kind permission of Officina Stellare.
\label{fig:ATUS_flexure}}  
\end{figure*}

This triggered a comprehensive system characterization, suggesting that an inadequate mechanical design caused an excessive amount of tube flexure. The findings prompted the manufacturer to develop an entirely new OTA in close collaboration with the authors, a decision we wish to greatly honor. An iterative design process supported with finite-element analysis (FEA) led to a considerably more rigid telescope structure, which maintained the relative alignment of both mirrors in all directions. \added{At the same time, the weight of the OTA could be reduced by $\approx 10\%$}. A ``Mark II'' (Mk~II) telescope, provided as a replacement at no additional cost, was eventually installed at SRO in May 2015. As it can be seen in Figure~\ref{fig:ATUS_flexure}, and as summarized in Table~\ref{tab:OTA_comparison}, the Mk~II OTA differed significantly in every aspect from the original Mk~I design.  


\added{The primary mirror offered major potential for reducing overall weight and structural loads. In the Mk~I, a 27-point whiffle tree provided axial flotation to a flat-back mirror substrate, while six brackets constrained the mirror radially. Considering that an 18-point whiffle tree supported the SOFIA 2.7-meter M1, this design appeared overly conservative, and radially overconstrained. The mass of the Mk~II's M1 was reduced by about 23\% via a mirror substrate whose back has a narrow flat annulus near the center and tapers toward the edge. As mirror geometry and mount form an integrated system, the shift to a tapered backside dictated a central hub mount. Radial mirror support is applied at the center hole where most of the material, and thus stiffness is located; axial support is provided by a shoulder at the bottom of the hub. The decision to hold M1 at its center was a pragmatic, but not entirely unproblematic choice, as we will discuss later.} 

To counteract tilting moments, the Mk~II's M2 mirror cell has been designed such that its center of mass is coincident with the median plane of the front ring when M2 is at its nominal position --- an obvious, yet seemingly unique feature among similarly sized, commercially available telescopes even today. This also shortened the length of the front truss significantly. A newly selected linear actuator for axial M2 movements enables a much higher resolution, thanks to an extremely fine pitch of its lead screw and its stepper motor with a 0.9\degr~step angle that uses feedback from an optical incremental encoder, able to resolve 8000 pulses per revolution (PPR). Thanks to the encoder feedback, the focus position has excellent accuracy and reproducibility, although a backlash of $\approx\,400$~steps must be accounted for in software. The jointly developed solution for reliable M2 adjustment got eventually incorporated into the manufacturer's product portfolio. 

Diameter and wall thickness of the carbon fiber-reinforced plastic (CFRP) truss tubes were considerably increased, just as panel thickness of the front ring, center ring, and backplane. These panels are hybrid laminate sheets of a glass fiber-reinforced plastic (GFRP) sheet sandwiched by two CFRP sheets. To counteract the increasing structural weight, the truss ball joints were made of a common titanium alloy. 


\begin{figure}[htb]
\centering
\includegraphics[width=\plotwidth]{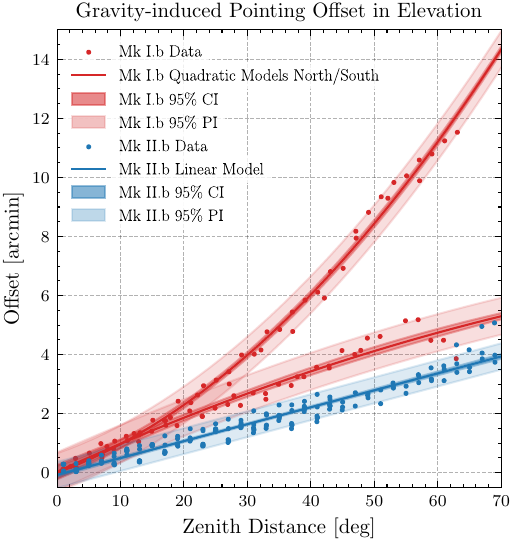} 
\caption 
{Pointing offsets of the Mark-I and Mark-II OTA due to gravity-induced flexure, measured after mirror cell modifications in the respective OTA design (revision ``b''). Images were acquired along the meridian in 2\degr~steps in declination; once in 2014 for the Mk~I.b, and three times in 2022 and 2024 for the Mk~II.b. The shaded areas represent the confidence interval (CI) and the prediction interval (PI) of the respective function fits. See text for more information.
\label{fig:Flexure_Plot} }
\end{figure} 


The Mk~II OTA telescope could now be reliably focused via M2, and image quality appeared consistent over the full field and elevation range. Figure~\ref{fig:Flexure_Plot} aims at a first quantification of this major redesign effort, comparing on-sky measurements of tube flexure of the Mk~I and Mk~II OTA over elevation. The telescope was commanded to point at the meridian in steps of 2\degr~in declination. Orthogonality and alignment errors of a German-equatorial mount cause pointing offsets that overlap with those introduced by instrument flexure, but along the meridian, any gravity-induced pointing offset directly projects itself on declination. The acquired images were astrometrically solved to compare the difference between the commanded, refraction-corrected pointing coordinate \citep[see e.g.][]{Meeus1998}, and the calculated coordinates of the image center (i.e.\ the achieved pointing). The Mk~I design showed large amounts of flexure with a strong North/South asymmetry, likely due to an insufficiently stiff M1 mirror support, in addition to its weak truss (cf.\ Figure~\ref{fig:ATUS_flexure}). In comparison, the Mk~II OTA shows a much reduced and consistent pointing offset of about 4~arcmin~at 70\degr~zenith distance. Although this number turned out slightly larger than predicted by the FEA simulation, it can be easily corrected with a pointing model. However, note that the pointing offset close to the horizon already approaches half of the main camera's FoV. 


\subsection{Image quality} \label{sec:IQ}

A perfectly aligned RC telescope is aplanatic, i.e.\ free of coma and spherical aberration over its entire field. In this ideal case, and assuming perfect mirrors, images would be free of aberrations at the field center, yet present themselves with a rotational-symmetric astigmatism that increases quadratically with field angle \citep{Gitton1998}. Real telescopes are never perfect, yet misalignment-induced aberrations need to be reduced to a level that they are no longer limiting image quality. This milestone was eventually accomplished after further mechanical modifications, correcting mirror spacing, and aligning mirrors with the help of wavefront measurements.


\subsubsection{FWHM as a function of elevation} \label{sec:image_size_vs_elevation}

Seeing degrades with increasing airmass; according to the Kolmogorov turbulence model, the full width at half maximum (FWHM) of the point spread function (PSF) follows $\epsilon_0 = 0.976 {\lambda}/{r_0}$ \citep{Martinez2010a}. The Fried parameter $r_0$ itself depends on wavelength $\lambda$ and zenith angle $z$; for the latter, $r_0 \propto (\sec{z})^{\nicefrac{3}{5}} = 1/(\cos{z})^{\nicefrac{3}{5}}$ \citep[see comprehensive discussions in e.g.][]{Quirrenbach2006,Martinez2010b,WilsonRTO2}. With a properly working telescope, and during stable seeing conditions, this trend should be directly measurable from images taken at different elevations. Images obtained during pointing model runs, sampling the entire sky in azimuth and in elevation, provide an ideal dataset for this endeavor. 

\begin{figure*}[htb]
\centering
\plotone{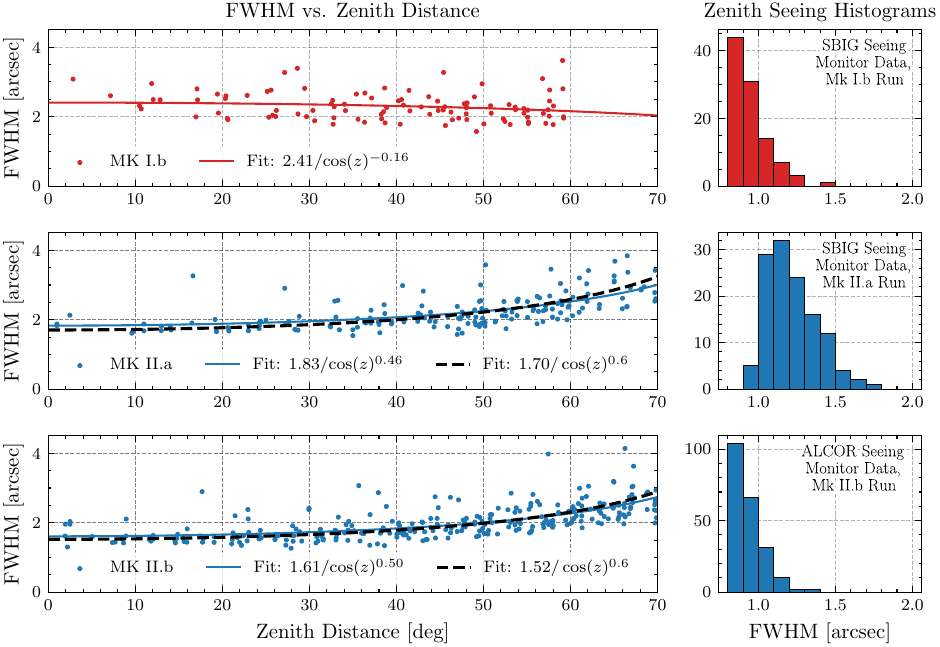} 
\caption 
{Average FWHM values derived from Gaussian-profile PSF fits of images taken at various elevations during pointing model runs in February 2014 (Mk~I.b, n=103, $\geq 30\degr$ el., $t_\mathrm{exp} = 15~\mathrm{s}$), August 2015 (Mk~II.a, n=166, $\geq 20\degr$ el., $t_\mathrm{exp} = 12~\mathrm{s}$), and September 2024 (Mk~II.b, n=265, $\geq 20\degr$ el., $t_\mathrm{exp} = 12~\mathrm{s}$). The histograms on the right quantify the site seeing monitor readings during the time of data acquisition, indicating very good to near-ideal conditions. See text for details.
\label{fig:FWHM_vs_ZD}}
\end{figure*}

The average FWHM of all stellar intensity profiles thus got measured in all images from various pointing model runs. Results are illustrated in Figure~\ref{fig:FWHM_vs_ZD}; seeing conditions during these acquisitions were reasonably stable (see panels on the right side), and exposure times $t_\mathrm{exp}$ sufficiently long to integrate over seeing effects. As seen in the upper panel, the Mk~I.b OTA did not exhibit the anticipated trend at all, despite near-ideal conditions, indicating that its image quality was certainly not seeing limited. It is critical to note that at that time, the Mk~I.b OTA had only been collimated by classic geometric means in its park position pointing at the celestial pole (37\degr~el.). Flexure degraded mirror alignment once the telescope pointed at any different azimuth and altitude. This degraded image quality with increasing elevation, contradicting the trend that image size decreases with airmass. In contrast, both the unaltered, ``as delivered'' Mk~II.a OTA (middle), and the further modified Mk~II.b OTA (lower panel) show the anticipated trend in FWHM over elevation reasonably well. Both Mk~II panels include function fits with the exponent either kept variable (blue) or fixed to its theoretical value (black, $k=0.6$). The entirely data-driven power-law fits are in very good agreement with the expected correlation; this provided confidence that the redesigned Mk~II OTA was able to perform significantly better than the Mk~I, and that the redesign effort had paid off. The minor discrepancy of the exponent may originate, e.g., from strong low-altitude turbulence in the atmospheric ground layer, given the observatory's location on the western slope of the Sierra Nevada. It may also originate from systematic errors in the determined image center coordinates, as atmospheric dispersion increases with airmass, and images during pointing runs were always taken without any bandpass filter.


However, image size still fell short of expectations, triggering further investigations. FWHM measurements corrected for zenith angle had a consistent offset of about 0.7 to 0.8~arcsec to readings from the site's Polaris seeing monitors; a discussion of this discrepancy and final image size follows in Section~\ref{sec:final_IQ}. Of greater concern was that the intensity distribution in defocused star images was asymmetric, hinting at coma due to misalignment and potential other aberrations. Both indicated that optical aberrations were still limiting image quality, and that traditional geometric means to collimate the telescope did not lead to a satisfying result --- at least not at a site where good to excellent seeing conditions prevail.

After several unsuccessful attempts to improve mirror collimation geometrically, and further inspection of the PSF in- and out-of-focus, two issues were identified: a) mirrors were ``pinched'', i.e.\ their cells introduced stresses into both mirror substrates, and b) mirror spacing differed quite substantially from the design value of the optical prescription.


\subsubsection{Quantifying mirror spacing} \label{sec:mirror_spacing}

During commissioning of the Mk~II.a OTA, the nominal flange focal distance (FFD) and astrometrically estimated focal length $f$ did not match the ``as-built'' optical model. An offset in mirror spacing appeared to be a plausible cause, which would also introduce spherical aberration, contributing to image degradation. Thus, the geometrical parameters of the telescope came under scrutiny early on.

A Leica Disto X310 laser distance meter provided a differential measurement of the distance from the M2 vertex to a reference surface above M1 at multiple M2 positions (see Appendix~\ref{sec:app_spacing} for more details). Several measurements were taken and averaged at each M2 position. Readings, displayed at an increment of 0.1~mm, were extremely reproducible, and deviated only on the order of the values provided in the calibration certificate\footnote{A manufacturer-issued calibration certificate lists deviations from $+0.1~\mathrm{mm}$ to $-0.2~\mathrm{mm}$ for reference distances ranging from 0.1055~m to 7.8363~m ($\pm 2 \sigma$ statistical confidence level, $23 \pm 3 \degr \mathrm{C}$ temperature, against a surface with maximum diffuse reflectivity, i.e.\ target plate albedo of 1).}. Measurements with a calibrated Mitutoyo digital caliper then linked the position of the reference surface to the M1 vertex and to the OTA's backplane, allowing calculation of mirror spacing. Star field images were then taken at identical M2 positions using a temporarily installed, computer-controlled rack-and-pinion focuser with a large travel range, and the QSI-632ws8 imaging camera from the refractor. At each M2 position, the camera would then be focused via the focuser at the backplane, the distance from the camera's face plate to the backplane measured, and images astrometrically solved, allowing to determine focal length  from the estimated plate scale. Knowing the CCDs back focus position inside the camera housing, and accounting for optical path length differences caused by filters and windows in the converging beam, the optically relevant back focus distance (BFD, measured from the M2 vertex) and the mechanically important FFD can then be calculated. 

\begin{figure}[htb]
\centering
\includegraphics[width=\plotwidth]{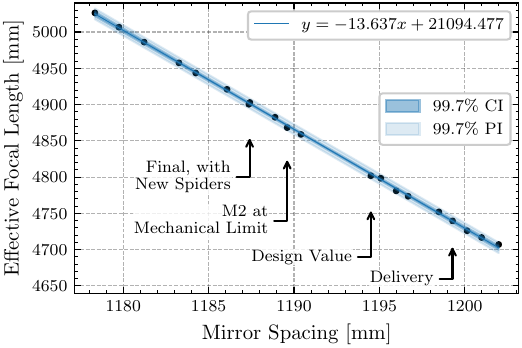} 
\caption 
{Effective focal length as a function of mirror spacing, established via geometrical measurements, exhibiting excellent linear correlation with a coefficient of determination of $R^2=0.99961$ (CI: Confidence Interval; PI: Prediction Interval). Mirror spacing was initially reduced to its nominal design value after an offset had been confirmed. Wavefront measurements with a Shack-Hartmann sensor later established the relationship of spherical aberration as a function of mirror spacing (see Figure~\ref{fig:SphericalAberration}), and eventually allowed ATUS to operate at its spherical-aberration-free point after replacing its M2 spiders. See text for more details.
\label{fig:f_vs_s}}
\end{figure}

Such measurements were initially conducted on the ``as-delivered'' Mk~II.a OTA, i.e.\ before modifying its mirror cells (see next section), and were then routinely repeated on various occasions, i.e.\ after mechanical modifications of the OTA or the image train. Additional data points were obtained from the main camera's spacing to the backplane, its images, and by measuring the mirror spacing at which the main camera obtained best focus during previous night time observations. Figure~\ref{fig:f_vs_s} illustrates how consistent these measurements are, and provides a linear regression model of the astrometrically derived focal length vs.\ the calculated mirror spacing. The arrow markers depict various spacings at which ATUS successively operated until spherical aberration was eventually eliminated (see also next section). A welcome side-effect of this effort was the independent verification of the M2 stepper motor resolution, which was initially limited by a firmware bug. Mirror spacing and focal length as functions of FFD can be found in Appendix~\ref{sec:app_spacing}, Figure~\ref{fig:FFD}.


While it quickly became clear that the telescope was not operating at its spherical-aberration-free point as specified by its optical design, image quality did not notably improve after adjusting the mirror spacing to its design value. Wavefront measurements played a crucial role to confirm the underlying issues, and eventually to find the right mirror spacing and alignment, as described in the next two sections.


\subsubsection{Optimizing mirror mounting, spacing, and alignment} \label{sec:optimizing_mirror_alignment}

After star testing had qualitatively hinted at pinched mirrors \citep{Suiter2009}, the ``Mark-II'' OTA received a newly designed and manufactured M1 cell, while its M2 cell got reworked --- again with direct support from the manufacturer, which the authors would like to gratefully acknowledge once more. \added{The iterated center hub design uses a stack of circumferential O-rings separated by spacer rings; a threaded flange compresses the stack axially and causes the O-rings to bulge into contact with the mirror substrate. The elastomer's compliance provides thermal decoupling between the aluminum hub and the mirror, while the radial pressure from the compressed O-rings provides the required radial constraint and tilt stiffness.}

\begin{figure*}[htb]
\centering
\includegraphics[width=0.95\linewidth]{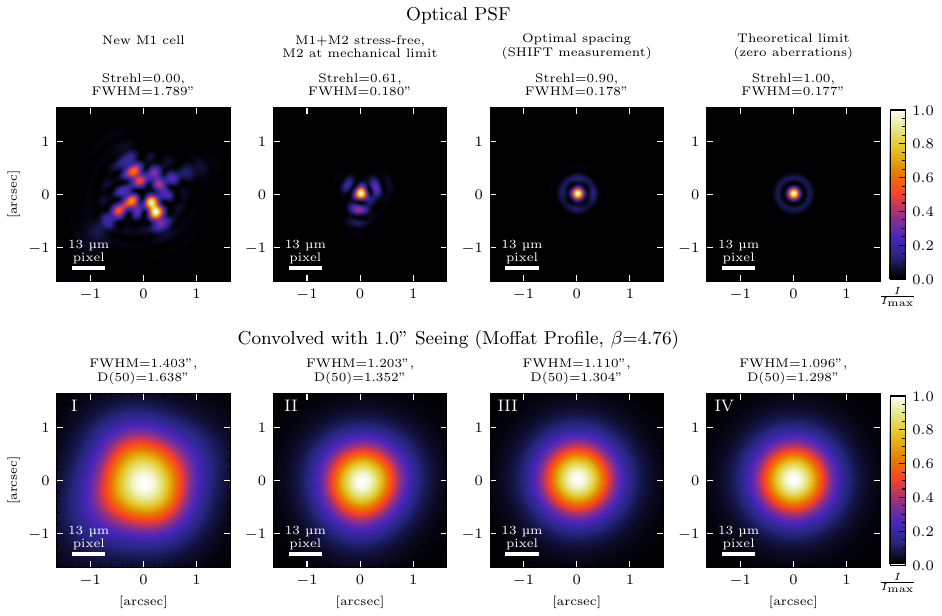} 
\caption 
{The optical point spread function (PSF) of the Mk~II.b OTA, reconstructed from Shack-Hartmann wavefront measurements conducted at different evolutionary stages (top row), and convolved with a 1.0~arcsec seeing represented by a Moffat profile (bottom row). Each panel has a size of 3.3~\texttimes~3.3~arcsec\textsuperscript{2}, i.e.\ 6~\texttimes~6 pixels. Radial profiles of the seeing-convolved PSFs are illustrated in Figure~\ref{fig:ATUS_PSF_radial_profiles}. See text for more details.
\label{fig:ATUS_PSF}}
\end{figure*}

We refer to this mechanically iterated design as the ``Mk~II.b''. Shack-Hartmann wavefront measurements, conducted successively after the M1 and M2 mirror cell modification, ultimately confirmed that both mirrors originally had stresses introduced by their respective original mirror cell, and that the mechanical modifications were effective in eliminating them (cf.\ Figure~\ref{fig:ATUS_PSF} and Section~{\ref{sec:final_IQ}}). This turned out to be a crucial step not only towards optimal image quality, but also towards fully reliable pointing, as we will later discuss in Section~\ref{sec:pointing}. With the stress-induced aberrations removed, and both mirrors collimated (see details in Section~\ref{sec:ll_collimation}), the discrepancy between FWHM measurements and seeing monitor readings shrunk to about 0.3 to 0.4~arcsec.


\begin{figure}[t]
\centering
\includegraphics[width=\plotwidth]{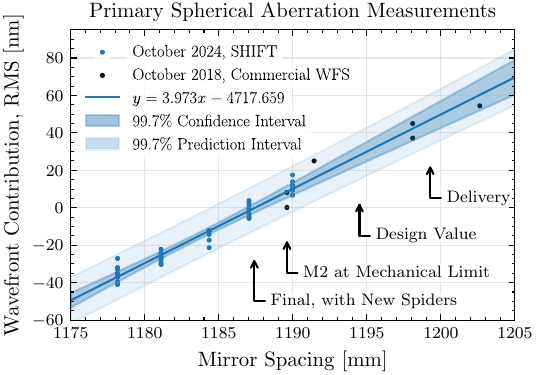} 
\caption 
{Correlation between the Zernike coefficient representing the \added{primary} spherical aberration's RMS wavefront contribution and mirror spacing. Due to a limited travel range of M2, the spherical-aberration-free point was initially extrapolated from wavefront measurements with a commercial Shack-Hartmann wavefront sensor in 2018. A verification and validation campaign of the SHIFT instrument, a wavefront sensor conceived and built for SOFIA (see Section~\ref{sec:SHIFT}), provided an opportunity to validate the optimal mirror spacing after a set of new spider vanes was integrated. See text for more details.
\label{fig:SphericalAberration}}
\end{figure} 


Being able to reference the commanded M2 position to a geometrical mirror distance (see previous section), wavefront measurements now also allowed to identify a linear trend of spherical aberration with mirror spacing (see Figure~\ref{fig:SphericalAberration}). Unfortunately, the estimated spherical-aberration-free point was outside the travel range of the M2 actuator. It was thus decided to move and fix M2 near its mechanical limit, reducing the remaining mirror spacing error to about 4~mm. As a workaround, a mechanical focuser got permanently installed at the telescope's backplane. The M1 tip/tilt mechanism provided an additional axial travel of 2~mm, further reducing the spacing offset to about 2~mm. 

To restore the ability to focus via M2, the manufacturer agreed to fabricate new spider vanes, shifting the entire M2 assembly closer to M1. By then, a Shack-Hartmann test instrument and software package for wavefront analysis was already under development at DSI for SOFIA (SHIFT; see Section~\ref{sec:SHIFT}). With ATUS as a testbed, the SHIFT instrument underwent first light, verification and validation once the new spider vanes were installed. Measurements were repeated at various M2 positions in larger quantity, and confirmed the trend  previously found with a commercial wavefront sensor (WFS) in 2018. With the new SHIFT measurements, the optimal mirror spacing was estimated at 1187.4~mm; an offset of 7.6~mm to the design value, and of almost 12~mm to the configuration at delivery. With an adjusted image train accommodating the larger FFD, ATUS can now operate free of spherical aberration, while a meaningful M2 travel range has been restored to focus the main camera.

Knowing this additional system parameter, the radii and conical constants of both mirrors can be calculated. We found that the radii match their design values extremely well, but that the conical constant of M2 appears to be about 5\% off. While this does not impact image quality, the corrected mirror spacing increased the FFD by about 100~mm, shifting the heavy instrument train significantly further back than initially planned, and leaving only a few millimeters of ground clearance on our pier at SRO when delaying or advancing the meridian flip. 


\clearpage

\subsubsection{Final image quality}\label{sec:final_IQ}

To accomplish optimal sensitivity, \added{misalignment-induced aberrations need to be controlled such that ATUS operates practically diffraction-limited, while the imaging detector needs to provide critical sampling of the seeing-convolved, diffraction-limited PSF.} The site was advertised with typical seeing values of 1.5 -- 2.0~arcsec during winter, and 1.0 -- 1.5~arcsec during summer months.

\begin{figure}[htb]
\centering
\includegraphics[width=\plotwidth]{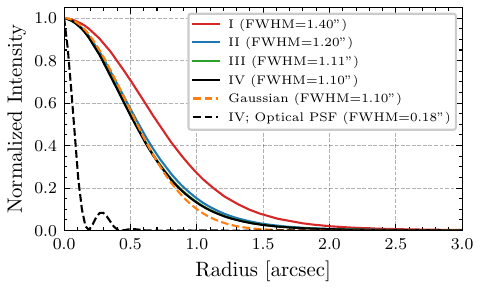} 
\caption 
{Radial profile plots of the four optical PSFs convolved with 1.0~arcsec seeing, as shown in the bottom row of Figure~\ref{fig:ATUS_PSF}. For direct comparison, the plot also includes a Gaussian profile with identical FWHM as the aberration-free, seeing-convolved PSF (orange dashed line; neglecting effects of the four spider vanes), as well as the aberration-free ideal optical PSF (black dashed line) that has a considerably lower peak intensity, and significantly more energy in its first diffraction ring than the PSF of a system with a small or absent central obscuration.
\label{fig:ATUS_PSF_radial_profiles}}
\end{figure}

Wavefront measurements obtained during iterations of the Mk~II OTA and to eliminate spherical aberration allow us to discuss the shape of the PSF; for this, we would like to revisit Figure~\ref{fig:ATUS_PSF} in greater detail. The top row illustrates the reconstructed optical PSF without atmospheric distortion (a) after integration of the new M1 mirror cell, (b) after eliminating stresses on both mirrors and operating the system with an improved, yet not optimal mirror spacing, (c) after achieving the optimal mirror spacing and carefully compensating coma via mirror tilt, and in comparison, (d) the theoretical limit of the system without any aberrations present. The bottom row convolves the optical PSF with a 1.0~arcsec seeing modeled by a Moffat function \citep{Trujillo2001,Moffat1969}. ATUS has a rather large central obstruction (cf.\ Table~\ref{tab:OTA_comparison}). Compared to a system with a small or even absent obstruction, the PSF has a significantly reduced peak intensity, has more energy in the first diffraction ring, and is of much broader extent (cf.\ Figure~\ref{fig:ATUS_PSF_radial_profiles}). Because of this, but also due to the Moffat seeing distribution, the often found ``rule of thumb'' of FWHM $\approx$ D(50) \citep[e.g.][]{Erickson2000} becomes inaccurate. The rather large central obscuration also negatively affects the Modulated Transfer Function (MTF), and thus impacts contrast; as ATUS is essentially built to observe unresolved point sources, this issue is not relevant for its use cases and beyond the scope of this paper, but should be kept in mind. Albeit small, the M2 spider vanes' impact on image quality has also been neglected in this discussion; both topics are addressed e.g.\ in \citet{Harvey1995}.


A Strehl ratio of \textgreater 80\% is widely recognized as a practical metric for an optical system operating diffraction-limited. While the reconstructed PSFs shown in Figure~\ref{fig:ATUS_PSF} are only based on the coefficients of the first 11 annular Zernike polynomials \citep{Mahajan1981}, i.e.\ neglect higher order terms of the reconstructed wavefronts, ATUS practically operates at the diffraction limit. It must be stressed that this could not have been accomplished without a Shack Hartmann sensor supporting mirror collimation. To our experience, the RC system is far too sensitive to minor mirror misalignment to accomplish a diffraction-limited PSF by a purely geometrical collimation procedure.

For a Gaussian PSF, sampling should be equal to its standard deviation describing its width; this corresponds to 2.355~pixels per FWHM \citep{Howell2006,Ofek2020}. For ATUS, this translates to a critical sampling at an FWHM of 1.3~arcsec, i.e.\ a lightly undersampled PSF during good summer nights. This appeared as an acceptable trade-off, considering the sensor's very narrow FoV. Little surprising, and confirmed by our practical experience with the system, FWHM measurements from images floor at 1.3~arcsec. Smaller PSFs are undersampled and can not be reliably measured via radial profile fits, i.e.\ during the SRO summer months, our image resolution was neither aberration- nor seeing-, but rather detector-limited; for the remainder of the year, it was appropriate.

Even with sufficiently small pixels, one should still be cautious about deducing seeing conditions from FWHM measurements in images: \citet{Martinez2010a,Martinez2010b} refuted the common assumption that the PSF's FWHM obtained from long exposures corresponds to the seeing, which usually gets measured with a Differential Image Motion Monitor (DIMM) as the ``gold standard''. It neglects the outer scale of the atmospheric turbulence, i.e.\ an FWHM measurement typically underestimates the true seeing, and overestimates a telescope's image quality. Thus, even if ATUS had sufficient detector sampling, a true assessment of seeing conditions at SRO would still not be that simple. Also, the site uses fixed Polaris seeing monitors instead of a DIMM; while these seeing monitors already indicate \added{airmass-corrected Zenith} seeing in FWHM, \added{such} values may as well underestimate the true seeing. 


\subsubsection{Elimination of stray light}
\label{sec:stray_light_elimination}

Another issue that plagued ATUS during much of its operational lifetime was a rotationally symmetric, central brightening visible in flat fields. This issue was present both with the Mk~I and Mk~II OTA. 

\begin{figure*}[tb]
\centering
\includegraphics[height=2.2in,valign=t]{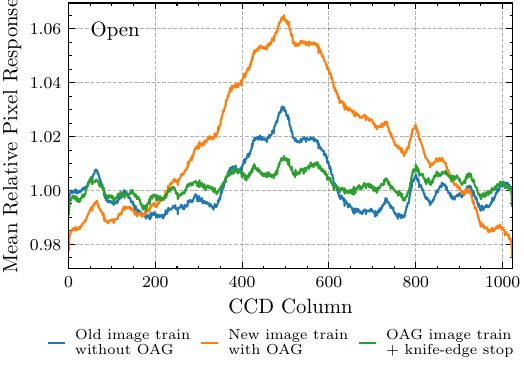} 
\hspace{0.2cm}
\fcolorbox{TableauOrange}{white}{\includegraphics[height=1.75in,valign=t]{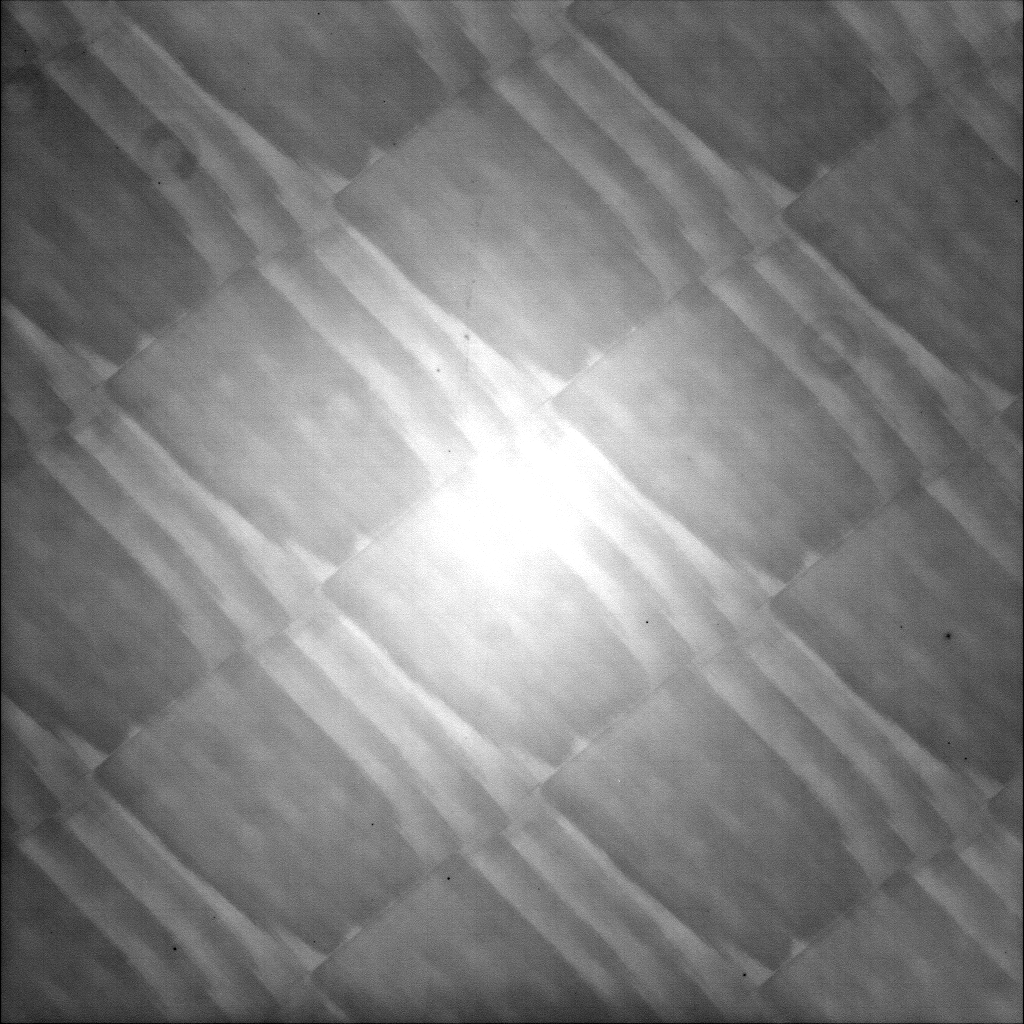}}
\hspace{0.2cm}
\fcolorbox{TableauGreen}{white}{\includegraphics[height=1.75in,valign=t]{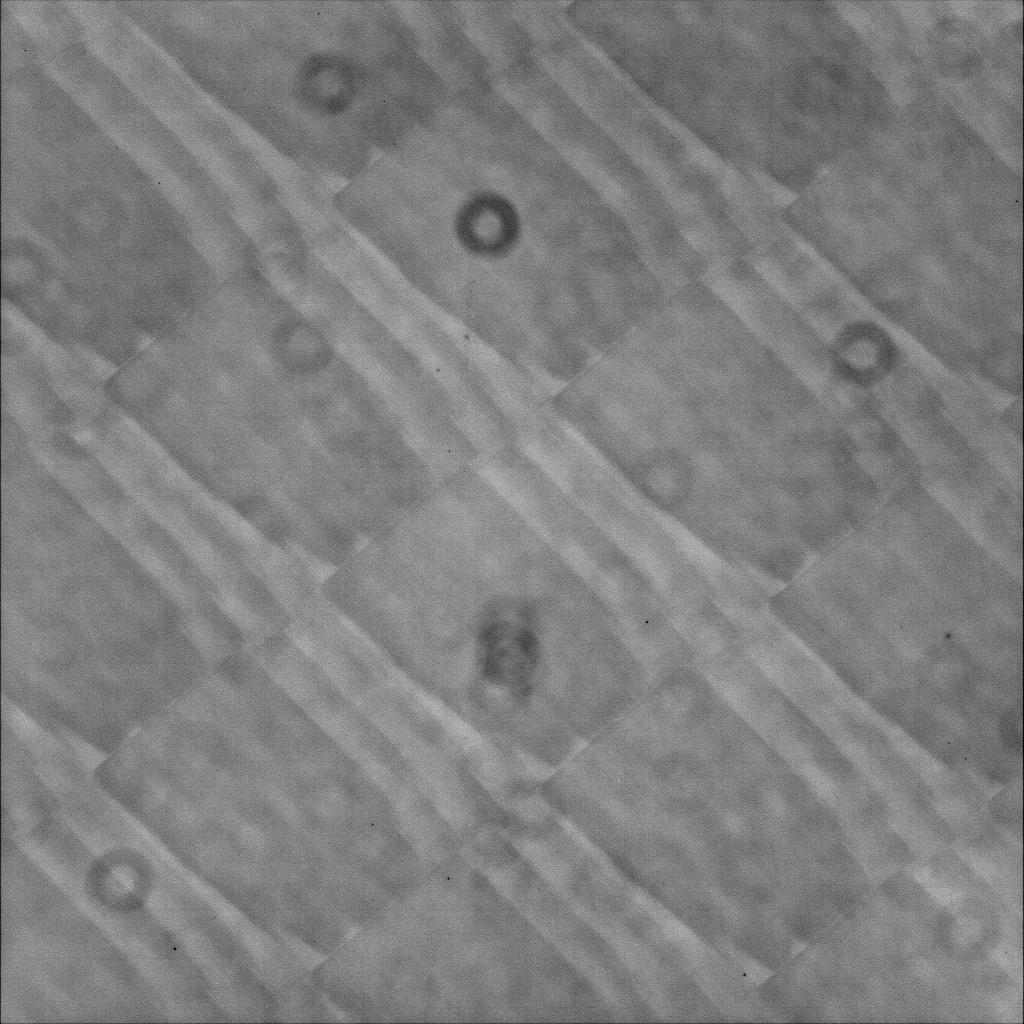}}

\caption 
{Horizontal line profiles of flat fields, averaged over sensor rows \#501-524, i.e.\ centered around the image center. Flat fields were acquired without a filter (open position) during twilight conditions, opposite in azimuth of the rising/setting sun. Flat fields revealed a stray light reflection appearing as a central bright spot (blue curve), which intensified as the image train interface diameter was increased (orange curve \& frame). A knife-edge stop above the EMCCD (cf.\ Figure~{\ref{fig:stray_light_aperture}}) resolved the issue (green curve \& frame). Both flat fields were scaled to a relative pixel response range of [0.95 -- 1.05]. The diagonal structure ($\approx \pm 3\%$) is a result of a laser annealing process of back-illuminated sensors, and caused by photons at the blue end of the spectrum. 
See text for more details.
\label{fig:flatfields}}
\vspace{1em}  
\end{figure*}

The camera/filter-wheel package was initially mounted to the telescope backplane flange via a stack of Astro-Physics 2.7-inch \texttimes~24~tpi extension rings, providing a sufficiently large clear aperture (CA) of 2.5-inch (63.5~mm). With the design and integration of the OAG (see Section~\ref{sec:OAG}), the CA of the backplane interface had to be notably increased to redirect light at the outer edge of the imaging field via a pick-off prism into a separate camera. A combination of Takahashi M98 extension tubes (CA $\approx$ 95~mm) and Optec-DSI\footnote{Optec uses the acronym DSI to refer to telescope manufacturer Deep Sky Instruments, not to the German SOFIA Institute.} 3.5-inch \texttimes~24~tpi spacer rings (CA~=~3.25-inch, or 82.5~mm) were used to accommodate the diameter of the converging beam (cf.\ Figure~\ref{fig:OAG+Image_Train}). Once the new image train was installed, the central bright spot in flat fields was even more pronounced (see Figure~\ref{fig:flatfields}). All internal surfaces were black and provided diffuse reflection, making it unlikely that something on the inside of the redesigned image train was the cause.


\setlength{\fboxrule}{2pt}
\setlength{\fboxsep}{0pt}

\begin{figure*}[htb!]
\centering

\newsavebox{\myimagebox}
\newlength{\imgwidth}
\newlength{\imgheight}
\savebox{\myimagebox}{\includegraphics[height=8.5cm]{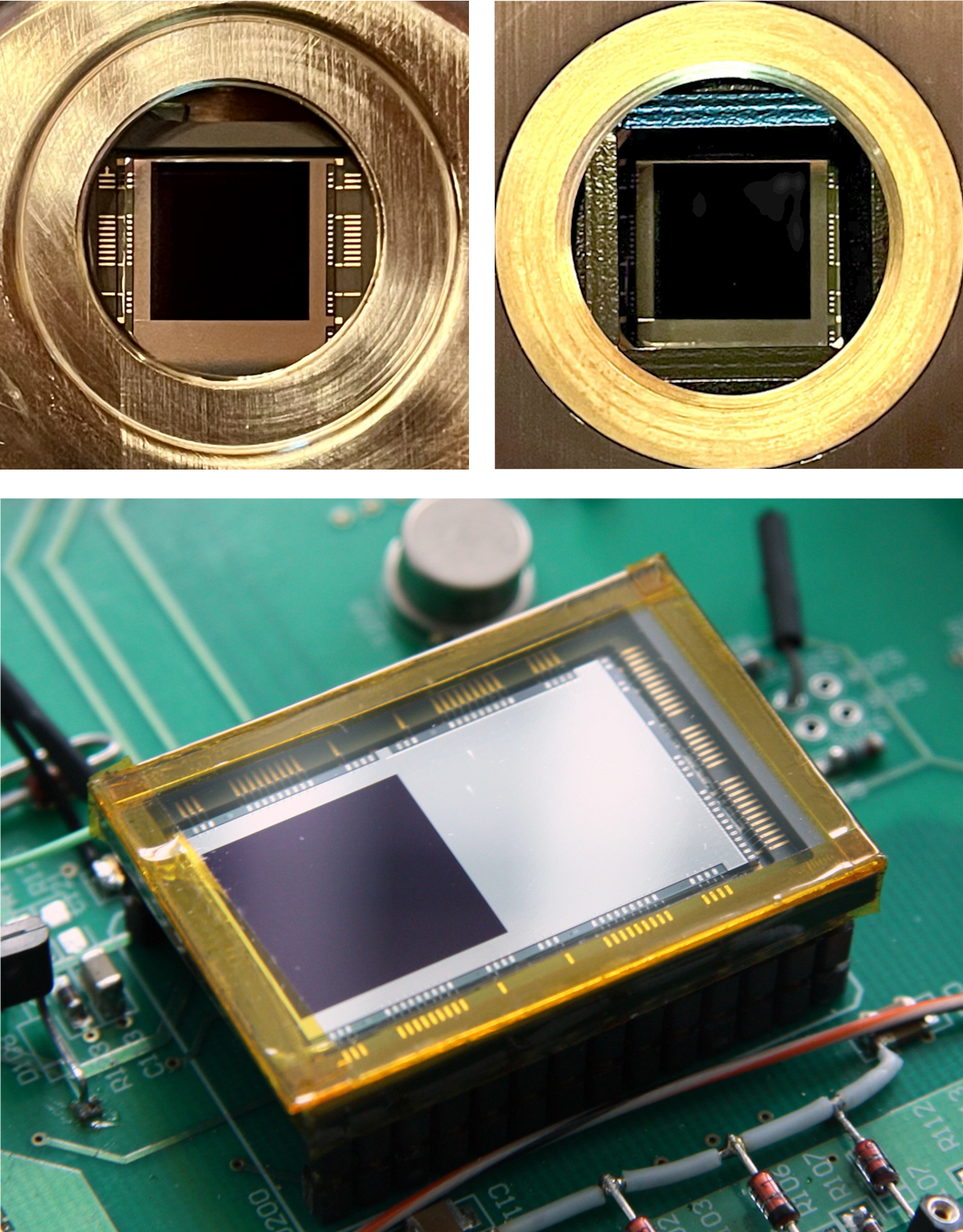}}
\settowidth{\imgwidth}{\usebox{\myimagebox}}
\settoheight{\imgheight}{\usebox{\myimagebox}}
\begin{tikzpicture}[baseline=(image.center)]
  \node[anchor=south west, inner sep=0] (image) at (0,0) {\usebox{\myimagebox}};
  \node[anchor=north west, text=white, font=\bfseries] 
    at ([shift={(0.012\imgwidth, -0.012\imgheight)}]image.north west) 
    {\contour{black}{\textcolor{white}{(a)}}};
  \node[anchor=north west, text=white, font=\bfseries] 
    at ([shift={(0.512\imgwidth, -0.012\imgheight)}]image.north west) 
    {\contour{black}{\textcolor{white}{(b)}}};  
  \node[anchor=north west, text=white, font=\bfseries] 
    at ([shift={(0.012\imgwidth, -0.418\imgheight)}]image.north west) 
    {\contour{black}{\textcolor{white}{(c)}}};
\end{tikzpicture}%
\hspace{0.2cm}%
\includegraphics[valign=c]{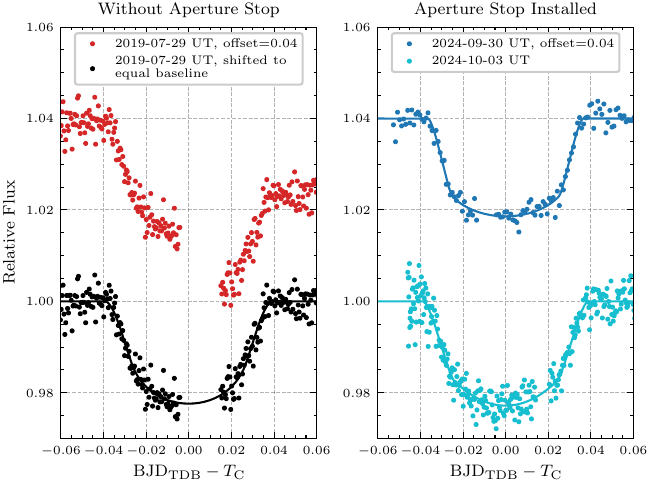} 

\caption 
{Left: (a) View of the EMCCD's active image area through the sensor-chamber window, with the camera's front plate removed. (b) A 17.5~mm square knife-edge stop, mounted on the inside of the camera's front plate, sits concentric to the active image area directly above the window. It suppresses stray light reflections originating from the EMCCDs metalized surface areas and surrounding bond pads. (c) The e2v CCD201-20 EMCCD on an unrelated prototype board. The metalized mask shields the frame-transfer sensor's image storage section and dark reference columns from light, yet resembles a perfect mirror. After installation of the quadratic stop, flat fields did no longer exhibit a central bright spot (cf.\ Figure~\ref{fig:flatfields}). Right: The improvement in data quality is demonstrated by transit light curves of the extrasolar planet TrES-5 b that required a meridian flip. Because of systematics caused by stray-light contamination, data points collected before and after the flip would usually not align and had to be shifted based on baseline offsets. Once stray light reflections were suppressed, the data aligned instantly. All three transits of TrES-5 b shown here were measured without interfering moonlight. 
\label{fig:stray_light_aperture}}
\end{figure*}
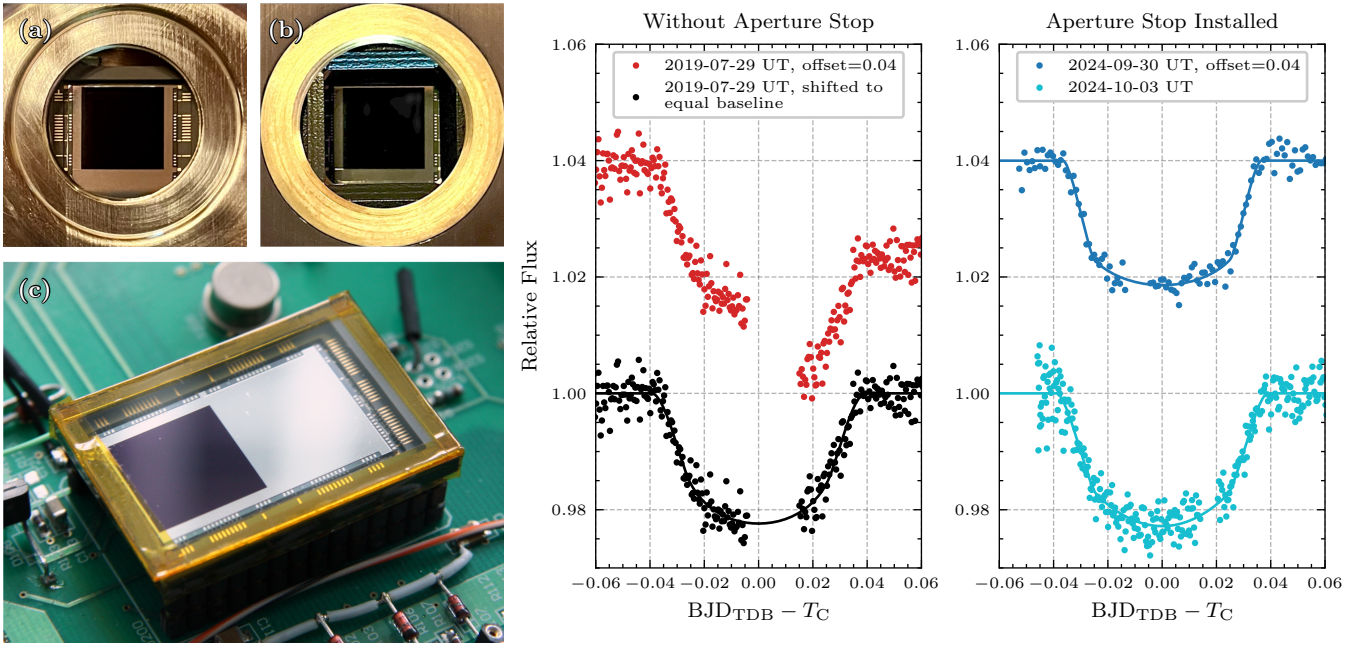 


Off-axis light rays that were previously cut off now passed through the system. The problem was even more puzzling given that the filter wheel, bolted to the camera's front plate, used 1.25-inch (31.75~mm) diameter filter cells (CA $\approx$~26~mm) in M28.5 threads (= CA in open position), acting as an aperture stop. The metalized mask covering the EMCCD's frame-transfer storage area and dark reference columns was eventually identified as the culprit. The deposited, thin aluminum film resembles a perfect mirror and reflects off-axis light rays, which find their way onto the active sensor area via rereflections. The solution was a 17.5 \texttimes~17.5~mm\textsuperscript{2} square knife-edge stop mounted directly above the camera's sensor-chamber window (cf.\ Figure~\ref{fig:stray_light_aperture}), effectively eliminating the central bright spot in flat fields. Having this source of stray light suppressed, flat-field systematics and image calibration errors were significantly reduced. This was impressively demonstrated by exoplanet transit light curves that no longer presented themselves with a relative flux offset after a meridian flip got executed.

In this context, it must be noted that ATUS is extremely sensitive to stray light originating from the moon or the surroundings once the fabric front truss shroud is removed. This raises questions if a closed tube design would have been more practical, which is further discussed in Section~\ref{sec:ll_closedtube}.

\subsubsection{Sensitivity}

\begin{figure*}[htb]
\centering
\plotone{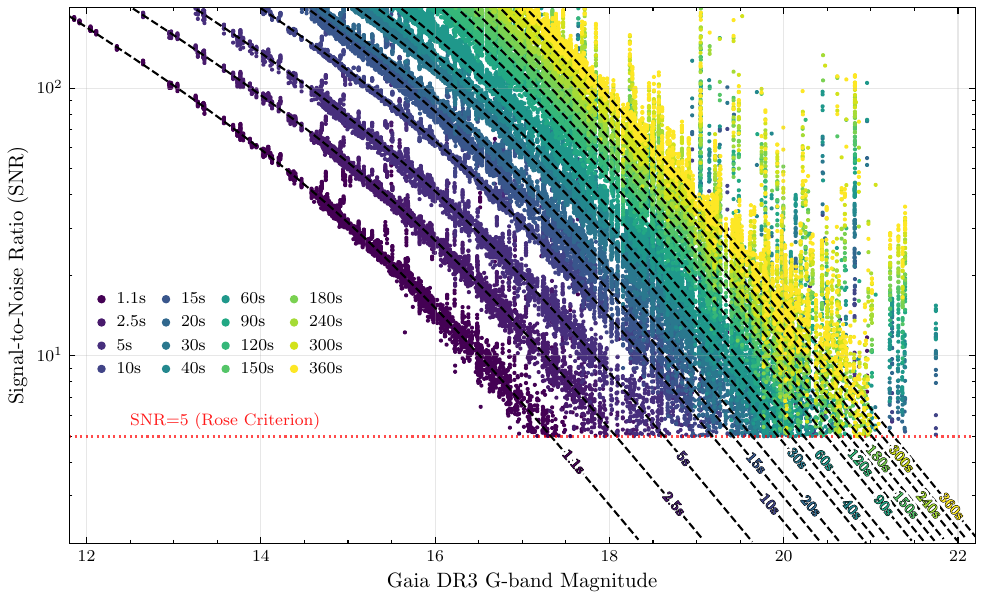}  
\caption 
{Photometric sensitivity as of September 2024, derived from images of the open star cluster NGC~188, and conservatively scaled to 1.5~arcsec seeing. Measurements were acquired without any filter and at various integration times. As atmospheric and seeing conditions can vary greatly, and stars have different spectral characteristics, this plot can only provide a first-order exposure time estimate for a desired SNR at a given stellar magnitude. Considering the Rose criterion as the detection limit (SNR = 5), the limiting magnitude in a 360~s exposure is G=21.3~mag. See text for more information. 
\label{fig:ATUS_sensitivity}}
\end{figure*}

Figure~\ref{fig:ATUS_sensitivity} illustrates the system's sensitivity, measured without a filter on field stars in the open star cluster NGC~188. The EMCCD was read at 1 MHz via the conventional amplifier (2.5\texttimes~pre-amp gain), providing the lowest possible read noise (see Table~\ref{tab:readout_modes}). The fitted functions model the signal-to-noise ratio (SNR) at each exposure time following the ``CCD equation'' \citep{Howell2006}, i.e.\ consider Poisson-distributed shot and background noise as well as Gaussian read noise. Sources got detected and extracted using the \texttt{DAOStarFinder} routine of the \texttt{photutils} package, then cross-matched to the Gaia DR3 catalog using \texttt{astroquery} and \texttt{astropy}. The plot combines measurements obtained in September 2024 (15~s to 360~s exposures), shortly after a cleaning of the primary mirror\footnote{\added{CO\textsubscript{2} snow proved mostly ineffective given the sticky contaminants and high-humidity conditions present at SRO. We found that a heavily diluted mixture of a tiny amount of soap and very little 99.9\% electronics-grade isopropyl alcohol in distilled water worked best. The mirror got sprayed with distilled water first, then with the heavily diluted soap-solvent mixture, then generously rinsed with distilled water, and carefully blotted dry with Kimtech Kimwipes. Blotting without any wiping motion was found to be the only effective method for removing sticky contaminants.}}, and August 2015 (1.12~s to 60~s exposures), after roughly three months of operations of the Mk~II.a OTA. In both cases, observing conditions were nearly ideal, with an average zenith seeing of 1.06 and 1.34~arcsec reported by the site's Polaris seeing monitor, and no moon light interfering during data acquisition. 


Although image quality was notably improved over the lifetime of the instrument (see Section~\ref{sec:IQ}), SNR estimates from 2024 measurements were only marginally larger than estimates from 2015 images of identical exposure time. \added{A slight degradation of mirror coating reflectance and instrument throughput after about 9.5 years of operations in the field could possibly explain that.} 


\subsection{Pointing and tracking improvements}
\label{sec:pointing}
Pairing a small detector with a long focal length makes ATUS a rather unique system in its size class. A pointing error \added{in excess of} 4.7~arcmin (cf.\ Table~\ref{tab:imagers}) places a target already outside its FoV. Corrections from a pointing model were essential for reliable target acquisitions\added{ and unguided tracking,} enabling routine operations. The counterweight assembly and overall mass distribution were iterated to eliminate settle time and excitation at faster slew and tracking speeds. A custom OAG was designed and commissioned to enable arbitrary exposure times. Today, ATUS can reliably point with sub-arcminute accuracy, collect ultra-deep exposures, and even track LEO satellites without image deterioration. The following sections describe the journey to this achievement.

\subsubsection{Importance of the pointing model}
The mount's control and pointing model software \citep[\texttt{APCC Pro},][]{APCC} essentially employs two separate models for telescope positions west and east of the pier. Images are taken at dozens to hundreds of points (a ``pointing model run''), typically spaced equidistantly in hour angle and declination, and get astrometrically solved. The measured offsets between commanded and solved positions are then used as data points for an empirical model, attempting to estimate not only the pointing offset, but also tracking rate corrections at any given sky coordinate. Thus, the declination axis also makes very slow movements during sidereal tracking --- something that actually voids the basic idea of a German-equatorial mount, yet works in practice.


\begin{figure}[b]
\centering
\includegraphics[width=\plotwidth]{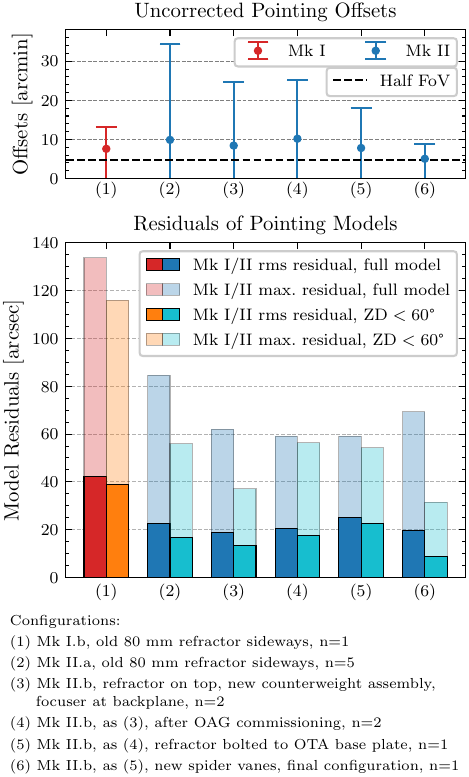} 
\caption 
{Top panel: \added{Measured} median (dot) and maximum (cap) pointing \added{offsets} \added{without pointing model corrections. These offsets serve as inputs for generating the pointing model.} Bottom panel: Maximum and root-mean-square (rms) \added{residuals} between modeled and measured pointing offsets, estimated from model points spanning the entire elevation range of operations (red/blue), and limited to points at a zenith distance (ZD) of less than 60\degr~(orange/cyan). 
All \added{data} were derived from \added{$n$ pointing model runs} conducted \added{across} various system configurations as ATUS evolved. See text for more details.
\label{fig:PM_evolution}}
\end{figure} 


After first light with the Mk~I OTA, the polar alignment error was minimized via the drift alignment method \citep[attributed to the German astrophysicist][]{Scheiner1892,Scheiner1897}. Pointing \added{offsets} of the Mk~I OTA were difficult to model due to complex flexure modes (cf.\ Figure~\ref{fig:PM_evolution}, lower panel), and plagued by sudden \added{shifts} originating from mechanical play in the M2 focus mechanism. The installation of the Mk~II\added{.a} OTA introduced a significant non-orthogonality of the OTA's optical axis relative to the declination axis, which considerably increased pointing offsets (Figure~\ref{fig:PM_evolution}, upper panel), yet error modeling and operations improved drastically. 

However, while pointing and tracking corrections were effective prior to a meridian flip, the pointing model degraded significantly after the telescope changed pier side, placing the target coordinate outside the FoV and requiring the observer to reestablish the zero-point for the model (``recalibration''). This was done by astrometrically ``blind-solving'' an image acquired after the slew, and using its center coordinate as a new reference point (see next section). A second slew would then successfully point to the desired target coordinate. This issue was traced to mechanical play in the Mk~II.a center hub support \added{(cf.\ Section~\ref{sec:optimizing_mirror_alignment})}. The pier-side-dependent pointing models were unable to model this erratic behavior, yet behaved acceptably after a recalibration --- though not at the levels ultimately achieved (Section~\ref{sec:tracking_pointing_performance}). As the model does not capture this play, the errors provided for the Mk~II.a in Figure~\ref{fig:PM_evolution} are not fully representative. The issue was eventually resolved with the installation of a new primary mirror cell. With the Mk~II.b OTA, ATUS could finally make a meridian flip without getting ``lost'' --- time-consuming reference-point calibrations into the respective sub-model were eliminated, and reliable all-sky pointing was finally accomplished. 

\added{Uncorrected pointing offsets} significantly exceeded half of the FoV \added{across} large parts of the sky during much of ATUS's lifetime. Only in its final configuration (labeled \#6) --- \added{incorporating} all improvements discussed in this article --- \added{did} roughly half of \added{the coordinates targeted} during a pointing model run \added{fall within} the FoV. \added{Notable decreases} in \added {uncorrected offsets (cf.\ Figure~\ref{fig:PM_evolution}, top panel)} trace to major modifications affecting the optical axis and its perpendicularity to the declination axis: removal, modification, and reintegration of both mirror cells with subsequent recollimation (\#2 $\rightarrow$ \#3); dismounting the OTA and removing its dovetail plate to add threaded holes for new mounting brackets (\#4 $\rightarrow$ \#5); and removal of the M2 mirror cell and its reintegration with new spider vanes (\#5 $\rightarrow$ \#6).


\subsubsection{Overcoming early software limitations} \label{sec:pointing_MK1}

Initially, the pointing model creation routine embedded in the mount's control software was not geared towards such small FoVs. Right after each image acquisition at a given sky coordinate, an astrometric plate solve was attempted using a classic triangle match algorithm \citep{Groth1986} via a third-party commercial software package \citep[\texttt{PinPoint},][]{PinPoint}. While this algorithm yields accurate solutions when given a reasonable initial guess, it requires the commanded center coordinate to be within the field --- a condition only met by a few dozen points at the beginning of a pointing model run, after establishing a reference point near the zenith. Already with the Mk~I OTA, pointing \added{offsets} exceeded this tolerance across much of the western sky, and essentially the entire eastern sky. Extending the plate solver's catalog search area offered little improvement, yet occasionally produced erroneous solutions skewing the sparsely sampled model further. The resulting incomplete models were impractical for ATUS. 

With the \texttt{astrometry.net} algorithm \citep{Lang2010}, images could get reliably ``blind solved'' regardless of their initial pointing \added{offset}. Originally developed for Linux, a \texttt{Cygwin} \citep{Cygwin} port enabled its use on Windows; a local server emulating the \texttt{astrometry.net} API became available with \texttt{ansvr} \citep{ansvr}. That enabled development of a custom tool that allowed plate-solving all acquired images: \added{initially} with \texttt{astrometry.net} to establish a World Coordinate System (WCS), which was subsequently refined via a triangle match for higher accuracy. At that time, the USNO B1.0 catalog was the only one deep enough to enable reliable solves of such small FoVs.
As the mount controller operates in apparent coordinates, precession and nutation corrections had to be applied to the ICRS image center coordinates from WCS headers. Ultimately, the tool reformatted the results into a pointing model input file compatible with the mount's control software --- an approach arguably not intended by its developer. 

Using this workaround, the Mk~I.b pointing model achieved sufficient accuracy to place targets in the field, even after a meridian flip --- unless M2 movement risked a shift of the optical axis, requiring recalibration. With reproducible, \added{yet much larger pointing offsets} after the installation of the Mk~II (cf.\ Figure~\ref{fig:PM_evolution}, \added{top panel}), the ability to complete pointing models without the mount's control software remained critically important for many years of operations. Blind solving via the \texttt{astrometry.net} API became eventually available in \texttt{APCC Pro} through \texttt{PinPoint} in 2017, subsequently enabling more robust, later often nearly complete models to be generated directly; plate solving tools migrated to Gaia catalog data once locally storable catalog index files for \texttt{astrometry.net} and \texttt{Pinpoint} became available. Our custom tool still remains in use, mainly to recover aborted and later finished pointing model runs, and to recover images that occasionally fail to solve with standard settings (e.g.\ with very few field stars). 

While born out of necessity for effective operations, routine ''blind-solving'' became a powerful demonstration of \texttt{astrometry.net}'s reliability using images that resembled SOFIA's FPI\textsuperscript{+}. That in turn motivated a spin-off project, ultimately leading to the implementation of \texttt{astrometry.net} on the telescope operator workstations onboard of SOFIA, streamlining telescope setup in flight (see Section~\ref{sec:astrometry.net}). 


\subsubsection{Improving observing efficiency and high-speed tracking} \label{sec:CW_modification}

\begin{figure}[htb]
\centering
\includegraphics[width=\plotwidth]{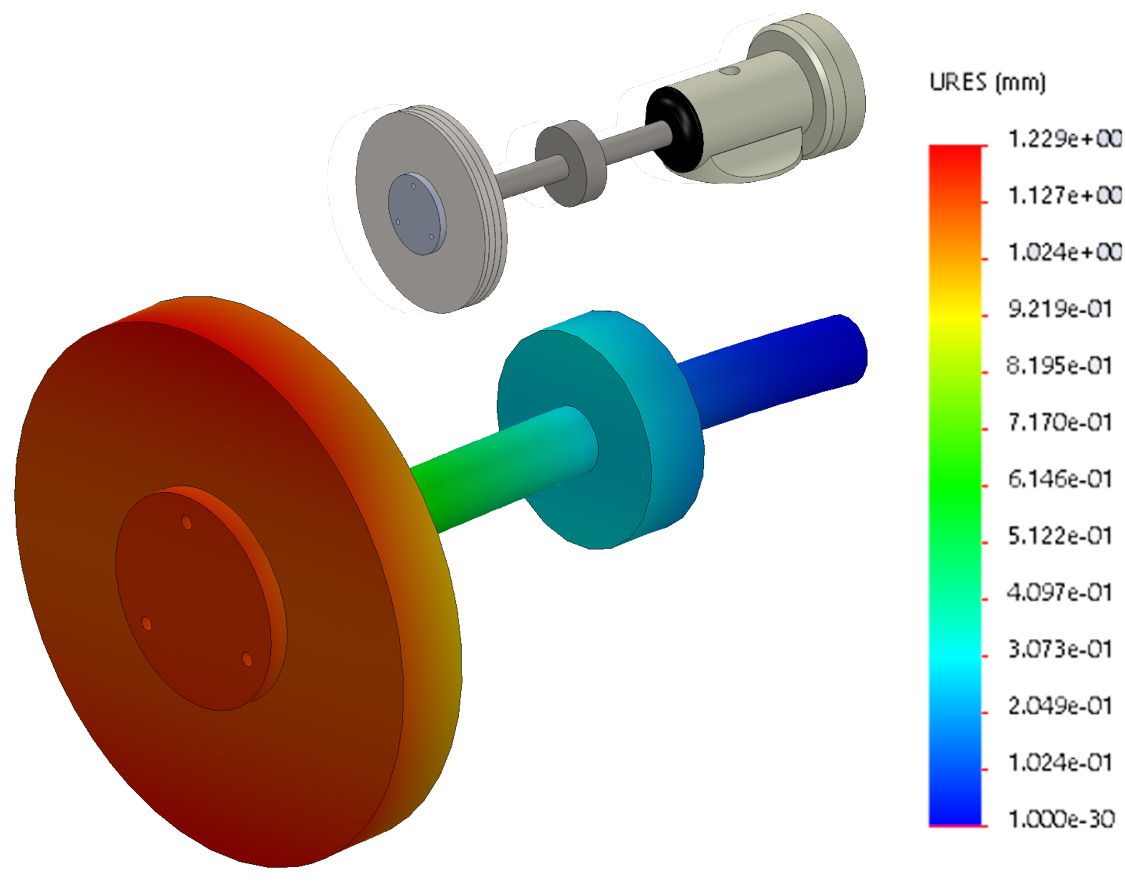} 
\caption{FEA simulation of the \added{former} 2.5-inch (63.5~mm) diameter, 30-inch (762~mm) long stock counterweight shaft, with a stack of steel plates ($\approx$~122~kg) bolted to its end, confirming its low stiffness and substantial bending in its horizontal position. 
\label{fig:mount_modified_cw}}
\end{figure}

Balancing ATUS with custom-made steel plates was a pragmatic decision to get it ``up and running'' (see Section~\ref{sec:mount}). Maximizing the moment arm allowed balancing the setup with the smallest counterweight, yet caused the largest moment of inertia. Consequently, the setup had a low natural frequency and began to swing like a pendulum in response to an angular acceleration. FEA simulations estimated the lowest natural frequency of the counterweight assembly at about 15.7~Hz, and a vertical displacement of 1.2~mm at the end of the counterweight shaft when positioned horizontal (see Figure~\ref{fig:mount_modified_cw}). That amount of bending was visible by resting a long steel ruler on the shaft.


To avoid streaked images, observations had to pause for about 10 to 12 seconds after each slew until oscillations had fully subsided. This was not recognized during initial indoor testing without access to the night sky. Aside from this delay, imaging of astronomical targets at sidereal tracking rates was unproblematic and reliable. However, attempts at open-loop satellite tracking revealed that even moderate, non-sidereal tracking rates caused image jitter that rendered images from the main camera useless. The continuous adjustments of the servo rates in both axes caused oscillations that were even visible in FFI and WFI images, severely limiting the system's ability to track man-made objects in Earth's orbit.

Issues arose after an upgrade of the mount's servo control unit in 2017. The new controller (GTOCP4) combined the functionality of the predecessor (GTOCP3) and its auxiliary encoder and limit switch controller (GTOELS) into a single, more capable unit. It provided a much faster encoder sampling rate and near-instant encoder feedback after slews --- fast enough to pick up and counteract polar axis oscillations before they subsided, inducing resonant servo motion. This made mount-specific changes to the servo control strategy necessary, implemented via a custom firmware. While this workaround enabled operations with the new controller, it precluded future firmware updates. As updates of the mount's control and pointing model software also require the latest firmware, software updates came effectively to a halt. 

Both limitations eventually motivated an upgrade to a 3.5-inch (88.9~mm) diameter, 15-inch (381~mm) long counterweight shaft (cf.\ Figure~\ref{fig:ATUS_evolution}, right side). The counterweight stack at the end of the shaft now consists of water-cut steel disks of 20.25-inch (514.4~mm) diameter, with a total weight of 196.5~kg in the final instrument configuration. A movable 8.125-inch (206.4~mm) diameter, 2.5-inch (63.5~mm) wide counterweight of 17~kg allows fine balancing (see Appendix~\ref{sec:app_weight_breakdown} for more details). This upgrade raised the lowest natural frequency of the counterweight assembly to about 80~Hz, lowered its moment of inertia by about 40\%, and reduced the vertical displacement at the end of the shaft by about 95\% \added{when the shaft is horizontal}. 

Since then, no oscillations have been detected, and individual adjustments to servo control are no longer necessary. The controller firmware has been restored to the official, unmodified release \added{version}. As a result, updates of the actively maintained control software are possible again. The telescope can now start imaging immediately after a slew, and has demonstrated its capability to track satellites in low Earth orbit (see Section~\ref{sec:SSA}). Observing efficiency has improved significantly; the duration of a pointing model run has roughly halved. 

The telescope's weight distribution was further optimized by mounting the auxiliary refractor and WFI on separate dovetail bars bolted directly to the OTA's base plate. Moving both imagers closer to the polar axis reduced the required counterweight mass by about 30~kg, reduced the moment of inertia of the setup further, and eliminated external loads on the OTA truss structure. The complete weight breakdown of the setup is provided in Table~\ref{tab:weight}. 


\subsubsection{Custom off-axis guider} \label{sec:OAG}

W\added{hen} observing transients with mmag variability, steady tracking over hours is crucial to avoid systematic errors in the photometric reduction. A practical solution was to guide directly on the main camera's images, which was routinely done with exposure times as long as 120~s. Beyond this, the time constant of the guider loop became too long, and the field started to drift slowly. Unguided imaging based on tracking rate corrections derived from the pointing model allowed exposure times up to about 240~s before star images became elongated. Differential flexure between the auxiliary refractor and the OTA's optical axis prevented its use for guiding. To achieve deep exposures and lower-cadence transient observations while \added{maintaining} steady pointing, an off-axis guider (OAG) in the same beam path as the main camera was a natural solution. The OAG would also streamline future automation: with a sufficiently large FoV, \added{it} would enable guiding regardless of the observer's intent with the main camera.

The few commercial OAGs that existed \added{could neither} carry our heavy image train free of flexure nor fit our setup optically and mechanically. Thus, a custom design was developed. With the main camera only using a small central fraction of the OTA's image circle (cf.\ Table~\ref{tab:imagecircle}), a pick-off prism can reflect unused light off-axis into a separate camera without vignetting the main FoV. The small filter diameter (cf.\ Table~\ref{tab:instrumentation_overview}) did not allow mounting a pick-off prism between camera and filter wheel. \added{Although} the ability to focus via M2 was later restored, the design also had to \added{accommodate} a mechanical focuser with limited payload capacity at the backplane. Finally, the OAG \added{needed to} provide a sufficiently large FoV to guarantee availability of a suitable guide star in any possible pointing. A compact camera (QSI 616s) with a cooled KAF-1603ME front-illuminated CCD was already available, able to provide a field of 9.7 \texttimes~6.5~arcmin\textsuperscript{2}. An internal study for SOFIA concluded that an 8.7~arcmin~square FoV would contain a 17~mag guide star all-sky, so even at $\approx 35\%$ lower sensitivity than the main camera in terms of QE, a sufficiently bright guide star would \added{always} be available at a reasonable image cadence. This eliminated the requirement to position the main camera field to find a guide star in the OAG, \added{allowing observers} to ``set it and forget it''. The size and position of the OAG camera's FoV in relation to the main camera are illustrated in Figure~\ref{fig:OAG_FoV}.


\begin{figure*}[htb]
\begin{center}
\resizebox{0.815\linewidth}{!}{%
\includegraphics[height=1cm]{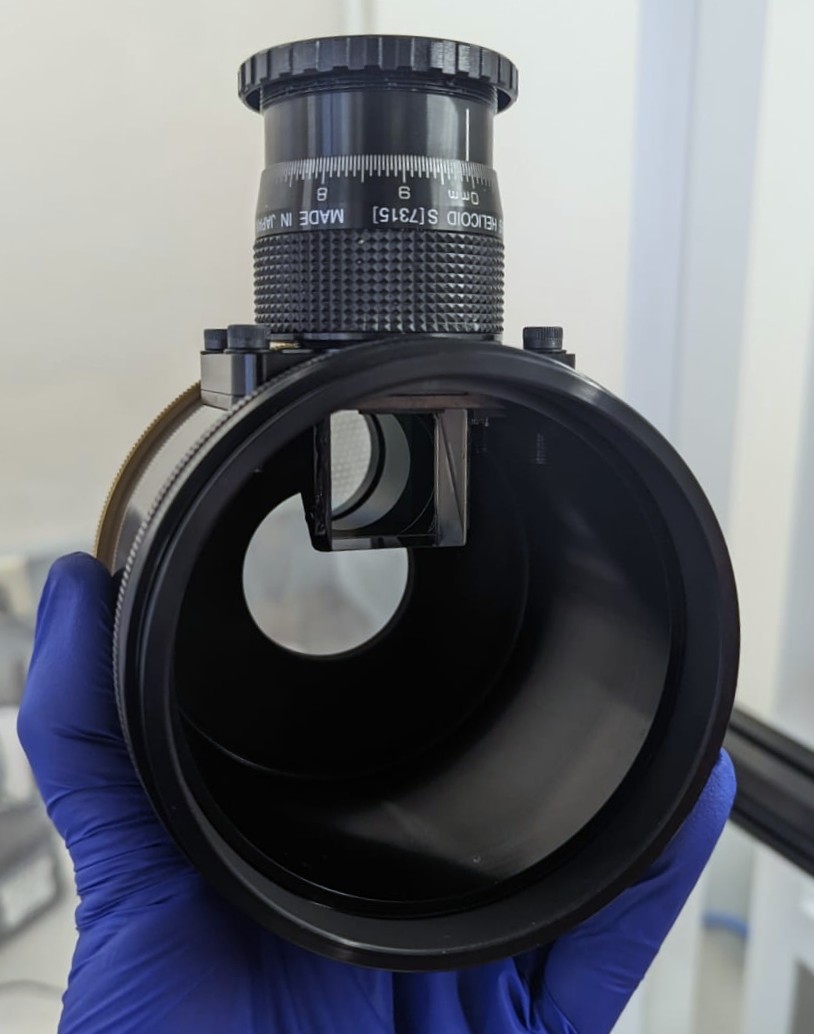} 
\includegraphics[height=1cm]{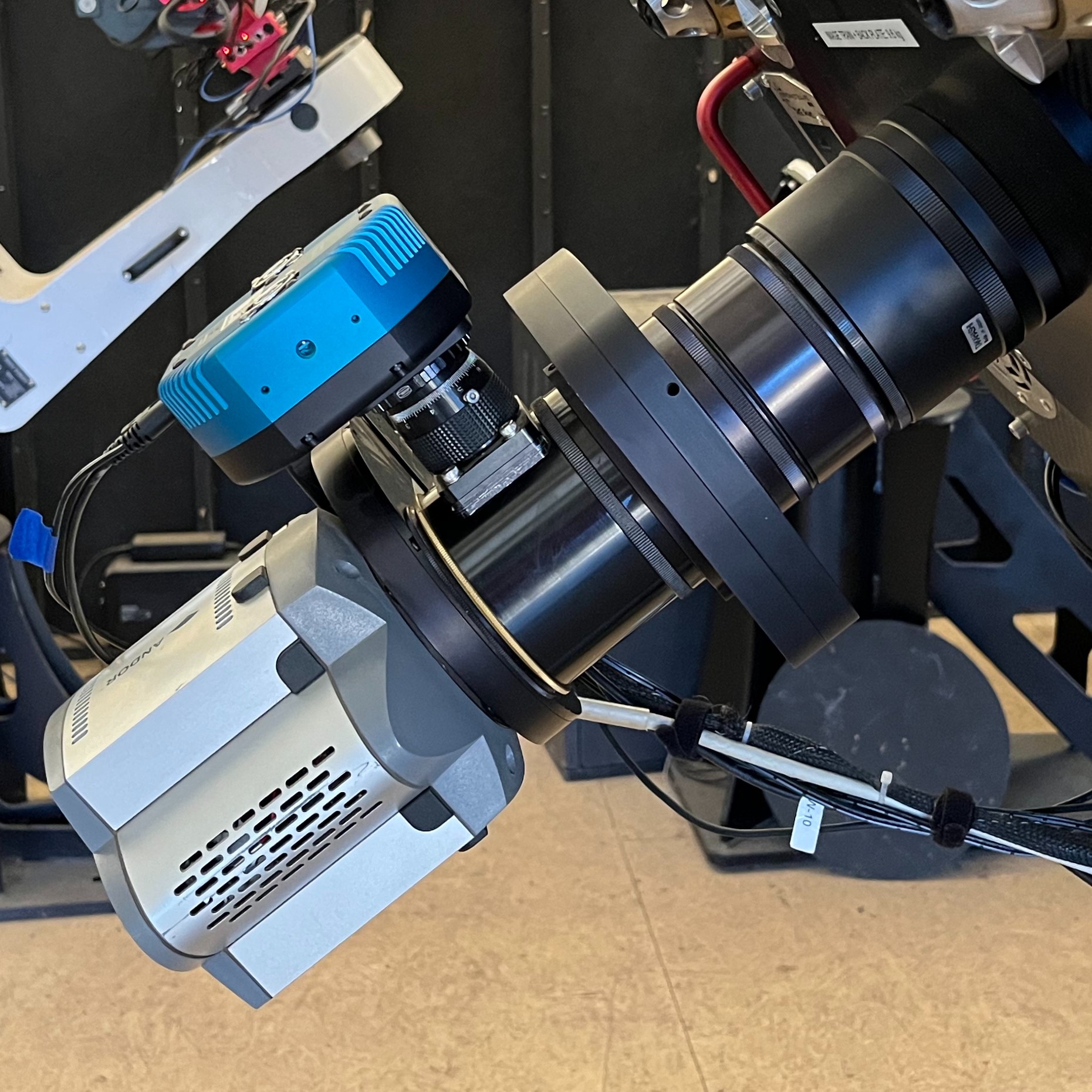} 
}
\end{center}
\caption 
{Left: The custom-built Off-Axis-Guider (OAG) with a 25~mm right-angle prism and helical focuser after assembly. Black felt was added before integration for additional stray light control. Right: Close-up of the image train with the FLI Atlas focuser, a QSI 616s camera mounted to the OAG, and the Andor iXon 888 camera behind an SBIG CFW-10 10-position filter wheel.
\label{fig:OAG+Image_Train}}  
\end{figure*} 


The iterative design \citep{Frueh2023} converged to a 25~mm right-angle pick-off prism from OptoSigma made out of BK7 glass\added{. The prism got} bonded to a holder made of stainless steel grade 416 (\added{UNS S41600 /} 1.4005 / X12CrS13 alloy), \added{offering a coefficient of thermal expansion (CTE) of 9.9~\texttimes~10\textsuperscript{-6}/K \citep{AtlasSteels2021}, excellent machinability, and a black oxide surface finish. This material was selected as it offered the best trade-off considering three requirements: a) a CTE close to BK7 glass \citep[CTE = 7.1~\texttimes~10\textsuperscript{-6}/K, ][]{SCHOTT2023}, b) the thickness and cost of a black surface finish, and c) the overall procurement and production cost. While Kovar and Titanium would have provided slightly smaller CTE mismatches, they had significantly higher material and manufacturing costs; in addition, Titanium would have required a powder coating finish.} At its position in the converging beam, a prism of that size uses almost the full width of the annulus of rays not received by the main camera. The prism was specified at an uncoated surface flatness of $\lambda/10$; the manufacturer kindly agreed on a shared-risk / best-effort basis to add an anti-reflective coating on the prism's faces with a post-coating flatness of $\lambda/8$, utilizing a coating run of other optical parts to minimize cost. Commissioning tests confirmed excellent optical performance of the prism. Note that the prism extends the optical path length to the OAG camera by 8.52~mm.

Space and weight constraints did not allow integration of a motorized focuser in front of the OAG camera; focusing is done manually via a non-rotating, lockable helical focuser (Hutech BORG 7315). The fully integrated OAG is shown in Figure~\ref{fig:OAG+Image_Train}. As filters in the converging beam shift the focal plane depending on their refractive index and thickness, the main camera has filter-dependent focus positions. The OAG camera was therefore focused approximately halfway between the two positions where the main camera obtains focus when imaging without a filter or through a Sloan filter. The OAG camera is thus never fully in focus, yet oversampling ensures robust calculation of guide star centroids and can be reduced via binned readout. 


\subsubsection{Final pointing and tracking performance} \label{sec:tracking_pointing_performance}

\begin{figure}[h]
\centering
\includegraphics[width=\plotwidth]{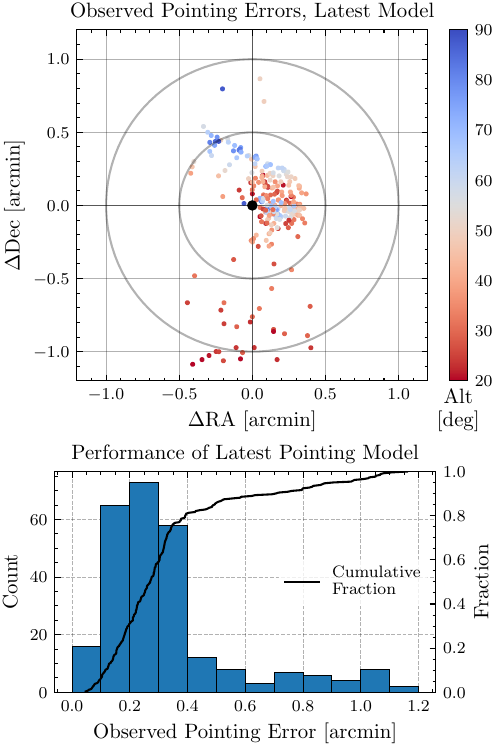} 
\caption 
{All-sky pointing performance with the latest pointing model, covering an elevation range from 20\degr~to zenith over the entire hemisphere.
\label{fig:PM_performance}} 
\end{figure}

Typical pointing errors \added{achieved with the pointing model in} the final instrument configuration are illustrated in Figure~\ref{fig:PM_performance}. To acquire these measurements, the mount was commanded to slew to the same pointing coordinates as during the creation of the pointing model. Roughly 85\% of all slews resulted in pointing errors \textless~0.5~arcmin; 95\% of slews were accurate to \textless~1.0~arcmin. Note that all slews were accurate to $\Delta\mathrm{RA} = \pm 0.5~\mathrm{arcmin}$; \added{slews} exceeding a total error of 0.5~arcmin were dominated by declination offsets, with the largest errors \added{occurring} at the lowest elevations (20 -- 30\degr). This may hint at errors in the applied refraction correction or at unaccounted-for atmospheric dispersion. Also, near the 20\degr~elevation limit, partial obstruction from adjacent telescope installations may have occurred, depending on where those instruments were pointed at the time the data \added{were} acquired. Overall, the pointing model \added{was} well-behaved, and enabled ATUS to point with sub-arcminute accuracy.


\begin{figure}[htb]
\centering
\includegraphics[width=\plotwidth]{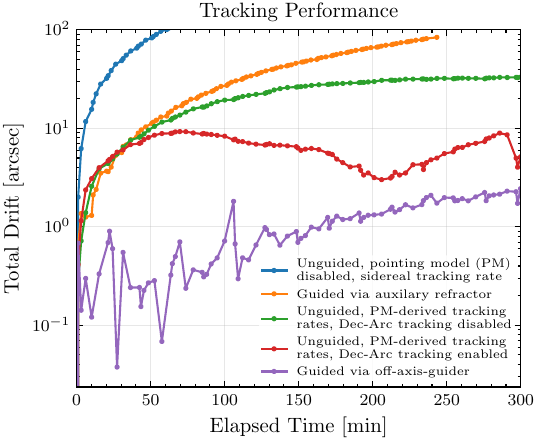} 
\caption 
{Comparison of various approaches to tracking. Not shown is on-axis guiding, i.e.\ deriving tracking corrections from the science images itself, as the drift is near zero. PM = Pointing Model.
\label{fig:tracking_methods}} 
\end{figure}

Figure~\ref{fig:tracking_methods} compares the tracking accuracy achieved by various means\added{, showing drift on a logarithmic scale over time}. To allow comparison, all data were acquired on an identical arc over the sky. Without pointing-model-derived tracking rate corrections, and without guiding, the field quickly drifted away (blue line). \added{Remarkably}, purely model-derived tracking rate corrections (green) \added{outperform} guiding with the auxiliary refractor (orange), \added {demonstrating how well} the pointing model \added{captures system behavior}. Coupling the refractor directly to the OTA's dovetail plate via brackets did not notably improve its performance as a guide scope. This illustrates that the statically overconstrained rear truss of the OTA is very rigid (see Section~\ref{sec:ll_closedtube}), \added{while both} mirrors undergo controlled flexure, as indicated by the FEM simulation (Figure~\ref{fig:ATUS_flexure}).

A new ``Dec-Arc'' tracking algorithm (red) was introduced in the mount's control software in 2021, and outperforms tracking rate corrections derived from the all-sky \added{pointing} model --- the drift remains below 10~arcsec over a duration of 5~hours. This mode derives tracking rates from a best-fit calculation of pointing errors along each declination arc instead of the full all-sky model, and interpolates them at the declination of the target. While the all-sky model aims at minimizing residuals everywhere, fits to points along a declination arc are optimized independently and can better capture localized behavior. The improvement thus results from a localized rather than a global optimization. Yet, the highest tracking accuracy is achieved by the off-axis guider (violet), which limits drift to about 4 pixels over the 5~hour test interval. This residual drift is negligible for all practical purposes; with the rise of satellite mega-constellations, practical exposure times are now \added{governed more} by potential light contamination from satellite trails \citep{Borlaff2025, Hainaut2020} than by the ability of ATUS to track steadily. Absolute pointing errors and pointing model performance also remained unchanged after commissioning of the OAG (cf.\ Figure~\ref{fig:PM_evolution}), indicating that its design goal of high stiffness was fully met.

Until the OAG was available, ATUS most commonly guided on-axis. For time series data with sufficiently short exposure times, guiding can be done directly on the main camera's images. This was demonstrated for exposure times up to 120~seconds; longer exposures are not practical, as the guider loop's time constant becomes too long --- guiding requires about 10 -- 12~exposures to stabilize, i.e.\ up to about 20 minutes. On-axis guiding was routinely used for observations of exoplanet transits (see Section~\ref{sec:exoplanet_transits}) with exposure times up to 100~seconds as a practical limit. Since drift is corrected to near-zero, on-axis guiding was not included in Figure~\ref{fig:tracking_methods}.


\section{Use cases}


\subsection{A test platform for SOFIA}

Initially motivated to support SOFIA, ATUS served as a test platform to evaluate hardware and software of interest for the airborne observatory. Three examples shall be briefly presented in this section. ATUS also provided data to support flight planning for SOFIA missions that targeted Solar System objects and transients, e.g.\ observing comets with highly variable comae shortly before scheduled missions to assist in selecting a suitable tracking strategy. Its data \added{were} also used to verify exposure time calculations and filter selection for FPI\textsuperscript{+} observations, as the spectral and instrumental characteristics of ATUS were transferable to the FPI\textsuperscript{+}. Operational procedures and FPI\textsuperscript{+} observing modes on SOFIA could be discussed and demonstrated with ATUS for training and educational purposes. 


\subsubsection{Automation of telescope pointing calibrations} \label{sec:astrometry.net}

Perhaps the most significant contribution to SOFIA was the successful field demonstration and extensive test campaign of the \texttt{astrometry.net} algorithm \citep{Lang2010, ASCL_Lang2012}, enabling a robust, reliable, and fully ``blind'' star field recognition from image data alone, without requiring any information about telescope orientation, imaging scale, or any other a-priori information. Early operations with the Mk~I and Mk~II.a OTA provided proof that its pointing could be reliably determined with \texttt{astrometry.net} with all three available cameras, i.e.\ even in a field as small as $\approx$~9~arcmin squared --- and also with a severely degraded image quality, e.g.\ due to passing clouds, or purposely induced tracking errors, defocus, or even vibrations. Its distribution as open-source software and independence from observatory housekeeping data made it unnecessary to modify SOFIA's complex and unique Mission Communications and Control System (MCCS). This greatly simplified its adoption and integration. \texttt{astrometry.net} eventually became available as a stand-alone tool on the SOFIA telescope operator workstations in March 2016 \citep{Schindler2016}. 


\subsubsection{Shack-Hartmann test instrument} \label{sec:SHIFT}

\begin{figure*}[htb]

\begin{center}
\resizebox{0.9\linewidth}{!}{%
\includegraphics[height=1cm]{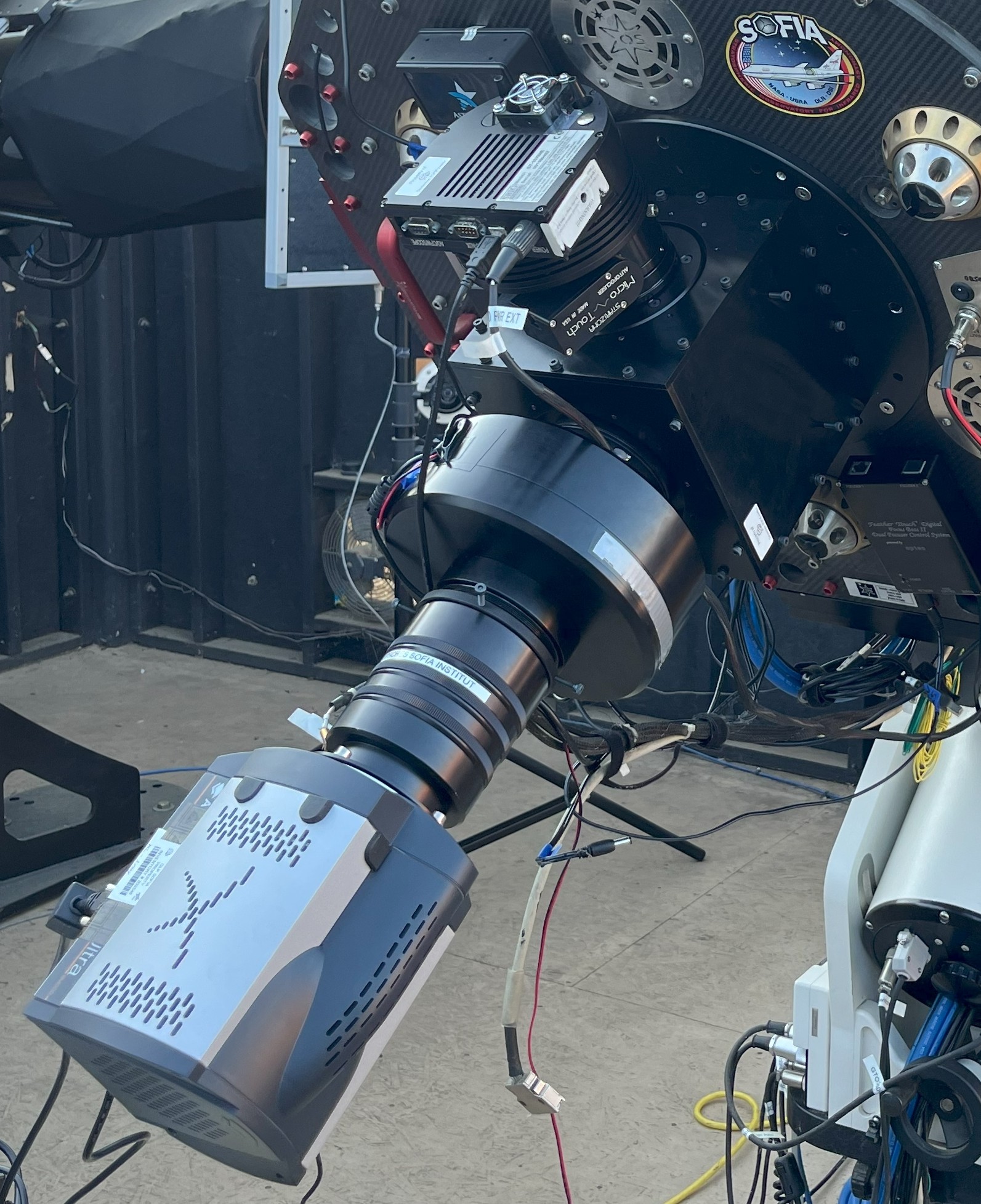} 
\includegraphics[height=1cm]{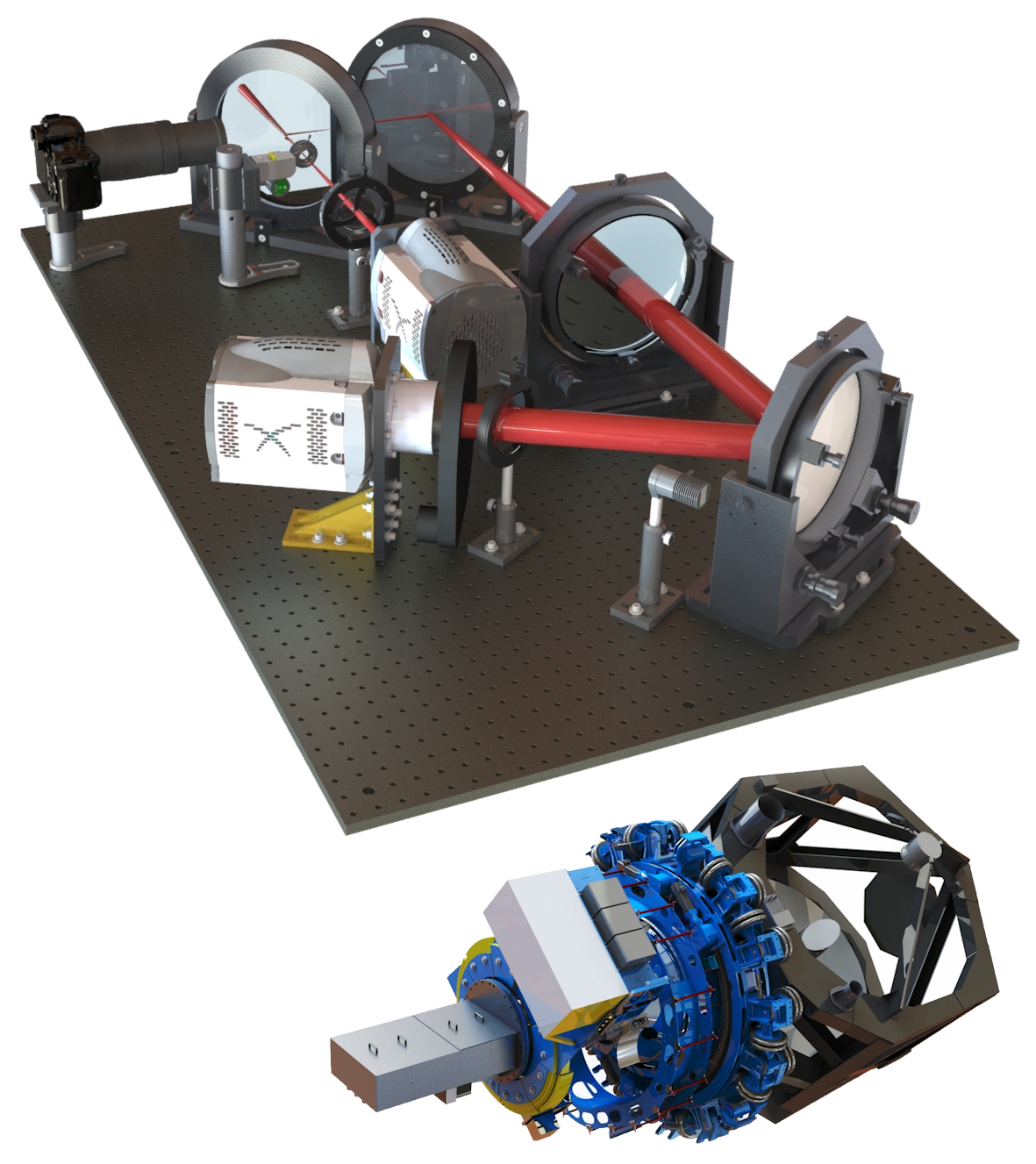} }
\end{center}
\caption 
{Left: SHIFT mounted on ATUS for verification and validation tests. SHIFT was a descoped, simplified iteration of a test instrument already undergoing integration when the SOFIA project was canceled. Its original design, occupying an optical bench, had three cameras simultaneously providing a measurement of the SOFIA telescope's optical wavefront, an image of its entrance pupil, and of the sky \citep[right side; image adopted from][]{Zabel2020}.
\label{fig:SHIFT+SHT}}  
\end{figure*}

With the decommissioning of HIPO, the {High Speed Imaging Photometer for Occultations} \citep{Dunham2004} in late 2017, SOFIA also lost its first-light test camera at visible wavelengths that had the ability to quantify aberrations of the optical wavefront via a Shack-Hartmann lenslet array. Wavefront sensing and pupil imaging would have become crucial once any of the three mirrors of SOFIA's telescope required realignment, i.e.\ after removal for recoating, or maintenance and repairs of SOFIA's one-of-a-kind M2 mechanism. To restore this capability, a new, dual-use test instrument had been designed at DSI (see Figure~\ref{fig:SHIFT+SHT}, right side): an optical bench allowing routine checks of image quality during ground tests (line ops) or even airborne (ferry flights), integrated on an easy-to-install science instrument mass dummy to balance the SOFIA telescope, e.g.\ during aircraft maintenance periods. Three cameras would have provided simultaneous \added{Shack-Hartmann measurements of the optical wavefront}, an image of the entrance pupil of the SOFIA telescope, and \added{an image} of the sky \added{as a pointing} reference during operations. Having passed its critical design review, the instrument was already undergoing integration when the cancellation of the SOFIA project was announced in late April 2022.

To salvage at least some procured components as well as the development spent on the data analysis pipeline, a descoped version dubbed the {Shack-Hartmann Instrument Fast-Tracked (SHIFT)} was conceived and assembled during the ramp-down phase of the SOFIA project \citep{Frueh2023}, and successfully commissioned with the Mk~II.b OTA. With a weight of about 16.5~kg, SHIFT was the heaviest instrument mounted on ATUS so far (Figure~\ref{fig:SHIFT+SHT}, left side). 


Light from the telescope first falls on a pellicle beamsplitter with 8\% reflectance, 92\% transmission, and a 101.6~mm CA, mounted under a 45\degr~angle to the incoming, converging beam. The reflected light is directed into a CCD camera with a motorized focuser providing an image of the sky, assisting in positioning and recovery of a target star onto the SH sensor. With a thickness of 2~\textmu m, the pellicle's impact on the wavefront is negligible for measurements in transmission. At the focal plane, a motorized two-position filter flip mount can move an LED-illuminated pinhole into the beam. This allows measurement of a reference wavefront to correct for aberrations introduced by the remaining optical elements down the optical path. An achromatic 1-inch (25.4~mm) lens with an effective focal length of 88.9~mm acts as a collimator, i.e.\ it is mounted such that its focal plane coaligns with the one of the telescope. A 1-inch (25.4~mm) Astrodon Sloan r'2 filter (562 -- 695~nm bandpass) is mounted into a filter cell that is screwed into a C-mount thread above the microlens array (MLA). This filter was chosen to limit chromatic aberrations, as it matches the design wavelengths of the achromat (587.6~nm) and of the MLA (635 nm); in addition, the EMCCD's quantum efficiency exceeds 90\% over the entire band. The MLA is a custom-gradient index (GRIN) array made of soda lime glass ($f = 18.6~\mathrm{mm}$, 300~\textmu m pitch); the interspaces between the round microlenses remained masked. In addition, three microlenses remained fully masked for calibration purposes. The spot pattern is recorded by an Andor iXon 888 Ultra camera, due to its FPI\textsuperscript{+} legacy, compatibility, and integration into the flying observatory's software architecture. 

Different mirror spacings could be measured by adding 3.5-inch (88.9~mm) OPTEC-DSI spacer rings of various width; the wavefront's defocus term then acts as a reference that M2 is positioned such that the telescope's focal plane coincides with that of the collimator. Results of the SHIFT validation and verification campaign have been previously discussed in Sections~\ref{sec:mirror_spacing} and \ref{sec:final_IQ}. The instrument can be easily adapted for use on other telescope optics.


\subsubsection[Near-infrared channel for the FPI+]{Near-infrared channel for the FPI\textsuperscript{+}}

\begin{figure*}[htb]
\plottwo{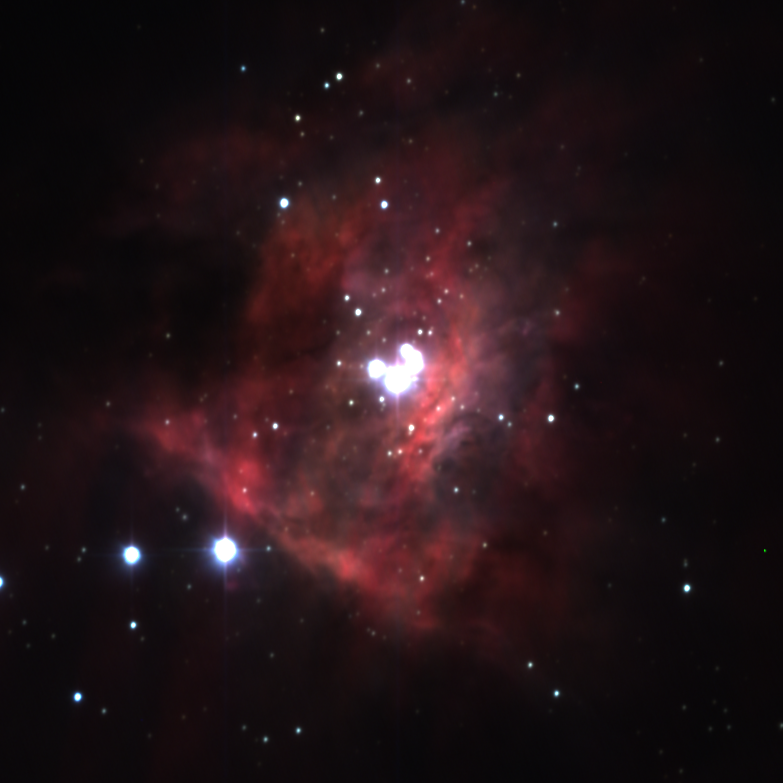}{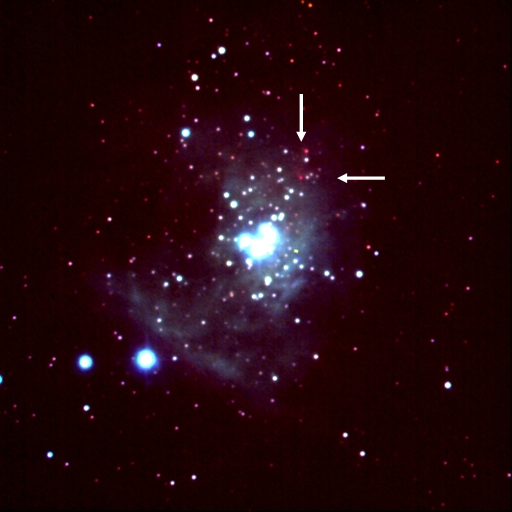}
\caption 
{False-color composite images of the Orion Nebula (M42) at wavelengths near the ``red'' sensitivity limit of the silicon-based EMCCD (cf.\ Table~\ref{tab:instrumentation_overview}) of the ATUS main camera (left), and over the full spectral range accessible to an InGaAs CMOS image sensor (right). On the left side, the color bands correspond to blue = 695 -- 844~nm (Sloan i'2), green = 826–920~nm (Sloan z\_s2), and red = 950 -- 1058~nm (Sloan Y2); on the right side, to blue = 975 -- 1075~nm (aligned with Sloan Y2), green = 1155 -- 1325~nm (aligned with J-band), and red = 1525 -- 1660~nm (short end of H-band). The white arrows indicate the location of the Becklin-Neugebauer object (BN / V2254 Ori), which is undetectable at visible wavelengths. Images were obtained during tests of a commercially available InGaAs camera in March 2018, once under consideration for a second NIR channel for SOFIA's FPI\textsuperscript{+}; CCD-band images were acquired in April 2018.  
\label{fig:M42}} 
\end{figure*}

Its fully reflective optical configuration \added{enabled ATUS to observe} beyond the silicon band-gap cutoff wavelength ($\approx$~1~\textmu m). After initial characterization measurements in the lab, two commercial cameras with InGaAs image sensors were eventually tested on-sky to determine sensitivity and noise characteristics. Figure~\ref{fig:M42} provides composite images of the Orion Nebula (M42) in various Sloan (left) and near-infrared bandpasses (right) as an illustrative example from that field test. The InGaAs-camera image successfully detected the Becklin-Neugebauer object (BN) at its longest wavelengths, which is likely a medium-mass protostar deeply embedded in the Orion Nebula's star-forming region. It is not visible in the spectral range of \added{silicon} CCD and CMOS sensors, as the surrounding dust has such a high density that it completely scatters and absorbs the star's light emission at shorter wavelengths. Yet, photons at longer wavelengths pass through the dust cloud, also illustrated by numerous other red-tinted stars visible in the right panel. The respective color channels in both images were weighted according to the filter transmission, spectral sensitivity of the sensor, and exposure time in order to obtain a representative impression of the relative amount of light that was emitted in the respective spectral band. Data for both composite images were obtained with the Mk~II.a OTA.


The findings from these tests provided the foundation for a joint proposal between the Massachusetts Institute of Technology (MIT) and DSI to extend the FPI\textsuperscript{+} with a second 1.0 -- 1.6~\textmu m channel, a spectral range not impacted by its ``warm'' relay optics \citep{Schindler2021,Pfueller2018a,Person2018}. After FLITECAM's decommissioning in late 2017, this NIR channel could have partially restored lost scientific capabilities that were particularly relevant for stellar occultation studies, e.g.\ as demonstrated in \citet{Person2021} to detect haze in Pluto's atmosphere and to derive its particle size. SOFIA's observations also did not suffer from telluric absorption between the Y/J/H bands. Operationally, SOFIA could have potentially benefited from an NIR guide camera providing an improved availability of guide stars during early twilight, in close proximity to the moon, and in regions of high extinction (i.e.\ dust clouds and nebulae).


\subsection{Scientific Research and Teaching}

Scientifically, ATUS was primarily used to observe transits of extrasolar planets and stellar occultations. On some occasions, it was also possible to record relevant data in parallel with SOFIA.

At University of Stuttgart, ATUS offered graduate students in Aerospace Engineering the opportunity to gain practical experience with a telescope, and provided a representative example of a complex system that needs to be controlled remotely in all aspects. Numerous students used the telescope to learn the basics of observational astronomy and astronomical data reduction as part of an annual seminar and lab course, and to conduct BSc and MSc thesis work on various topics in engineering, instrumentation and software development, and astronomy.  


\subsubsection{Stellar Occultations}

High timing accuracy is a key enabling requirement for stellar occultation observations. Ingress and egress timing errors propagate directly into uncertainties of albedo, shape, and size estimates. While stellar occultations have provided the best available estimates of these properties for bodies that were not visited by space probes to date --- and spatial sampling of the occultation shadow is another important factor as well --- event time uncertainties continue to plague observations up to the present. The time referencing system employed by ATUS (cf.\ Section~\ref{sec:time_referencing}) eliminates all systematic timing errors; the remaining uncertainty is entirely governed by sampling rate and SNR.

Over the years, ATUS participated in more than 40 coordinated observing campaigns attempting to record stellar occultations by trans-Neptunian objects (TNOs), Centaurs, asteroids, and satellites of gas giants from multiple observing sites. During its lifetime, astronomy entered the era of the Gaia star catalogs, improving stellar positions and proper motions by orders of magnitude. Alongside, high-cadence, wide-field sky surveys came online that repeatedly measured Solar system body positions against the Gaia reference frame and made vast datasets publicly available, yielding substantially improved ephemerides \citep{Knieling2024}. 

An early, pre-Gaia highlight was the first successful multi-chord occultation by a detached trans-Neptunian object, (229762)~G!k\'{u}n\textdoublepipe{}’\`{h}òmd\'{i}m\`{a}\footnote{The name officially recognized by the International Astronomical Union (IAU) is derived from the mythology of the Ju\textpipe{}’hoansi people of Namibia, and incorporates phonetic notation for click consonants and diacritics. Further details on pronunciation and origin can be found at: \url{http://www2.lowell.edu/users/grundy/tnbs/229762_how_to_say.html}} in November 2014 \citep[][]{Schindler2017}, providing the first accurate geometric albedo estimate for an object of this dynamical class. Geometrical constraints from this occultation enabled improved modeling of the TNO's thermal emission to match \textit{Herschel}/PACS flux measurements, constraining its volume and equivalent diameter.  
In March 2017, ATUS delivered a decisive negative measurement of an occultation by the binary TNO~(90482)~Orcus \citep{Sickafoose2019}. Stations in Texas and Hawaii detected occultations with differing normalized flux levels that summed to approximately 1.0, indicating the same body must have occulted two separate, unresolved stars, confirmed by subsequent speckle imaging at Gemini North. Ephemeris comparisons revealed that Orcus's satellite Vanth had occulted both stars. ATUS missed Vanth's shadow by less than 21~km, enabling a tight size constraint of $442.5 \pm 10.2$ km --- the first documented direct size determination of a TNO satellite from a multi-chord occultation other than Charon, Pluto's largest moon.

A grazing observation of a bright G~=~13.0~mag star by (134340)~Pluto's atmosphere in August 2018 contributed to a long-term study spanning 2017 to 2023 \citep{Sickafoose2019a,Sickafoose2023,Sickafoose2026}, revealing that its size and pressure had entered decline in 2022. Additional noteworthy contributions to occultation measurements include: (174567)~Varda, \added{observed in} parallel to SOFIA in September 2018 \citep{Schindler2019}, providing the closest negative westward of Varda's shadow; (28978)~Ixion on two consecutive nights in May 2021 (A.~M.~Colclasure, 2026, in prep.); Titan occulting a rare bright star (8.7~mag) in July 2022; (50000) Quaoar with multiple negatives over the years and a solid-body chord in August 2022; and a non-detection of Quaoar's moon Weywot in June 2023, providing yet another important size constraint \citep{FernandezValenzuela2026}.


\subsubsection{Exoplanet Research} \label{sec:exoplanet_transits}

In total, more than 90 transits of 46 different extrasolar planets have been observed with ATUS. Many were acquired by MSc-students during an annual lab course on observational astronomy, challenging them to plan an observing run, reduce the data, extract a light curve, and determine the properties of the observed exoplanet. Popular targets, due to their frequency and magnitude drop, have been transits by ``ultra-hot Jupiters'' (UHJ) --- giant planets orbiting at very close distances with periods less than 3-days. These planets experience significant tidal forces that may cause orbital decay until annihilation. Their lifetimes and decay rates are poorly understood, yet stars with UHJs tend to be younger, and perhaps half of all stars could have already swallowed an UHJ in their first billion years of life \citep[see discussion in][]{Adams2024}.

Being minuscule, the detection of orbital decay rates requires long observational baselines. Strong evidence only exists for WASP-12~b and potentially Kepler-1658~b so far. A decreasing orbital period has been suggested for several others of the $\approx$ 100 known UHJs. ATUS data helped scrutinize a postulated orbital decay of TrES-5~b --- one of only four UHJs where multiple studies supported this claim --- observing 13~transits between 2019 and 2024. \citet{Rothmeier2025} combined these with other previously unpublished ground-based data, Transiting Exoplanet Survey Satellite (TESS) observations, and carefully validated published midtimes. Having compiled a total of 280 TrES-5 b transit midtimes over 17 years --- nearly twice as many as earlier studies --- a linear ephemeris model provides the best fit, not justifying a detection of orbital decay. The study concluded that ground-based telescopes, especially those with accurate time referencing like ATUS, can achieve better transit timing precision than TESS, with smaller median midtime uncertainties. Even a few transits acquired annually with ATUS-like telescopes can be as constraining as an entire TESS season, yet avoid periods with sparse data coverage prone to misinterpration. 

\begin{figure*}[htb]
\centering
\plotone{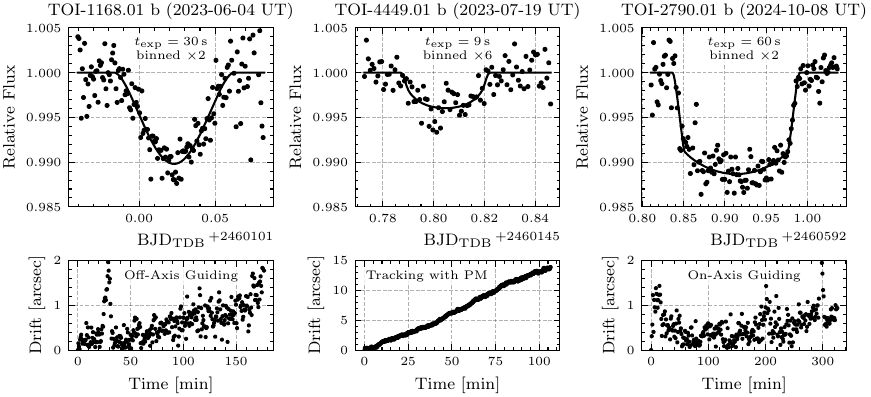}
\caption 
{ 
Light curves of transits of various extrasolar planet candidates found by the TESS mission, illustrating that ATUS is capable of resolving mmag-level flux changes. The black line represents a fitted transit model by \citet{Mandel2002} considering quadratic limb darkening. The three examples  also showcase three different approaches to tracking. Left: TOI-1168.01~b, using the OAG, keeping the pointing \added{drift} under 2~arcsec over nearly 3~h. This was the first transit measured after commissioning of the OAG. \added{Middle:} TOI-4449.01 b, using tracking rate corrections derived from the pointing model without Dec-arc tracking. This transit was observed ``live'' during a public lecture that was scheduled in the midst of on-site maintenance work, with the OAG camera unavailable that night. \added{Right:} TOI-2790.01 b, guided on-axis using the science images itself, which has been the default observing mode for transients with exposure times under $\approx$ 100~s. This was also the final transit observed with ATUS from SRO. 
\label{fig:Exoplanet_Lightcurves}}
\end{figure*}

Figure~\ref{fig:Exoplanet_Lightcurves} further illustrates the performance of ATUS, using three transits by extrasolar planet candidates identified by TESS as an example. The many other, yet-unpublished transits allow potentially valuable contributions to similar long-term studies of other systems. Further work in this area is planned, making observations that date back several years all the more interesting. 


\subsubsection{Space Situational Awareness} \label{sec:SSA}

The exponential increase of man-made satellites and space debris has spurred an increasing interest in detection and tracking capabilities of objects in Earth orbit to prevent collisions, plan re-entries, and protect space-based assets. While ATUS employs a mount that offers sufficiently fast slew speeds to track even very low orbiting satellites, its German-equatorial design requiring a meridian flip essentially limits its observing window to the time before or after a satellite has crossed the meridian ($\pm$ applicable advance/delay). With satellite passes lasting only a few minutes, the target is usually no longer observable once the telescope has changed pier sides, so the observer usually needs to choose to start or end an observation near the meridian. An astrographic pier or a paralactic fork mount would overcome this limitation.

\begin{figure}[htb]
\centering
\includegraphics[width=\plotwidth]{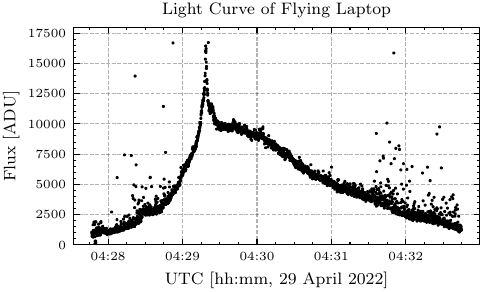}  
\caption 
{ 
Light curve (single aperture) of the Flying Laptop satellite (\added{SATCAT} 42831, COSPAR ID 2017-042-G) during its pass over SRO on 29 April 2022 UT. Images were acquired in open-loop tracking at a frame rate of 8.5~Hz (0.114~s exposures) in 2\,\texttimes\,2 binning mode without a filter.
\label{fig:FLP_lightcurve}}
\end{figure}

\begin{figure}[htb]
\centering
\includegraphics[width=0.5\plotwidth]{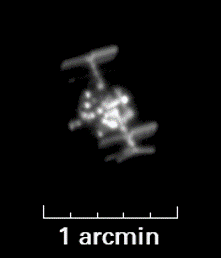}  
\caption 
{ 
The International Space Station (ISS) during a pass \added{over Sierra Remote Observatories} on July 14, 2022 at 11:34:14.749 UT \added{at maximum elevation (77.2\degr), and thus closest distance to ATUS ($\approx 430\,\mathrm{km}$). The angular extent of the ISS is 0.98~arcmin; the image was acquired during open-loop tracking.} The average orbital altitude of the ISS at that time was \added{417~km; its apogee was at 420~km}. Due to its extremely high brightness, a very narrow {H-\textalpha} filter (3~nm bandpass) was used to avoid overexposure despite the very short exposure time of 64~ms. The camera was operating unbinned for maximum resolution, reading a window of 1024~\texttimes~585 pixels to enable a frame rate of 15.4~Hz. 
\label{fig:ISS}}
\end{figure}

Satellite tracking has been considered the ultimate test of the pointing and tracking capabilities of ATUS; it also demonstrates the effectiveness of the mechanical modifications described earlier. Figure~\ref{fig:FLP_lightcurve} showcases a single-aperture-photometry lightcurve of the Flying Laptop satellite \citep{Eickhoff2016}, built at University of Stuttgart's Institute of Space Systems. It has a size of 60~\texttimes~70~\texttimes~85 cm\textsuperscript{3} and a mass of about 110~kg, was launched into a low polar orbit in 2017, and is still operational at the time of writing. The favorable geometry of this particular pass allowed open-loop tracking for over five minutes past the meridian. Figure~\ref{fig:ISS} captures a still frame of the International Space System during a pass near-zenith on July 14, 2022, at a distance of only 430~km. While image resolution is limited by the camera's pixel size and the telescope's modulated transfer function (MTF), it illustrates what ATUS can accomplish. The host PC communicates with the mount controller via TCP. In both instances, open-loop tracking was based on Two-Line Element sets (TLEs) obtained from CelesTrak\footnote{\url{https://celestrak.org}}; the continuous calculation and  commanding of required RA/Dec slew rates was done with \texttt{SkyTrack} \citep[FLP: v1.5.7, ISS: v1.6.2;][]{SkyTrack}.


\newpage


\section{Lessons learned} \label{sec:lessons_learned}


\subsection{OTA design decisions} \label{sec:ll_closedtube}


\paragraph{Open truss instead of a closed tube}

Most smaller telescopes with an open truss structure void the principle of the Serrurier truss \citep{Serrurier1938,Yoder2006}, and ATUS is no exception. To keep both mirrors collimated, the front and rear truss must flex equally, which requires mounting the OTA at the center frame. Without a tertiary mirror reflecting the beam into a Nasmyth port, the maximum instrumentation weight must be defined ahead of operations. In practice, smaller mounts and telescopes are built by different manufacturers, thus need some standardized mechanical mating interface. The ubiquitous solution is a dovetail saddle plate on the mount that clamps a plate mounted on the OTA, both requiring sufficient length for adequate clamping force. For essentially all commercially available open-truss telescopes, the interface plate is bolted to the center and rear frame, leading to a statically overconstrained system. With the rear truss clamped, the front truss must be overengineered to be fully rigid. This comes at the expense of a much increased weight, and is particularly relevant for RC designs that are very sensitive to small mirror misalignment. 

The weight of the OTA's stand-alone truss structure (without the M1 cell and baffle, M2 assembly, spiders, dovetail, etc.) is on the order of 50 -- 55~kg (cf.\ Table~\ref{tab:weight}). In contrast, a closed CFRP tube with an inner diameter of 720~mm and a wall thickness of 5~mm provides a fully rigid OTA of similar length, yet weighs only about 30~kg. A closed CFRP tube would thus provide not only considerable weight savings, but also an intrinsic stiffness that could have alleviated many issues addressed in this paper. Given these advantages, a closed tube was suggested to the manufacturer at the time of the Mk~II OTA redesign, but did not receive its approval, citing convective tube currents as a strong counterargument, but also an aesthetic preference for an open truss from a marketing perspective. 

A significant source of turbulence is the boundary layer above the primary mirror that is warmer than ambient; despite its lower mass, the Mk~II M1 still has a significant thermal capacity. Tube currents can likely be mitigated by a careful tube and mirror cell design with sufficient clearance and forced ventilation. From a stray light perspective, the closed tube also appears superior to a fabric truss shroud; tests at SRO have shown that ATUS can not operate without a shroud covering its front truss, and essentially all commercial open truss telescopes have at least a shrouded rear section. Future instrumentation projects should carefully assess the benefits of a closed tube and the potential drawbacks of the pseudo-Serrurier truss exhibited by most, if not all \added{open-truss} OTAs that are on the market today.  


\paragraph{Weight savings on mirror substrates}

With the manufacturer gaining experience in CNC-milling of large glass-ceramic substrates and the subsequent polishing of contoured mirrors, the conical back of the M1 was a novel approach at the time of the Mk~II redesign. \added{Back then}, a straight taper was a conservative choice to mitigate manufacturing risk, but also to limit further fabrication cost. At a moderate $f/3$ focal ratio, the mirror's center of mass is still well inside the substrate's center hole, where the center hub supports it, reducing the risk of astigmatism from self-deflection at low elevations. Since load is introduced into the substrate at its greatest thickness, a well-designed hub minimizes mirror-mount-induced stresses and distortion, as \added{eventually demonstrated with the iterated} Mk~II.b hub design. 

A higher stiffness-to-weight ratio can be achieved with a parabolic taper, with the parabola's vertex at the mirror edge, but is more difficult to manufacture \citep{Vukobratovich2017}. All mirrors with a contoured back are sensitive to rapid temperature changes; due to their variable thickness, a temperature gradient develops in the substrate causing surface distortion. While this is a significant concern for air- and spaceborne telescopes, it is usually not a concern for smaller ground-based astronomical telescopes; the ATUS M1 did not present any image quality issues during its cool down at nightfall. We estimate that about 2 -- 3 kg of additional mass savings could be achieved with a parabolic taper on future, otherwise similar OTA designs. Even more weight can be removed by milling out pockets, should the project budget allow for it\added{, or by a closer match of the center hole diameter to the central obscuration}. 

Mirror designs with further optimized stiffness-to-weight ratios (such as double-arch backs and sandwich structures) exist, but drive manufacturing cost and require more complex mirror cell designs. A tapered single-arch remains the best compromise between weight and cost savings, also owing to its relative ease of mounting. 


\paragraph{Focusing via single M2 actuator}

Perhaps unsurprisingly, yet readily apparent from wavefront sensing, is that M2 does not move exactly along the optical axis. In retrospect, this appears to be a requirement that is nearly impossible to fulfill --- a small angular offset between the axis of movement and the optical axis will always remain. For routine focusing, the changes in M2 position result in minor, acceptable coma; a larger change in M2 position requires a recollimation. RC telescopes that operate at various mirror spacings should therefore avoid an M2 driven by a single actuator.


\subsection{System design} 


\paragraph{Astrographic pier}

Building height constraints at SRO and the length of the Mk~I OTA dictated a very low pier height of just 18-inch (45.7~cm); the base surface of the polar fork assembly was just 58.7~cm above floor level. This prevented the use of an astrographic pier. Given the relaxed constraints at a tentative future site, an astropgrahic pier should now be reconsidered, as it would eliminate the meridian flip completely. The mount controller supports this operational mode --- treating it as an equatorial fork mount --- and prevents crossing the meridian in the nadir direction to avert cable twist. A taller pier will also provide a welcome increase in ground clearance to accommodate the enlarged FFD. A downside is the significantly increased weight of the pier, requiring a lifting device that was not needed at SRO. 


\paragraph{Operational envelope}

German-equatorial mounts are known for their large swept volume. In its final configuration, ATUS requires an operational envelope of about 3.2~m (east-west) by 3.0~m (north-south). This required careful planning to avoid collisions with neighboring instruments at SRO; smaller setups were installed on the piers near ATUS to accommodate this constraint. 


\paragraph{System architecture}

The communication interfaces of all peripheral devices allowed for sufficiently long ($\approx$ 5~m) passive cable connections to place the control PC in a small server enclosure near the telescope pier, together with other ancillary equipment. This avoided adding weight and a heat source to the OTA; the main camera requires a PCIe interface card, precluding a more compact PC. The enclosure also provided some protection from dust and very low temperatures. 

Low data bandwidth requirements of all serial devices allowed shared connections via industrial-grade, ruggedized 7-port USB 2.0 and 8-port RS232 hubs mounted underneath the OTA, minimizing cabling through the hollow shafts of the mount's axes. However, modern sCMOS cameras with larger image sensors and bandwidth requirements --- most often requiring USB3.2/4 connectivity --- will likely require moving the PC to the OTA, as newer protocols support significantly shorter passive cable lengths, and active USB extension cables are a frequent source of failures in the field. An instrumentation upgrade would likely requires a revision of the system architecture. 


\paragraph{Software}

We considerably underestimated the required software development effort. The main camera in particular presented challenges for integration with common astronomical equipment control software. Several components of ATUS are rarely found on smaller telescopes, or lacked manufacturer-provided software support. On Windows PCs, the ASCOM Platform \citep[Astronomy Common Object Model,][]{ASCOM_Platform} provides a software layer that allows control of equipment via universal interfaces. Numerous ASCOM drivers were developed at DSI with the help of students. The most relevant development was an ASCOM camera driver that directly processes the event time tags broadcasted by the TM-4, and provides the GPS time stamp with the image data so it can be written directly in the FITS header. 

After significant efforts, all components could be controlled via ASCOM by late 2020. This made the setup fully scriptable (e.g., via Python), and provided compatibility with all open source and commercial software packages available on Windows. Operations via ASCOM Remote were successfully demonstrated from Stuttgart, but proved impractical due to severe transatlantic bandwidth limitations (often \textless 2 Mbps), and the strict sequential operation of the protocol (e.g.\ an image needs to be fully received before the next exposure starts), introducing lag and significantly reducing shutter-open time. We plan to address automated operation with emerging software packages. 


\paragraph{Reliability}

Operating conditions at SRO were harsh, as components had to cope with significant dew and humidity especially in spring and fall, very low temperatures during winter nights, and very high temperatures in summer under a closed roll-off roof. The setup was exposed to the night sky whenever conditions allowed observations. Operations in the Sierra Nevada forest exposed the setup to dust, pollen, insects, and even smoke and ash from the 2021 Creek Fire.

Whenever possible, industrial-grade or enterprise-grade devices were selected for ancillary equipment, when available with extended temperature rating. This decision paid off --- none of the auxiliary components ever failed in the field, with the exception of an unbranded USB hub that erroneously promised to be ruggedized. A full list of the ancillary equipment is provided in Appendix~\ref{tab:periphery}.

After five years of use in the lab and on SOFIA, followed by almost eight years in the field, the main camera's cooling ability slowly degraded. \added{As night time temperatures rose during early summer of 2021, the camera's TEC cooler was unable to stabilize the sensor temperature at our standard setting of $-60\mathrm{\degr C}$, indicating a degradation of the hermetic sensor chamber seal. We did not observe ice crystals forming on the sensor; we envision that the vacuum quality degraded very slowly, and that the vacuum chamber pressure – and thus, partial water vapor pressure -- was still too low to form visible condensation. It is also conceivable that any water vapor froze out near the cold top of the TEC-stack at the back of the sensor before the sensor surface itself reached sub-zero temperatures. We continued regular operations at $-50\mathrm{\degr C}$ sensor temperature until fall that year without issues.} Despite its age, the manufacturer agreed to service the camera, renewed the sensor chamber's vacuum and seal, and restored its original cooling performance. Note that the sensor chamber exceeded its seven year warranty by almost a factor of two.

Most reliability issues \added{originated from two} electronic boards on the \added{OTA}. \added{The ``ATC-02'' PCB connects a microcontroller and distributes power to the M2 actuator, M1 fans, temperature sensor, and a daughter board for M1 shutter control. This board failed twice in identical fashion. We traced the issue to the failure of a Texas Instruments LM22676/-Q1 step-down voltage regulator (``buck converter'') that converts the 12~VDC input to 5~VDC for the microcontroller and temperature sensors. Once the buck converter fails, its output voltage exceeds 10~VDC and destroys the components it supplies. The buck converter is an automotive-grade component with a wide temperature range rating.} The potential cause was moisture, \added{presumably} causing a short circuit on the board. \added{Visual inspection of the PCB under a microscope revealed traces of humidity-induced corrosion and oxidation.} During repair, a protective coating was applied; the board is mounted without any enclosure or packaging under the primary mirror. Fortunately, instrument downtime was avoided during both failures, as a separate focuser had already been installed at the backplane, and the 12~VDC supply for the mirror shutter motors was unaffected. A second problematic component was \added{the daughter board --- an intermediate controller for the mirror shutter motors --- which did not fail electronically, but was} unable to reliably open \added{or close} all four mirror shutters \added{after receiving a command from the microcontroller}. This issue was resolved via a direct RS485 interface to the motor bus, \added{circumventing the problematic daughter board entirely.} 


\subsection{Adopted collimation procedure} \label{sec:ll_collimation}

We arrived at the conclusion that geometric methods do not sufficiently align mirrors to enable a seeing-limited image quality\footnote{\added{Image quality is seeing-limited when the PSF is dominated by atmospheric turbulence, and optical aberrations are sufficiently small that they do not contribute meaningfully to image degradation.}}, at least not at sites with 1-arcsecond seeing conditions or better --- the RC design is too sensitive to small mirror misalignments. Wavefront measurements on a bright star were crucial to measure and eliminate field coma. We will briefly recap the procedure we adopted that led to an image quality limited by seeing and detector sampling.

The RC's hyperbolic mirrors correct for spherical aberration and coma when properly aligned. Spherical aberration depends entirely on mirror spacing, which was optimized via systematic wavefront measurements (see Section~\ref{sec:optimizing_mirror_alignment}). However, the RC telescope can provide coma-free images even when the optical axes of M1 and M2 are not congruent. The coma-free condition is met when both mirror axes intersect at the coma-free point (CFP). 

In a coma-free RC, the remaining dominant aberration is third-order astigmatism. If both mirror axes are congruent, astigmatism is zero on-axis, and increases quadratically with field angle. If both mirror axes merely intersect at the CFP, the astigmatism's center of symmetry is no longer at the field center, resulting in unbalanced off-axis astigmatism. 

For large professional telescopes, where M1 tilt adjustments \added{are not always possible}, the typical approach is to tilt M2 around its CFP both to eliminate coma and balance astigmatism \citep{Gitton1998, Noethe2000}. This usually requires coordinated translation and rotation, e.g.\ via a hexapod. For small RC telescopes like ATUS, where M2 tip/tilt is achieved via simple adjustment screws on the back of the M2 mirror cell, a precise M2 movement around the CFP would be extremely difficult. 

Unlike large telescopes, ATUS is also equipped with M1 tip/tilt screws. If we disregard true axis congruence and primarily aim at the coma-free condition --- i.e.\, that both mirror axes intersect at the CFP --- it does not matter in principle whether M1 or M2 gets tilted. However, coma correction via M1 tilt provides the advantage that we can first establish M2 as a reference in the system, i.e.\ align the M2 axis with the mechanical axis of the instrument flange, and thus in good approximation with the detector center. 

\begin{figure*}[b]
\begin{center}
\resizebox{0.9\linewidth}{!}{%
\includegraphics[height=1cm]{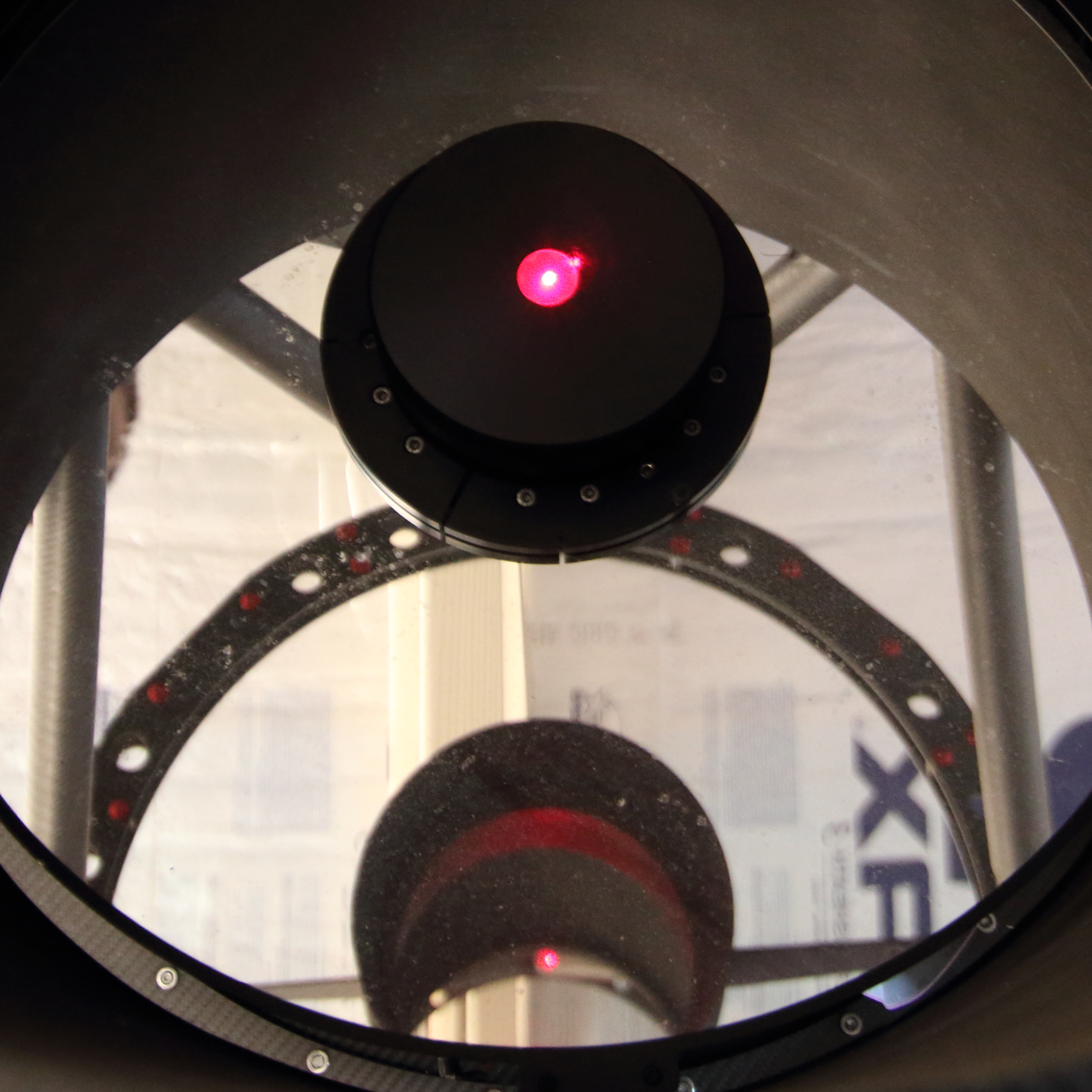} 
\includegraphics[height=1cm]{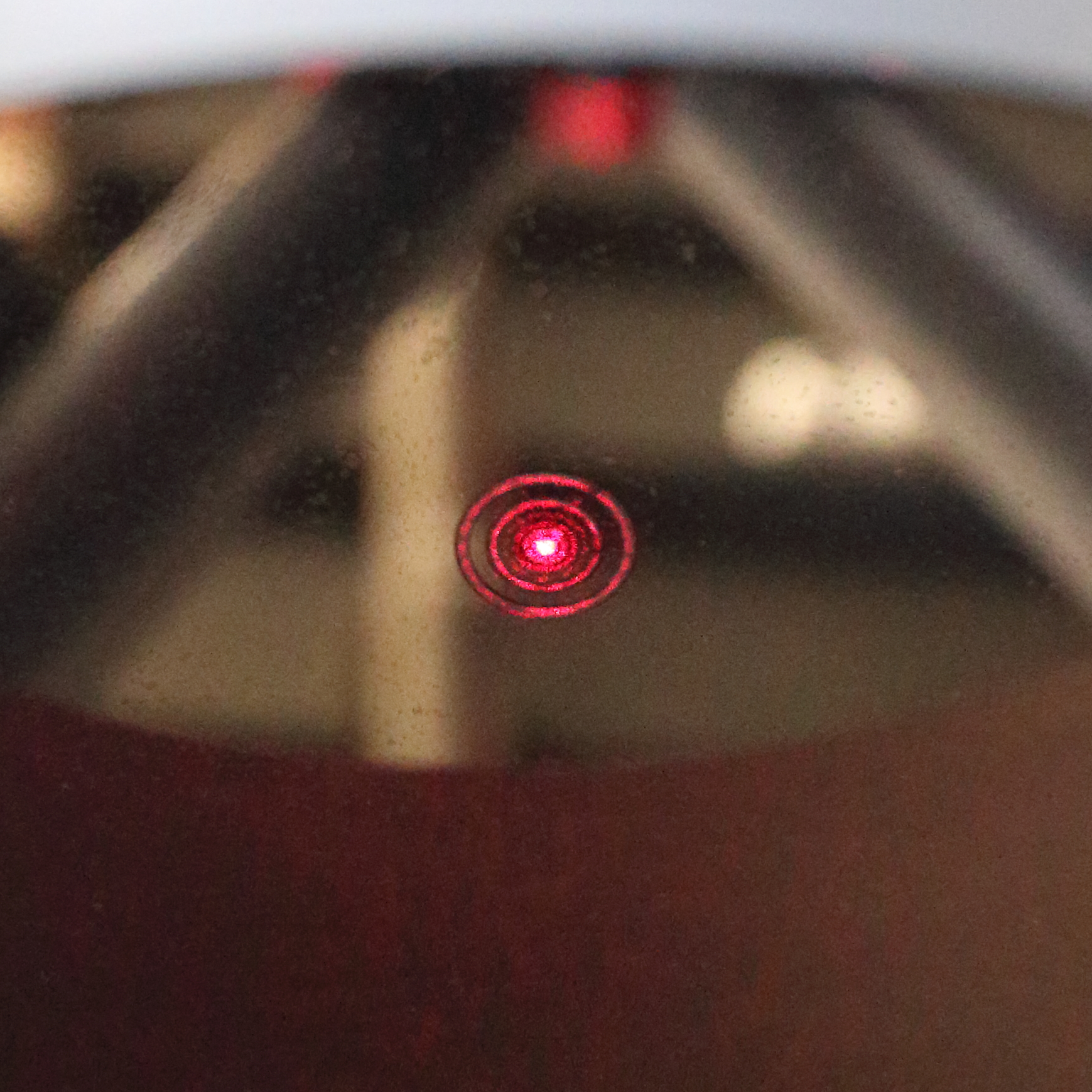} }
\end{center}
\caption 
{Collimation targets allowing initial mirror alignment with a laser mounted at the instrument flange. Left: A metal cap with a diffuse screen and pinhole at its center, screwed on top of the M1 baffle interface thread; the laser beam passes its pinhole and is reflected back from M2 on the edge of the screen. M2 tip/tilt adjustments bring the reflected beam in congruence with itself, effectively coaligning the backplane flange axis with the M2 axis. Right: The collimation laser hitting the M2 center mark (three concentric rings of 3, 7, and 11~mm diameter) after lateral M2 adjustment via its spider vanes. See text for more details.
\label{fig:collimation_target}}  
\end{figure*}

We accomplished this by using a collimation laser (2-inch 650~nm laser collimator with adjustable brightness, by Howie Glatter), installed in a self-centering, high-quality 2-inch eyepiece holder at the backplane flange. The M1 baffle was removed, a collimation target with a diffusing screen and a pinhole at its center screwed onto the M1 baffle's mounting thread (cf.\ Figure~\ref{fig:collimation_target}, left), and the OTA pointed to the zenith. Thanks to sufficient manufacturing tolerances, the laser beam is concentric with the collimation target's screen, passes its pinhole, and gets reflected at M2. The spider vanes are then adjusted such that the laser is centered on the M2 center mark (cf.\ Figure~\ref{fig:collimation_target}, right). Once that is achieved, the M2 is adjusted via its three tip/tilt screws (spaced 120\degr~on its backside) until the laser beam gets reflected back on itself, i.e.\ back on the target's center pinhole (note the non-centered reflection in Figure~\ref{fig:collimation_target}, left). 


Once the laser beam is perfectly centered on the M2 center mark and reflected back on itself, the M2 axis is aligned with the instrument flange axis. In a second step, a Shack-Hartmann WFS such as SHIFT is mounted at the backplane flange, and a bright star without a close companion is selected as a target. We then tilt M1 to null the coma terms of the wavefront measurement. Once this is accomplished, the M1 axis intersects the M2 axis at the CFP.

If the initial misalignment of the mirror axes after pre-alignment with the laser collimator was small, the angle between the M1 and M2 axes at the CFP should also be small, and the center of symmetry of the unbalanced astigmatism field should be close to the detector center.

Following this procedure, field-dependent astigmatism was not notable in our small FoV, indicating that the intersection angle at the CFP is sufficiently small not to cause a large displacement of the astigmatism's center of symmetry. 

For ATUS, this pragmatic collimation strategy was sufficient, given site conditions, detector size, and PSF sampling. While WFS measurements were only conducted on-axis, they indicated that the mirror alignment enables diffraction-limited image quality (cf.\ Figure~\ref{fig:ATUS_PSF}). Larger image sensors may require an adapted collimation procedure, iteratively balancing astigmatism via M2 tilt and subsequently eliminating coma via M1 tilt, ideally supported by additional off-axis WFS measurements.


\section{Conclusions and outlook}

This article documents the continuous, rigorous optimization of the ATUS telescope over its eleven-year lifetime. Few telescopes in this size class have been scrutinized at this level, and possibly even fewer have had this effort documented. We hope this article will guide other instrument builders to avoid pitfalls or to improve existing telescopes.

We achieved a mechanically-stiff Ritchey-Chrétien system with a stable collimation, a seeing- or detector-limited image quality, and without image degradation due to vibrations after a slew or during high-speed tracking. Its heavily customized mount and OTA can carry heavy instrumentation. The M1 is a center-hub mounted single-arch mirror, providing a good compromise between stiffness, weight, and cost of manufacturing. The M2 assembly is held at its center of gravity. Target acquisition is reliable at the sub-arcminute level, and a custom-built off-axis guider enables long-term tracking with the highest accuracy to obtain ultra-deep exposures. Stray light reflections from the frame-transfer-sensor's metalized surface area have been carefully mitigated.

Another unique aspect of ATUS is its very accurate and precise time referencing system, often not available at this level even on instruments at much larger telescopes. This capability, together with its ability to take images at high cadence and without dead time, enables optimal photometric measurements of transient events such as stellar occultations, exoplanet transits, and variable stars. ATUS is therefore ideally positioned for future tasks in time-resolved astronomy. Its reliable pointing and ability to instantly capture images upon reaching its target coordinates make it attractive for instant follow-up, such as alert-triggered observations of gamma-ray bursts in the optical domain. 

ATUS operated at Sierra Remote Observatories for more than 11 years until it was moved due to the end of the SOFIA project and the closure of DSI's offices in California. The overlap with hardware and software development activities for SOFIA provided mutual benefits. A new operational concept is being developed to continue operating ATUS at a new location, and we hope to report about the next decade of operations in a future paper.


\begin{acknowledgments}

{{SOFIA}}, the {{Stratospheric Observatory for Infrared Astronomy}}, was a joint project of the Deutsches Zentrum für Luft- und Raumfahrt e.V. (DLR; German Aerospace Center, grants: 50OK0901, 50OK1301, 50OK1701, and 50OK2002) and the National Aeronautics and Space Administration (NASA). It was funded on behalf of DLR by the Federal Ministry for Economic Affairs and Climate Action based on legislation by the German Parliament, the State of Baden-Württemberg, and the University of Stuttgart. {{SOFIA}} activities were coordinated on the German side by the German Space Agency at DLR and carried out by the German SOFIA Institute (Deutsches SOFIA Institut, DSI) at the University of Stuttgart, and on the U.S.\ side by NASA and the Universities Space Research Association (USRA). ATUS received funding from University of Stuttgart's budgetary funds for SOFIA, the State of Baden-Wuerttemberg's quality assurance funds, and DLR. 

We would like to thank Officina Stellare for the significant effort in redesigning and improving the OTA. Our special thanks go to Gianluca Carotta (design and structural mechanical simulation), Massimo Riccardi (optical design and manufacturing), Alessio Galliazzo (electronics and motor control), Paolo Spano, Walter Girardin (improvements of the mirror mount and imaging quality), Giovanni Dal Lago and Gino Bucciol at Officina Stellare for the good cooperation and long-term support, which have been essential for the development and continuous improvement of the Mk~II OTA.

The extensive modifications of the 3600GTOPE mount would not have been possible without the exceptional support and commitment of the manufacturer, willing to tackle new challenges even several years into the lifetime of its products. We would like to extend our appreciation to the team at Astro-Physics, in particular Howard Hedlund, Ray Gralak and Mike Hanson for their commitment and effort in continuously improving the controller firmware and software, even of older mounts, and their attention to reported issues and feature requests. Some new firmware and software functions were proudly tested in advance with ATUS before they were adopted in official releases to the benefit of all customers. We also would like to thank Brent Boshart for his many years of work on SkyTrack and the many emails, iterations and test runs we exchanged until achieving reliable tracking of objects in low Earth orbit; Ray Gralak was again instrumental to implement the necessary functionality and bug fixes in the mount's control software.

We also extend our gratitude to Robert C.\ Brummett, Michael T.\ Gaunce and Emmett A.\ Quigley at NASA Ames Research Center for their help in procurement of parts from local workshops, and Craig Binford of Warren Associates for donating one of the Bosch IP cameras for operational safety and site monitoring. 

Many colleagues and students contributed to the evolution and success of ATUS. We would like to highlight and thank in particular (all from University of Stuttgart): Clemens Berger (ASCOM driver for Andor iXon 888 with TM-4 time tag support \& code maintenance of other drivers), Andreas Dörr (focus and aperture control of the WFI's Canon EF lens with an Arduino microcontroller \& ASCOM driver), Jonas Früh (mechanical OAG design, support of SHIFT first light \& measurements), Timon Genthner (ASCOM driver for SBIG CFW-10 filter wheel \& mirror cover motors), and Bastian Knieling (SHIFT pipeline \& data analysis, supporting field measurements). 
Setup, maintenance, modifications, countless field campaigns, and ultimately the disassembly and shipment via sea cargo would not have been possible without the DSI team at Ames: Manuel Wiedemann, Michael Lachenmann, Enrico Pfüller, Jonas Früh \& Bastian Knieling, as well as many students coming along while visiting ARC for their master/bachelor thesis work and internships. The administrative support from our DSI colleagues in Stuttgart --- Thomas Keilig, Katja Paterson, and Monika Rößler --- was as crucial to successfully operate a telescope 9228~km away from our home base.

Last but not least, we would like to thank Samuel A.\ Miller and Evan Cornelsen for their on-site support during our eleven years at SRO, and Keith B.\ Quattrocchi and Larry Van Vleet for hosting us at SRO. In memory of Melvin R.\ Helm Jr.\ (\textdagger~December 24, 2021), co-founder of Sierra Remote Observatories, and Hans-Peter Röser (\textdagger~December 8, 2015), former department head of the Institute of Space Systems and an essential supporter who helped that ATUS became a reality.

This work has made use of data from the European Space Agency (ESA) mission \href{https://www.cosmos.esa.int/gaia}{\it Gaia}, processed by the \href{https://www.cosmos.esa.int/web/gaia/dpac/consortium}{{\it Gaia} Data Processing and Analysis Consortium (DPAC)}. Funding for the DPAC has been provided by national institutions, in particular the institutions participating in the {\it Gaia} Multilateral Agreement. \added{A figure in the appendix made use of Digitized Sky Survey (DSS2) data, produced at the Space Telescope Science Institute (STScI) under U.S.\ Government grant NAG W-2166. The underlying photographic plates were obtained with the Oschin Schmidt Telescope at Palomar Observatory.} This research also made use of \href{https://ui.adsabs.harvard.edu/}{NASA’s Astrophysics Data System Bibliographic Services} \added{and of Aladin, developed at the Strasbourg astronomical Data Center (CDS), France.}

\end{acknowledgments}

\vspace{5mm}
\facilities{ATUS, SOFIA}

\software{pandas \citep{McKinney2010,pandasdevelopmentteam2024},
          Astropy \citep{AstropyCollaboration2013,AstropyCollaboration2022, ASCL_Greenfield2013}, 
          photutils \citep{Bradley2025},
          astroquery \citep{Ginsburg2019},
          Matplotlib \citep{Hunter2007},
          \added{SciencePlots \citep{Garrett2021}},
          Numpy \citep{Harris2020},
          scikit-learn \citep{Pedregosa2011},
          SciPy \citep{Virtanen2020},
          AstroImageJ \citep{Collins2017, ASCL_Collins2013},
          \added{Aladin \citep{Bonnarel2000}},
          astrometry.net \citep{Lang2010, ASCL_Lang2012},
          ASCOM Platform \citep{ASCOM_Platform},
          Astro-Physics Command Center (APCC) Pro \citep{APCC},
          PinPoint \citep{PinPoint},
          Cygwin \citep{Cygwin},
          ansvr \citep{ansvr},
          SkyTrack \citep{SkyTrack},
          PHD2 \citep{PHD2}
          }

\newpage

\appendix
\counterwithin{table}{section}
\counterwithin{figure}{section}
\renewcommand{\thefigure}{\thesection\arabic{figure}}
\renewcommand{\thetable}{\thesection\arabic{table}}

\section{Technical details}
\subsection{Imager configuration} \label{sec:app_imager_configuration}


\begin{table}[htb]
\centering
\caption{Imager configuration as of October 2024.}
\label{tab:instrumentation_overview}

\begin{tabular}{p{3.7cm} Z{2.0cm} Z{2.0cm} Z{4cm} Z{4cm}}
\tablewidth{0pt}
\hline
\strut                       & \multicolumn{2}{c}{\Centerstack{Main instrument\\``FPI''}}                         & \Centerstack{Refractor\\``FFI''}                                      & \Centerstack{Wide-field imager\\``WFI''} \\
\hline\hline
Optical configuration       & \multicolumn{2}{c}{Ritchey-Chr\'etien}                      & Apochromatic refractor                            & Commercial telephoto lens \\
\hline
Model                       & \multicolumn{2}{c}{\Centerstack{Officina Stellare\\ProRC 600}}   & \Centerstack{Officina Stellare (LZOS)\\Hiper APO 130 (Triplet)}     & \Centerstack{Canon\\EF 135~mm $f/2.8$ Soft Focus} \\
\hline
Aperture \& focal ratio    & \multicolumn{2}{c}{600~mm $f/8.17$}                               & 130~mm $f/6$                                        & 48.2~mm $f/2.8$ \\
\hline
Camera model 				& \multicolumn{2}{c}{\Centerstack{Andor iXon\textsuperscript{EM+}\\DU-888E-C00-\#BV}} & QSI 632ws-8                    & \Centerstack{FLI ProLine PL4720\\(custom build)} \\
\hline
Sensor type 				& \multicolumn{2}{c}{{e2v CCD201-20 EMCCD}}           & \Centerstack{ON Semiconductor\\KAF-3200ME CCD}      & {e2v CCD47-20 CCD} \\
\hline
Active area (pixels)		& \multicolumn{2}{c}{1024~\texttimes~1024}                    & 2184~\texttimes~1472                              & 1024~\texttimes~1024 \\
\hline
Pixel size \& plate scale	& \multicolumn{2}{c}{13~\textmu m (0.55~arcsec/pixel)}        & 6.8~\textmu m (1.80~arcsec/pixel)                 & 13~\textmu m (19.9~arcsec/pixel) \\
\hline
Field of view       		& \multicolumn{2}{c}{9.4~\texttimes~9.4~arcmin\textsuperscript{2}} & 65.4~\texttimes~44.1~arcmin\textsuperscript{2}& 5.65~\texttimes~5.65~deg\textsuperscript{2}\\
\hline
Mechanical shutter 			& \multicolumn{2}{c}{No} & Yes & No \\
\hline
Filter wheel 				& \multicolumn{2}{c}{\Centerstack{SBIG CFW-10-SA\\(10 pos., 1.25-inch filter cells)}} & \Centerstack{Integrated in camera\\(8 pos., 1.25-inch filter cells)} & \Centerstack{FLI CFW-2-7\\(7 pos., 49.7~mm unmounted)}\\
\hline
Available filters   & \multicolumn{2}{c}{\begin{minipage}[t]{4.0cm} \begin{itemize}[leftmargin=*,itemsep=-0.5ex,label=--] \item{Astrodon Sloan 2\textsuperscript{nd} Gen.:\newline g'2, r'2, i'2, z\_s2 \& Clear (\textgreater~395 nm); parfocal} \item{Astrodon Exoplanet\newline ($\geq$ 500~nm)} \item{Astrodon \mbox{H-\textalpha} (3~nm BP)} \item{``VR''; Omega Optical\\500-700~nm} \item{Open / closed positions} \end{itemize} \end{minipage}} & {\begin{minipage}[t]{4cm} \begin{itemize}[leftmargin=*,itemsep=-0.5ex,label=--] \item{Astrodon Sloan 1\textsuperscript{st} Gen.:\newline g', r', i', z' \& Clear\newline (\textgreater~395~nm); parfocal} \item{Schuler \mbox{H-\textalpha} (9~nm BP)} \item{Astrodon O-III, S-II\\(3~nm BP)} \end{itemize} \end{minipage}} & {\begin{minipage}[t]{4cm} \begin{itemize}[leftmargin=*,itemsep=-0.5ex,label=--] \item{Parfocal Johnson/Bessel\newline UBVRI filter set} \item{Closed position} \end{itemize} \end{minipage}} \\
\hline
Focuser 		    & \multicolumn{2}{c}{FLI Atlas or M2 mechanism} & Optec TCF-Si 2-inch & Integrated in telephoto lens \\
\hline
Readout speed (fps)         & Unbinned & 2\texttimes2 binning  & Unbinned & Unbinned \\
\hline
\hspace{0.5cm}Full frame                  & 8.9      & 17.4         & 0.36     & 1.5 \\
\hspace{0.5cm}128~\texttimes~128 pixel AOI            & 65.5     & 119          & 0.46     & 5.6 \\
\hline
\noalign{\vskip 6pt}
\cline{1-3}
\strut                        & \multicolumn{2}{c}{\Centerstack{Off-axis guider\\(OAG)}} & \multicolumn{2}{r}{\multirow{9}{*}{%
  \begin{minipage}{8cm}
    \includegraphics[width=\linewidth]{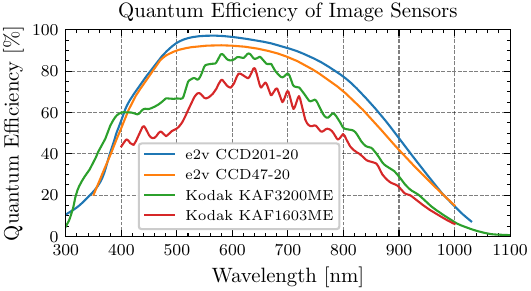}
    \captionof{figure}{Quantum efficiency of all image sensors.}
    \label{fig:QE}
  \end{minipage}%
}}    \\
\cline{1-3}\noalign{\vskip\doublerulesep}\cline{1-3}
Camera model 				& \multicolumn{2}{c}{QSI 616s} &&\\
\cline{1-3}
\Centerstack{Sensor type} 	& \multicolumn{2}{c}{\Centerstack{ON Semiconductor\\KAF-1603ME CCD}} \\
\cline{1-3}
Active area (pixels)		& \multicolumn{2}{c}{1536~\texttimes~1024} &&\\
\cline{1-3}
Pixel size \& plate scale	& \multicolumn{2}{c}{9~\textmu m (0.38~arcsec/pixel)} &&\\
\cline{1-3}
Field of view       		& \multicolumn{2}{c}{9.7~\texttimes~6.5~arcmin\textsuperscript{2}} && \\
\cline{1-3}
Mechanical shutter 			& \multicolumn{2}{c}{Yes} && \\
\cline{1-3}
Filter wheel 				& \multicolumn{2}{c}{none} && \\
\cline{1-3}\noalign{\vskip 2pt}
\parbox[c]{4cm}{\raggedright Angular separation of field center to main camera} & \multicolumn{2}{c}{21.5~arcmin} && \\
\noalign{\vskip 2pt}\cline{1-3}\noalign{\vskip 2pt}
\end{tabular}
\tablecomments{LZOS = Lytkarino Optical Glass Plant, Russia; QSI = Quantum Scientific Imaging; FLI = Finger Lakes Instrumentation; e2v --- now Teledyne e2v; ON Semiconductor --- formerly TrueSense/Kodak; BP = bandpass.}
\end{table}

\begin{figure}[htb]
\centering
\includegraphics[width=\plotwidth]{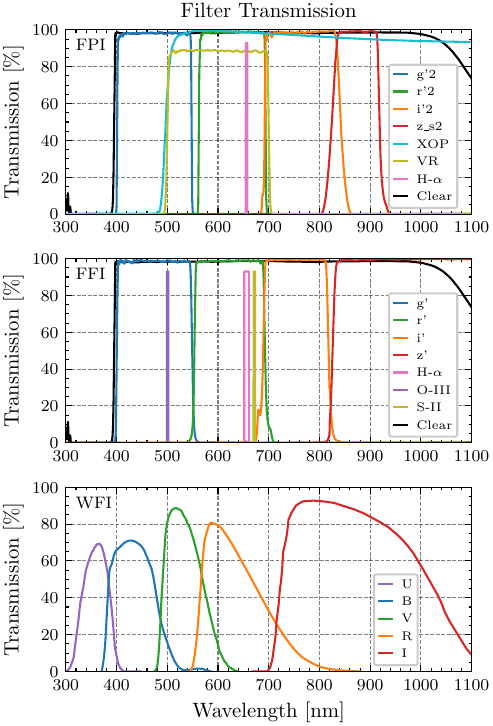}  
\caption 
{Transmission of all filters mounted in the respective imager's filter wheel, as of October 2024.
\label{fig:filter_transmission}}
\end{figure}

The properties of all imaging cameras on ATUS are documented in Table~\ref{tab:instrumentation_overview}; the quantum efficiency of their respective image sensors is shown in Figure~\ref{fig:QE}. Transmission curves of all currently mounted filters are documented in Figure~\ref{fig:filter_transmission}; other filters (e.g.\ Sloan u’2, Sloan Y’2, J, H) are available as well. Note that the filters of all three imagers are mounted in the converging optical beam. The effect on bandwidth, central wavelength, and peak transmission are negligible at an aperture ratio of $f/8.17$. 

While the FFI's Astrodon Sloan and WFI's Johnson/Bessel filter sets were designed parfocal (i.e.\ of identical filter substrate thickness), the refractive optics of the guide scope and telephoto lens have a wavelength-dependent, and thus filter-dependent focus position due to chromatic aberrations. 


\begin{figure*}[htb]
\centering
\includegraphics[width=.8\linewidth]{ 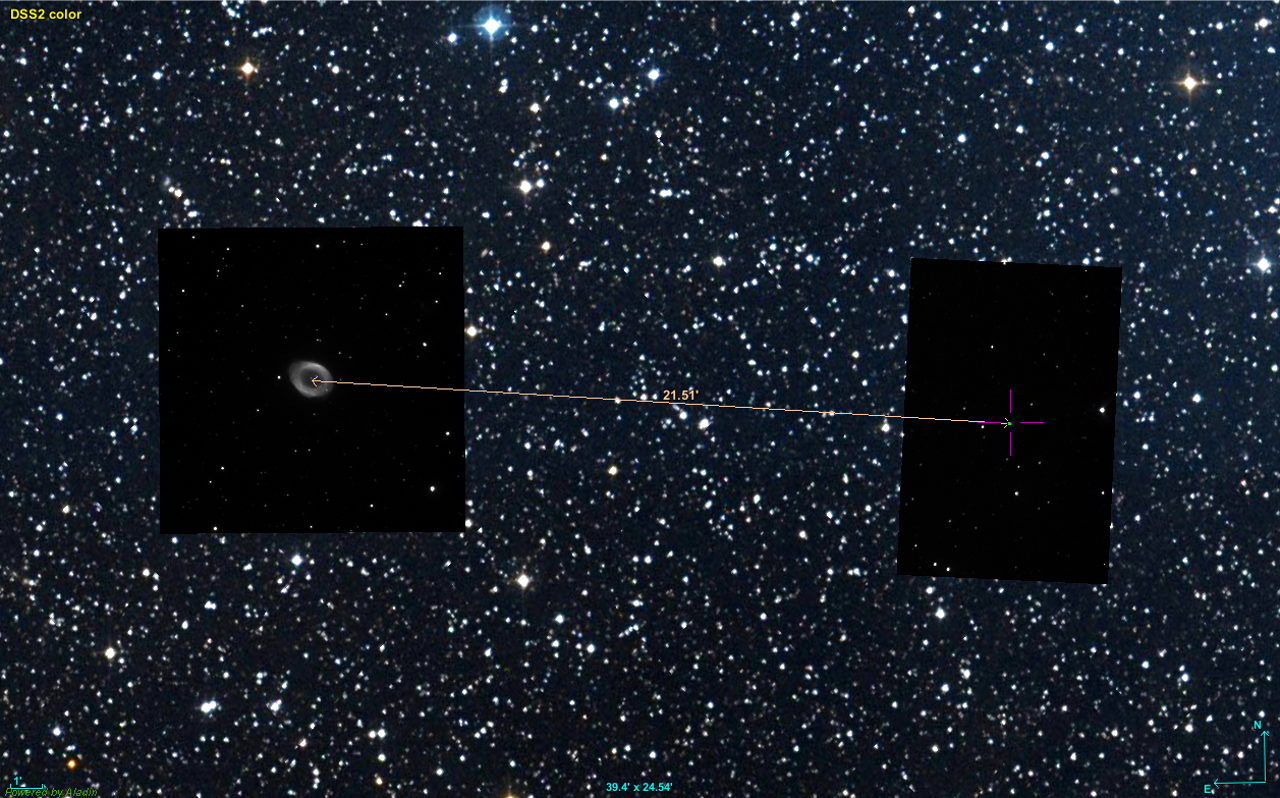 } 
\caption 
{Images of the main camera and OAG camera referenced on-sky, using the Ring Nebula (M57) as a target. Both field centers are separated by 21.5~arcmin. The OAG's FoV and cooled image sensor provide all-sky guide star coverage. \added{This figure was created using Aladin (Bonnarel et al. 2000); the false-color composite image in the background is based on the DSS2-red and -blue survey, with a synthetic green channel interpolated from red and blue data.}
\label{fig:OAG_FoV}}
\end{figure*}

The main camera has a large number of readout modes; see Table~\ref{tab:readout_modes}. Its EMCCD sensor has a full-well capacity of about 89000 $\mathrm{e^-}$. The slowest readout rate (1 MHz) provides the lowest possible read noise and provides maximum dynamic range with the 2.4\texttimes~preamplifier gain setting, making this the default readout mode for regular imaging operations. Faster readout rates have been routinely used for stellar occultation measurements as stellar brightness allowed, but are not meaningful for low-cadence imaging or deep exposures due to their high read noise. The angular separation between the field centers of the OAG camera and the main camera is about 21.5~arcmin, as illustrated in Figure~\ref{fig:OAG_FoV}.


\begin{table}[htb]
\centering
\caption{Readout modes of the main camera. The default readout mode for low-cadence imaging is marked in boldface.}
\label{tab:readout_modes}
\begin{tabular}{ccccccc}
\hline
Output Amplifier              & \Centerstack{Readout Rate\\[MHz]} & \Centerstack{ADC Resolution\\[bit]} & Pre-Amp Setting & \Centerstack{Gain\\[$\mathrm{e^-}$/ADU]} & \Centerstack{Read Noise\\[$\mathrm{e^-}$ rms]} & \Centerstack{Max. Pixel Value\\[ADU]}  \\
\hline
\hline
\multirow{3}{*}{\textbf{Conventional}} & \multirow{3}{*}{\textbf{1}}   & \multirow{3}{*}{\textbf{16}}  & 1     & 3.6           & 7.1   & 24518 \\
                              &                      &                      & \textbf{2.4}   & \textbf{1.5}           & \textbf{5.6}   & \textbf{60959} \\
                              &                      &                      & 5.1   & 0.7           & 5.8   & 65535\added{\textdagger} \\
\hline
\multirow{3}{*}{Conventional} & \multirow{3}{*}{3}   & \multirow{3}{*}{14}  & 1     & 9.4           & 12.9  & 9448  \\
                              &                      &                      & 2.4   & 3.8           & 9.7   & 16383\added{\textdagger} \\
                              &                      &                      & 5.1   & 1.7           & 9.1   & 16383\added{\textdagger} \\
\hline
\multirow{3}{*}{EM}           & \multirow{3}{*}{1}   & \multirow{3}{*}{16}  & 1     & 18.2          & 29.0  & 4885  \\
                              &                      &                      & 2.4   & 7.3           & 18.0  & 12158 \\
                              &                      &                      & 5.1   & 3.3           & 14.8  & 26888 \\
\hline
\multirow{3}{*}{EM}           & \multirow{3}{*}{3}   & \multirow{3}{*}{14}  & 1     & 44.0          & 47.1  & 2024  \\
                              &                      &                      & 2.5   & 18.2          & 30.1  & 4893  \\
                              &                      &                      & 5.2   & 8.3           & 24.7  & 10762 \\
\hline
\multirow{3}{*}{EM}           & \multirow{3}{*}{5}   & \multirow{3}{*}{14}  & 1     & 45.6          & 71.8  & 1950  \\
                              &                      &                      & 2.5   & 18.3          & 39.3  & 4858  \\
                              &                      &                      & 5.2   & 8.3           & 34.4  & 10710 \\
\hline
\multirow{2}{*}{EM}           & \multirow{2}{*}{10}  & \multirow{2}{*}{14}  & 2.5   & 21.1          & 47.1  & 4226  \\
                              &                      &                      & 5.2   & 9.7           & 42.2  & 9185  \\
\hline
\end{tabular}
\tablecomments{\added{\textdagger~=~Limit due to ADC, not due to full-well capacity ($\approx 89000\,\mathrm{e^-}$).}}
\end{table}



\subsection{Ancillary equipment} \label{sec:app_periphery}


\begin{table}[htb]
\centering
\caption{Ancillary equipment of ATUS.}
\label{tab:periphery}
\footnotesize

\begin{tabular}{Z{0.12\linewidth} Z{0.61\linewidth} Z{0.02\linewidth} Z{0.03\linewidth} Z{0.15\linewidth}}  
\hline
Function         & Device                                                                                                           & NM & Class & \parbox[c]{0.14\textwidth}{\centering Continuous Operation}                                                                             \\
\hline\hline
\multirow{3}{*}{IP camera}        & 3 -- 9~mm lens, IP66 (Bosch FLEXIDOME IP starlight 7000 VR, NIN-733-V03P)                                               & \checkmark & R     & since 09/2013                                                                                   \\
                 & 1.8 -- 3~mm lens, IP67 (Bosch FLEXIDOME IP starlight 7000 RD, NDN-733V02-P)                                             & \checkmark & R     & since 09/2013                                                                                   \\
                 & 2.5~mm, $f/2.8$ lens, for indoor use (Bosch Tinyon IP 2000 PIR, NPC-20012-F2L)                                                         & \checkmark & C     & since 04/2017                                                                                   \\
\hline
UPS~             & 1000 W Tower UPS with external
  battery module (DELL K788N, K806N) & \checkmark & E     & since 09/2013\textdaggerdbl \\
\hline
Network switch   & PoE managed 10-port gigabit switch (Cisco Small Business SG300-10P)                                                      & \checkmark & E     & since 09/2013                                                                                   \\
Network switch   & Unmanaged 5-port gigabit switch (Cisco Small Business SG100D-05)                                                         &   & E     & since 10/2015                                                                                   \\
\hline
PC               & \Centerstack{Former FDC control PC in ATR enclosure,\\ environmentally
  tested for flight; flown on SOFIA 2010-2012; Windows 7}            & \checkmark & R     & 09/2013 -- 10/2018\textdagger                                                                               \\
PC               & \Centerstack{Industrial PC with PCIe extension module (Advantech MIC-7500),\\extended temperature range; Windows 10/11}    & \checkmark & R     & since 10/2018                                                                                   \\
\hline
PDU              & Rack power distribution unit, 16 NEMA 5-15P
  outlets (APC AP7931)                           &\checkmark & E     & 09/2013 -- 05/2024\textdaggerdbl
                        \\
PDU              & 1U rack PDU, 8\texttimes~C13 outlets (Digital Loggers, Inc.\ Universal Smart PDU)              & \checkmark & E     & since 05/2024                                                                                   \\
PDU              & 8\texttimes~relay board with terminal
  blocks (Digital Loggers, Inc.\ DIN Relay IV)                                           & \checkmark & E     & since 06/2024                                                                                   \\
\hline
Serial hub       & 8-port RS232 hub (Moxa UPort 1610-8)                                                                             &   & R     & since 09/2013                                                                                   \\
Serial converter & 1-port RS232/422/485 USB converter (Moxa Uport 1150)                                                             &   & R     & since 10/2015                                                                                   \\
USB hub          & 7-Port USB 2.0 powered hub (no-name, metal enclosure)                                                            &   & C     & 09/2013 -- 05/2016                                                                                 \\
USB hub          & 7-port USB 2.0 Hub (Moxa UPort 407-T), wide temperature range                                                    &   & R     & since 05/2016                                                                                   \\
\hline
KVM              & KVM-over-IP Switch (Aten CN8000, VGA)                                                                            & \checkmark & C     & 10/2013 -- 03/2021\textdagger                                                     \\
KVM              & KVM-over-IP Switch (Aten CN9600, DVI)                                                                            & \checkmark & C     & since 03/2021                                                                                   \\
\hline
Env. sensor      & Temperature/humidity sensor (Practical Design Group, THUM)                                                       &   & n/a   & 10/2013 -- 03/2021\textdagger                                                     \\
Env. sensor      & Temperature/pressure/humidity/GPS sensor (Astromi.ch MGBox V2)                                                   &   & n/a   & since 03/2021 \\
\hline
\end{tabular}
\tablecomments{NM = Network manageable; Class: R = Ruggedized, E = Enterprise-grade, C = Consumer-grade; \textsuperscript{\textdagger} = Functioning, but obsolete / replaced with a higher-performance component; \textsuperscript{\textdaggerdbl} = Functioning, but decommissioned to prepare for 220 -- 240~VAC, 50~Hz power grid.}
\end{table}



\begin{figure*}[htb]
\begin{center}
\resizebox{0.9\linewidth}{!}{%
\includegraphics[height=1cm]{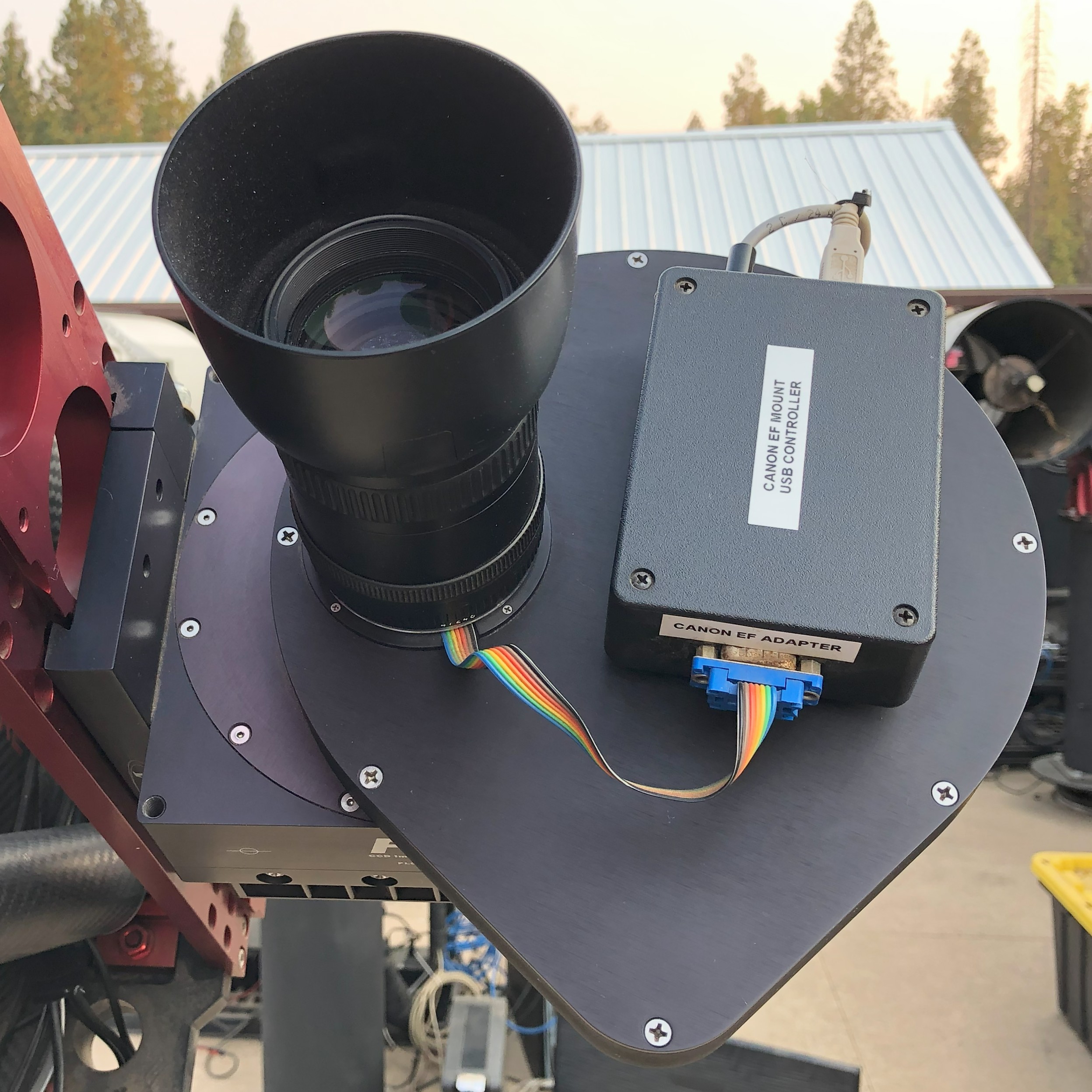} 
\includegraphics[height=1cm]{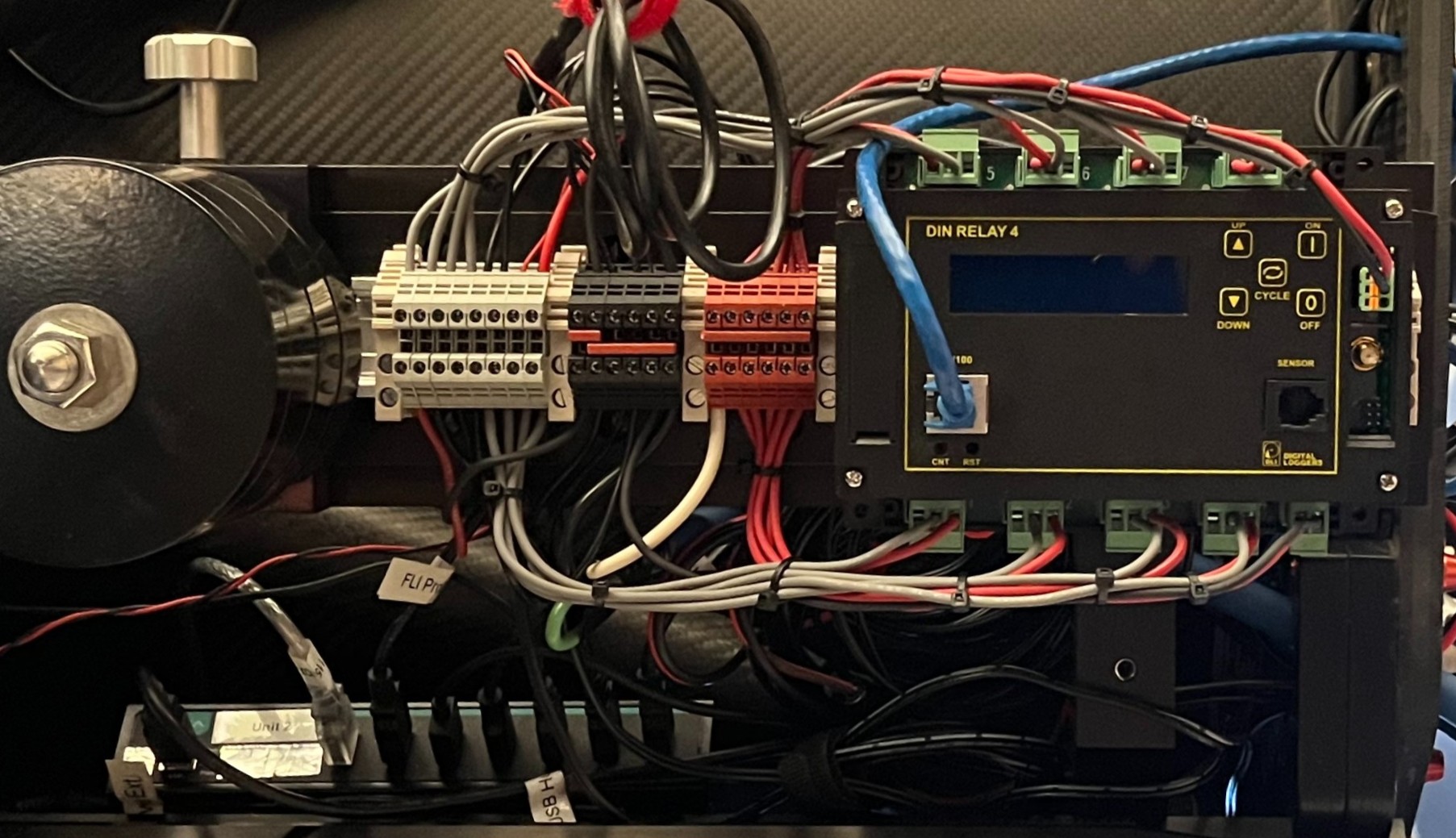} 
}
\end{center}
\caption 
{Left: The Wide Field Imager (WFI) uses a commercial Canon EF 135~mm $f/2.8$ Soft Focus telephoto lens whose focus motor is controlled by an Arduino microcontroller via a custom interface board with spring-loaded contacts. Right: A single 12~VDC power line is distributed to eight loads via a power distribution unit (PDU) with network-controlled relays.
\label{fig:WFI+PDU}}  
\end{figure*} 


Table~\ref{tab:periphery} documents all auxiliary devices used to operate ATUS. Cables run through the mount's axes shafts and are lead through the rear end of the dovetail saddle plate for distribution. Cabling was minimized to maintain a flexible and durable harness. With most auxiliary devices requiring 11 -- 14~VDC input, a single 14~AWG power line from a 20~A linear power supply runs through the mount and connects to a DIN rail terminal block mounted near the WFI. From there, power is distributed via a network-controlled relay card (Digital Loggers DIN Relay IV) that switches the respective loads as desired and also meters the input voltage (see Figure~\ref{fig:WFI+PDU}). To avoid stiff harness during cold winter nights, a cable with a jacket that remains flexible at very low temperatures and provides good environmental and abrasion resistance was used (Polar Wire Arctic Ultraflex). A network-managed, line-interactive, uninterruptible power supply (UPS) provided brownout protection and ample run time ($\approx$ 4 h) during power outages. The telescope PC would cut power to the mount and peripheral devices and perform an orderly shutdown once the UPS charge dropped below 10\%. During the summer of 2020, SRO installed a backup generator with a propane tank which limited power outages to less than a minute for switchover, making a large UPS for each individual setup obsolete. A solar array with battery storage, slated to provide true uninterruptible power site-wide for all customers, was eventually completed in spring 2025, i.e.\ after ATUS had already moved out of SRO. 

A pair of IP cameras with exceptional low-light performance (Bosch FLEXIDOME IP Starlight 7000 VR \& RD) provides \added{near-instant visual} feedback to remote operators. Initial issues with the intermediate controller board for the M1 shutters triggered the installation of a compact, yet less sensitive IP camera (Bosch Tinion 2000) via a custom 3D-printed mounting bracket at the OTA's front ring, providing \added{visual confirmation} that the mirror cover has fully opened \added{or closed}. A custom 850~nm, 1~W LED could be triggered remotely, illuminating the inside of the OTA light shroud at night. All three cameras are powered via Power over Ethernet (PoE) from a network switch. Reliable shutter motor control was eventually achieved via a direct RS-485 interface to the linear bus of the four DC motors. 

Table~\ref{tab:periphery} gives an anecdotal overview of all auxiliary devices and their exceptional reliability in the field. None of the image train components listed earlier in Table~\ref{tab:instrumentation_overview} experienced a catastrophic failure at SRO either. Devices that got decommissioned were either obsolete or tailored to the US power grid. 


\subsection{Weight breakdown} \label{sec:app_weight_breakdown}

The upgrade of the counterweight assembly initially increased its total mass from 156.6~kg to 258.8~kg. Later on, the telescope's weight distribution got optimized by mounting both auxiliary imagers on dovetail bars that are bolted directly to the OTA's dovetail interface plate. Now requiring less moment for balancing, the counterweight assembly's weight got reduced to 233.9~kg. This number includes the weights of the shaft (20.4~kg), a movable counterweight (17~kg) for fine balancing, the fixed counterweight stack at the end of the shaft (5~\texttimes~\nicefrac{3}{4}-inch, 1~\texttimes~\nicefrac{1}{2}-inch, 1~\texttimes~\nicefrac{1}{4}-inch thick \added{galvanized steel plates}, with individual weights of 31.5~kg, 21~kg, and 10.7~kg each), and a stainless steel plate that secures the counterweight stack to the end of the shaft. To carry this immense load, the end cap on the declination axis holding the shaft, previously a weight-optimized aluminum part (3.1~kg), was replaced by a solid part made of stainless steel (15.75~kg, which is included in the weight of the modified declination axis in Table~\ref{tab:weight}).
For balancing in declination, a ``D-Series'' dovetail saddle holding 3~\texttimes~3.5~lbs (1.59~kg) and one 1.75~lbs (0.79~kg) steel counterweight is mounted behind the WFI, compensating the weight of the refractor setup.


\begin{table}[htb]
\centering
\caption{System weight breakdown.} 
\label{tab:weight}
\begin{tabular}{c c r c r} 
\hline
                                 & Item & \multicolumn{3}{c|}{Weight [kg]}   \\
\hline\hline
\multirow{6}{*}{OTA}             & M2 assembly incl.\ focus mechanism                & 12.9  & \rdelim\}{6}{6pt} & \multirow{6}{*}{131.2} \\
                                 & Spider vanes                                    & 1.0   && \\
                                 & M1 baffle                                        & 1.2   && \\
                                 & M1\textsuperscript{\textdagger}                  & 37.1  && \\
                                 & Dovetail sliding bar (interface to mount)        & 10.9  && \\
                                 & Truss, M1 cell, fans, cabling, PCBs\textsuperscript{\textdaggerdbl}   & 68.1  && \\
\hline
\multirow{8}{*}{\Centerstack{Auxiliary\\equipment}} & Main image train with OAG     & 9.7   &\rdelim\}{8}{6pt} & \multirow{8}{*}{46.8} \\
                                & Refractor (incl.\ all auxiliary items)             & 13.6  && \\
                                & Wide Field Imager                                 & 5.4   && \\
                                & 2\texttimes~extra dovetail bars, D-series                  & 7.0   && \\
                                & Dovetail mounting brackets \& bolts               & 2.3   && \\
                                & OTA counterweights                                & 6.0   && \\
                                & 12~VDC power distribution unit (PDU)                & 0.9   && \\
                                & Auxiliary electronics                             & 1.9   && \\
\hline
\multirow{6}{*}{Mount}          & Dovetail saddle plate                             & 15.0  & \rdelim\}{6}{6pt} & \multirow{6}{*}{369.5} \\
                                & Declination axis with modified end cap (15.75 kg)                         & 52.0  && \\
                                & Right Ascension axis in modified polar fork       & 68.6  && \\
                                & Counterweight shaft (3.5-inch diameter)             & 20.4  && \\
                                & Movable counterweight for fine balancing          & 17.0  && \\
                                & Counterweight stack                               & 196.5 && \\                               
\hline
Pier                            & 18-inch tall, custom-built for SRO                  & 87.8  &&   \\
\hline\hline
\Centerstack{Total weight\\on foundation}  & \Centerstack{Excluding power / data cables,\\sand in pier ($\approx$~18 kg)}  & 635.3 &&\\
\hline
\end{tabular}
\tablecomments{All components have been weighed directly, with the exception of the primary mirror (\textsuperscript{\textdagger}; estimated from CAD model) and the bare OTA truss structure (\textsuperscript{\textdaggerdbl}, calculated by subtracting the weights of its subcomponents from the measured total OTA weight). Main image train: includes all components mounted on the backplane of the RC600 telescope (spacers and adapters, focuser, OAG with helical focuser and camera, main camera with filter wheel). Refractor: includes focuser, camera with filter wheel, spacers and adapters, tube rings, saddle plates for mounting on dovetail bar. Wide Field Imager: includes camera, filter wheel, telephoto lens with mechanical adapter, Arduino microcontroller in plastic housing. }
\end{table} 


\subsection{Geometric measurements of mirror spacing}\label {sec:app_spacing}

\begin{figure}[htb]
\centering
\includegraphics[width=\plotwidth]{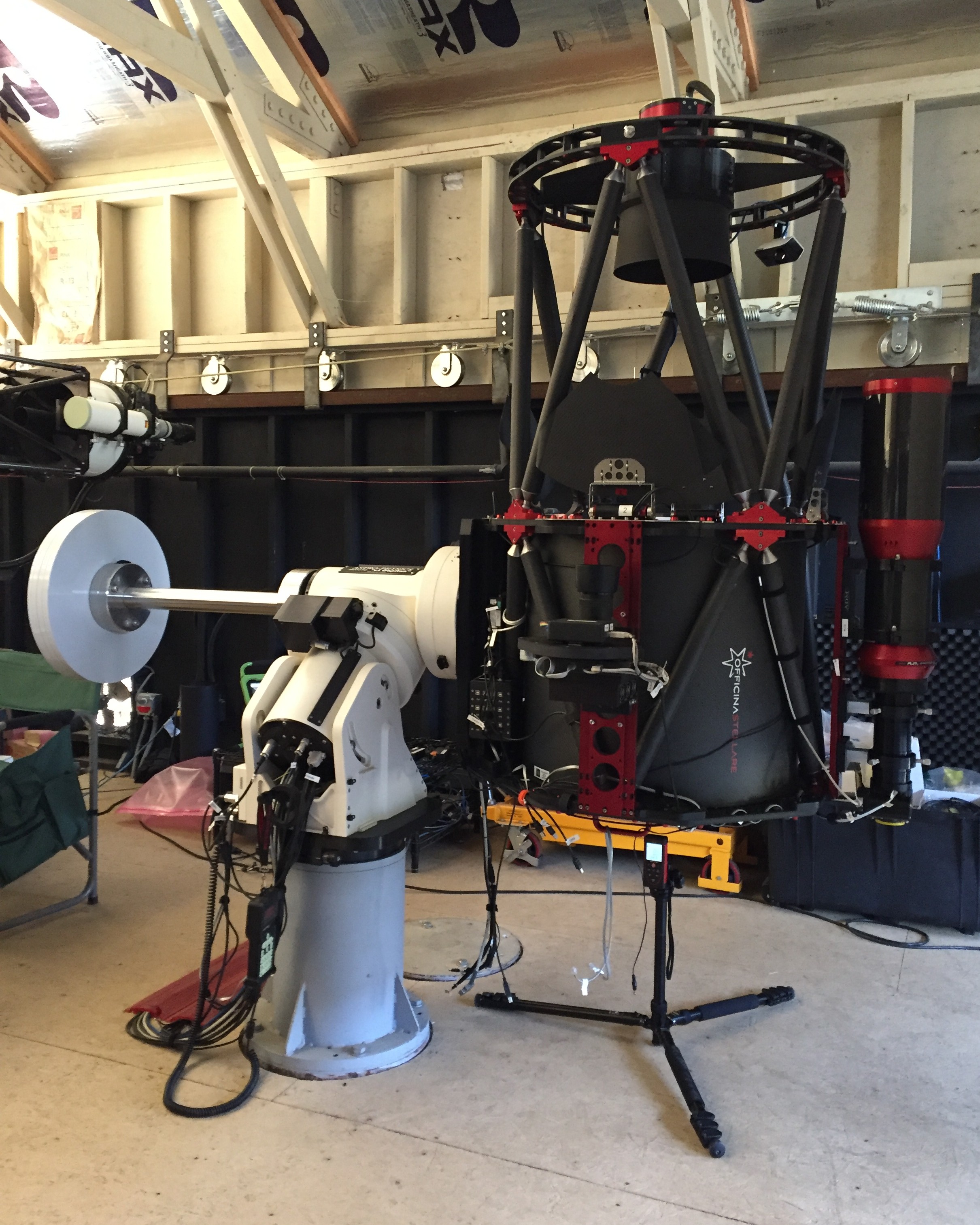} 
\caption 
{Using a Leica Disto X310 laser distance meter, the distance between M2 and a reference surface above M1 got measured at various M2 positions. These measurements were crucial to establish the relationship between mirror spacing, effective focal length, and flange focal distance (Figures~\ref{fig:f_vs_s} and \ref{fig:FFD}).
\label{fig:Leica}}  
\end{figure}

Figure~\ref{fig:Leica} visualizes how measurements with the laser distance meter were conducted. After detaching the image train, the telescope got pointed to the zenith, and the M1 stray light baffle --- screwed onto a large thread on top of the M1 cell's center hub --- removed and replaced with a calibration target disk containing a small diffuse screen with a pinhole at its center (same as in Figure~\ref{fig:collimation_target}, left). Mounted to a ball head on a sturdy tripod, the distance meter would then be placed under the telescope and iteratively adjusted, until the laser would pass the pinhole and hit the M2 center mark. The reflection of the laser beam falls on the diffuse screen to assess and reasonably adjust incidence angle. To prevent damage to the meter's receiver optics --- not designed to measure against a fully reflective surface --- the transparent adhesive portion of a $\approx$ 1-inch Post-It flag was placed onto the M2 center mark, creating a diffuse reflection, while the colored, non-adhesive tip of the flag allowed for an easy removal without any residue once measurements concluded. Anecdotally, in one instance we even found the method effective at removing organic residue from a deceased insect stuck on the M2. Measurements were repeated many times over the years without experiencing issues in removal or degradation of the silica-protected coating. Note that the M2's central portion does not receive rays from the sky due to its own obscuration. 

Once adjusted, the M1 calibration target was removed, and two distances were measured and subtracted from each other at each M2 position: (1) to the M2 vertex, (2) to a flat aluminum plate resting on the top rim of the threaded baffle flange at the center of the M1 cell. Each measurement was repeated several times and averaged. Measuring the spacing from the baffle thread to the M1's surface, and accounting for M1 hyperbolic curvature then allowed calculation of the M1 -- M2 spacing.


\begin{figure}[htb]
\centering
\includegraphics[width=\plotwidth]{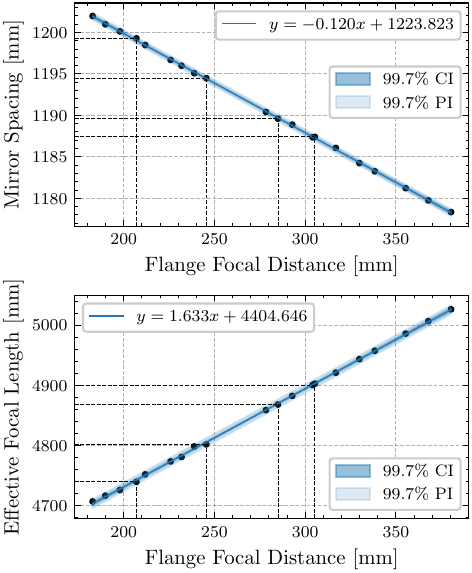} 
\caption 
{Effective mirror spacing and focal length as a function of flange focal distance that can be easily measured at the telescope backplane (CI: Confidence Interval; PI: Prediction Interval). Measurements correlate with a coefficient of determination of $R^2=0.99978$ and $R^2=0.99962$, respectively. The black dashed lines mark how the instrument evolved over time, and correspond to the markers in Figure~\ref{fig:f_vs_s}.
\label{fig:FFD}}
\end{figure}

Earlier, Figure~\ref{fig:f_vs_s} provided the correlation between focal length and mirror spacing (see Section~\ref{sec:mirror_spacing} for details). While this information is relevant from an optical design perspective, a more important metric for telescope operations is the flange focal distance (FFD). Figure~\ref{fig:FFD} provides the correlation of mirror spacing and focal length as a function of FFD. At its spherical-aberration-free point, ATUS operates at an FFD of 305~mm and a focal length of 4900~mm. This also dictates the setup's envelope, most notably the minimum pier height to provide sufficient ground clearance.


\begin{figure*}[htb]
\centering
\includegraphics[width=\plotwidth]{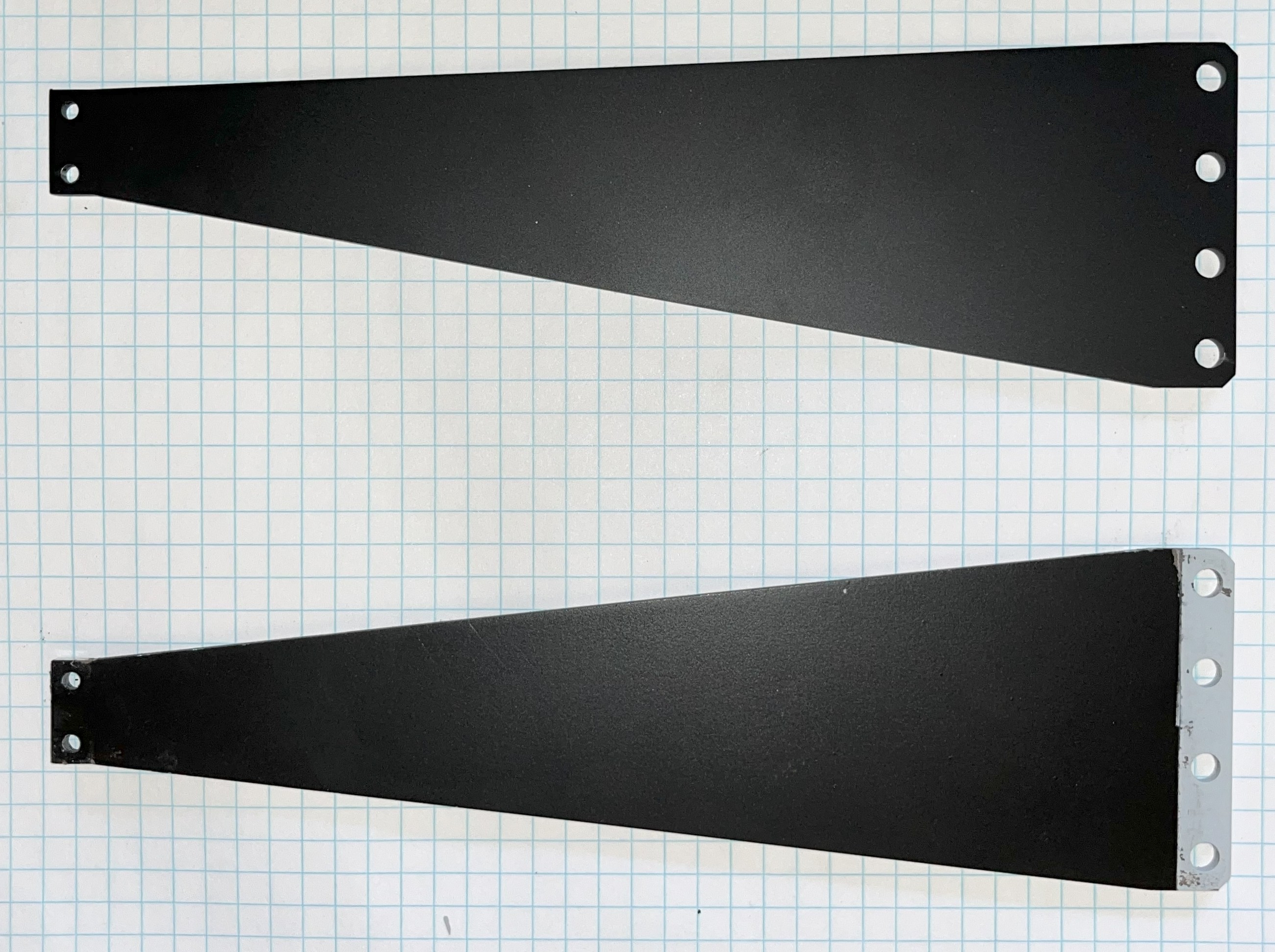} 
\caption 
{New (top) vs. old (bottom) spider vanes. The left side is attached to pull screws at the circumference of the \added{front} ring; the M2 assembly's mounting points are on the right. The new spider \added{vanes} moved the entire M2 assembly towards M1 by 14~mm. 
\label{fig:spider_vanes}}  
\end{figure*}

To decrease mirror spacing by $\approx$~12~mm, new spider vanes (Figure~\ref{fig:spider_vanes}) were required (see discussion in  Section~\ref{sec:optimizing_mirror_alignment}). Earlier WFS measurements from 2018 were sparse and limited by the M2 travel range (cf.\ Figure~\ref{fig:SphericalAberration}); they indicated zero spherical aberration at a mirror spacing of about 1186~mm. The new spider vanes were thus designed with a 14~mm offset of their M2 attachment points from the front ring's median plane. The considerably more detailed SHIFT verification campaign later led to a refined value of 1187.4~mm. The small discrepancy was easily compensated for via the M2 mechanism, still providing ample travel range for focusing. 


\section{Summary of modifications}\label {sec:summary_mod}

Table~\ref{tab:modifications_summary} summarizes all modifications described in this article.


\begin{table}[t]
\centering
\footnotesize
\caption{Summary of all modifications mentioned in this article.} 
\label{tab:modifications_summary}
\makebox[\textwidth][c]{%
\begin{tabular}{p{9.5cm} p{9.5cm}} 
\hline
Modification & Result \\
\hline\hline
Mark II OTA \\
\hspace{3mm}Overall FEA-assisted structural redesign & Preservation of mirror alignment, reduced flexure \\
\hspace{3mm}Proper M1 cell stiffness \& M2 cell load transfer & Preservation of mirror alignment \\
\hspace{3mm}Conical M1 backside, titanium truss joints & Weight reduction \\ 
\hspace{3mm}New M2 mechanism, high-resolution actuator, stepper motor \& encoder & Reproducible M2 movements while preserving mirror alignment \\
\hspace{3mm}Revised M2 stepper motor firmware & M2 moving at true actuator resolution \\
\hspace{3mm}Direct RS485 interface to linear bus of mirror shutter motors & Reliable shutter motor operation \\
\hspace{3mm}Laser collimation target for M1 cell & Improved coarse alignment of mirrors \& image quality \\
\hspace{3mm}M1 \& M2 cell modifications eliminating stress and M1 flop & Improved image quality \& pointing \\
\hspace{3mm}Corrected mirror spacing (after M2 spider vane replacement)  & Improved image quality (no spherical aberration), larger FFD \\
\hspace{3mm}WFS-assisted collimation & Improved fine alignment of mirrors \& image quality (coma compensated) \\
\hline
Mount \\
\hspace{3mm}Stiffening of polar fork assembly               & Larger load capacity, higher natural frequencies \\
\hspace{3mm}Original counterweight shaft with stack of steel plates     & Larger load capacity \& moment of inertia, lower natural frequencies \\
\hspace{3mm}Thicker \& shorter counterweight shaft, more steel plates & Higher natural frequencies, smaller moment of inertia \\
\hline
Overall instrument \\
\hspace{3mm}Addition of custom off-axis guider & Long exposures without image degradation \\
\hspace{3mm}Replacing individual 12~VDC power supplies (PSUs) with master PSU & Less cabling and weight, elimination of cable movement \& slack \\
\hspace{3mm}Knife-edge stop above camera sensor & Eliminated stray light reflections, improved photometric calibration \\
\hspace{3mm}Revised auxiliary imager placement & Smaller moment of inertia \& counterweight, no external loads on OTA \\
\hspace{3mm}Custom ASCOM camera driver incorporating TM-4 event time tags & All images time-referenced \textless 1~\textmu s, independent of control software \\
\hspace{3mm}ASCOM drivers for all auxiliary devices & Enabling script-based automation and broad compatibility to commercial and open-source software packages \\
\hline
\end{tabular}}
\end{table} 


\clearpage

\phantomsection
\addcontentsline{toc}{section}{References}
\bibliography{references}{}

@Article{Pfueller2018,
  author   = {{Pf{\"u}ller}, Enrico and {Wolf}, J{\"u}rgen and {Wiedemann}, Manuel},
  journal  = {Journal of Astronomical Instrumentation},
  title    = {{The SOFIA Focal Plane Imager: A Highly Sensitive and Fast Photometer for the Wavelength Range 0.4 to 1 Micron}},
  year     = {2018},
  month    = jan,
  number   = {4},
  pages    = {1840006},
  volume   = {7},
  adsurl   = {https://ui.adsabs.harvard.edu/abs/2018JAI.....740006P},
  doi      = {10.1142/S2251171718400068},
  eid      = {1840006},
}

@Article{Lang2010,
  author        = {{Lang}, D. and {Hogg}, D.~W. and {Mierle}, K. and {Blanton}, M. and {Roweis}, S.},
  journal       = {\aj},
  title         = {{Astrometry.net: Blind Astrometric Calibration of Arbitrary Astronomical Images}},
  year          = {2010},
  month         = may,
  pages         = {1782-1800},
  volume        = {139},
  adsurl        = {http://adsabs.harvard.edu/abs/2010AJ....139.1782L},
  archiveprefix = {arXiv},
  doi           = {10.1088/0004-6256/139/5/1782},
  eprint        = {0910.2233},
  primaryclass  = {astro-ph.IM},
}

@Book{WilsonRTO2,
  author    = {{Wilson}, Raymond N.},
  publisher = {{Springer Berlin, Heidelberg}},
  title     = {{Reflecting Telescope Optics II -- Manufacture, Testing, Alignment, Modern Techniques}},
  year      = {2001},
  edition   = {Corrected},
  series    = {{Astronomy and Astrophysics Library}},
  doi       = {10.1007/978-3-662-08488-5},
}

@Article{Temi2018,
  author   = {{Temi}, Pasquale and {Hoffman}, Douglas and {Ennico}, Kimberly and {Le}, Jeanette},
  journal  = {Journal of Astronomical Instrumentation},
  title    = {{SOFIA at Full Operation Capability: Technical Performance}},
  year     = {2018},
  month    = jan,
  number   = {4},
  pages    = {1840011-186},
  volume   = {7},
  adsurl   = {https://ui.adsabs.harvard.edu/abs/2018JAI.....740011T},
  doi      = {10.1142/S2251171718400111},
  eid      = {1840011-186},
}

@PhdThesis{Wiedemann2016,
  author = {{Wiedemann}, Manuel},
  school = {{Universität Stuttgart}},
  title  = {{Improving the Sensitivity of the SOFIA Target Acquisition and Tracking Cameras}},
  year   = {2016},
  type   = {Dissertation},
}

@InProceedings{Pfueller2018a,
  author    = {{Pf{\"u}ller}, Enrico and {Wolf}, J{\"u}rgen and {Schindler}, Karsten and {Person}, Michael J.},
  booktitle = {{Ground-based and Airborne Instrumentation for Astronomy VII}},
  title     = {{Adding a second spectral channel to the SOFIA FPI+ science instrument}},
  year      = {2018},
  month     = jul,
  note      = {Society of Photo-Optical Instrumentation Engineers (SPIE) Conference Series},
  pages     = {107022V},
  series    = {\procspie},
  volume    = {10702},
  adsurl    = {https://ui.adsabs.harvard.edu/abs/2018SPIE10702E..2VP},
  doi       = {10.1117/12.2313663},
  eid       = {107022V},
}

@InProceedings{Wiedemann2012,
  author    = {{Wiedemann}, Manuel and {Wolf}, J{\"u}rgen and {Roeser}, Hans-Peter},
  booktitle = {{Ground-based and Airborne Telescopes IV}},
  title     = {{Upgrade of the SOFIA target acquisition and tracking cameras}},
  year      = {2012},
  editor    = {{Stepp}, Larry M. and {Gilmozzi}, Roberto and {Hall}, Helen J.},
  month     = sep,
  note      = {Society of Photo-Optical Instrumentation Engineers (SPIE) Conference Series},
  pages     = {84442T},
  series    = {\procspie},
  volume    = {8444},
  adsurl    = {https://ui.adsabs.harvard.edu/abs/2012SPIE.8444E..2TW},
  doi       = {10.1117/12.925583},
  eid       = {84442T},
}

@Article{Souza2006,
  author   = {{Souza}, Steven P. and {Babcock}, Bryce A. and {Pasachoff}, Jay M. and {Gulbis}, Amanda A.~S. and {Elliot}, J.~L. and {Person}, Michael J. and {Gangestad}, Joseph W.},
  journal  = {\pasp},
  title    = {{POETS: Portable Occultation, Eclipse, and Transit System}},
  year     = {2006},
  month    = nov,
  number   = {849},
  pages    = {1550-1557},
  volume   = {118},
  adsurl   = {https://ui.adsabs.harvard.edu/abs/2006PASP..118.1550S},
  doi      = {10.1086/509665},
}

@MastersThesis{Doerr2014,
  author = {{Dörr}, Andreas},
  school = {{Universität Stuttgart}},
  title  = {{Thermal and structural design and characterization of electro-optical and mechanical components for the Imager Upgrade of the Stratospheric Observatory for Infrared Astronomy}},
  year   = {2014},
  month  = oct,
  note   = {IRS-14-S-012},
  type   = {{Pre-diploma thesis, IRS-14-S-012}},
}

@MastersThesis{Frueh2023,
  author = {{Früh}, Jonas},
  school = {{Universität Stuttgart}},
  title  = {{Opto-Mechanical Design of SOFIA’s Modular “Shack-Hartmann Instrument Fast Tracked” (SHIFT) and an Off-Axis Guider for ATUS}},
  year   = {2023},
  month  = mar,
  note   = {IRS-23-S-015},
  type   = {{Master's thesis, IRS-23-S-015}},
}

@InProceedings{Schindler2016,
  author    = {{Schindler}, Karsten and {Lang}, Dustin and {Moore}, Liz and {H{\"u}mmer}, Martin and {Wolf}, J{\"u}rgen and {Krabbe}, Alfred},
  booktitle = {{Software and Cyberinfrastructure for Astronomy IV}},
  title     = {{Computer-aided star pattern recognition with astrometry.net: in-flight support of telescope operations on SOFIA}},
  year      = {2016},
  editor    = {{Chiozzi}, Gianluca and {Guzman}, Juan C.},
  month     = aug,
  note      = {Society of Photo-Optical Instrumentation Engineers (SPIE) Conference Series},
  pages     = {991307},
  series    = {\procspie},
  volume    = {9913},
  adsurl    = {https://ui.adsabs.harvard.edu/abs/2016SPIE.9913E..07S},
  doi       = {10.1117/12.2231531},
  eid       = {991307},
}

@Article{Serrurier1938,
  author  = {{Serrurier}, Mark},
  journal = {{Civil Engineering}},
  title   = {{Structural Features of the 200-inch Telescope for Mt. Palomar Observatory}},
  year    = {1938},
  issn    = {0009-7853},
  month   = aug,
  number  = {8},
  pages   = {524-526},
  volume  = {8},
}

@Article{Sickafoose2019,
  author        = {{Sickafoose}, A.~A. and {Bosh}, A.~S. and {Levine}, S.~E. and {Zuluaga}, C.~A. and {Genade}, A. and {Schindler}, K. and {Lister}, T.~A. and {Person}, M.~J.},
  journal       = {\icarus},
  title         = {{A stellar occultation by Vanth, a satellite of (90482) Orcus}},
  year          = {2019},
  month         = feb,
  pages         = {657-668},
  volume        = {319},
  adsurl        = {https://ui.adsabs.harvard.edu/abs/2019Icar..319..657S},
  archiveprefix = {arXiv},
  doi           = {10.1016/j.icarus.2018.10.016},
  eprint        = {1810.08977},
  primaryclass  = {astro-ph.EP},
}

@Article{Schindler2017,
  author        = {{Schindler}, K. and {Wolf}, J. and {Bardecker}, J. and {Olsen}, A. and {M{\"u}ller}, T. and {Kiss}, C. and {Ortiz}, J.~L. and {Braga-Ribas}, F. and {Camargo}, J.~I.~B. and {Herald}, D. and {Krabbe}, A.},
  journal       = {\aap},
  title         = {{Results from a triple chord stellar occultation and far-infrared photometry of the trans-Neptunian object (229762) 2007 UK$_{126}$}},
  year          = {2017},
  month         = apr,
  pages         = {A12},
  volume        = {600},
  adsurl        = {https://ui.adsabs.harvard.edu/abs/2017A&A...600A..12S},
  archiveprefix = {arXiv},
  doi           = {10.1051/0004-6361/201628620},
  eid           = {A12},
  eprint        = {1611.02798},
  primaryclass  = {astro-ph.EP},
}

@InProceedings{Schindler2019,
  author    = {{Schindler}, K. and {Bosh}, A.~S. and {Levine}, S.~E. and {Person}, M.~J. and {Wolf}, J. and {Zuluaga}, C. and {Krabbe}, A.},
  booktitle = {{AGU Fall Meeting Abstracts}},
  title     = {{Results from a stellar occultation by KBO Varda}},
  year      = {2019},
  month     = dec,
  pages     = {P42C-08},
  volume    = {2019},
  adsurl    = {https://ui.adsabs.harvard.edu/abs/2019AGUFM.P42C..08S},
  eid       = {P42C-08},
}

@Article{Sickafoose2026,
  author  = {{Sickafoose}, Amanda A. and others},
  journal = {submitted to PSJ},
  title   = {{Changes in Pluto’s Atmosphere Based on Stellar Occultation Data from 2017–2023}},
  year    = {2026},
}

@Article{AstropyCollaboration2013,
  author        = {{Astropy Collaboration} and {Robitaille}, Thomas P. and {Tollerud}, Erik J. and {Greenfield}, Perry and {Droettboom}, Michael and {Bray}, Erik and {Aldcroft}, Tom and {Davis}, Matt and {Ginsburg}, Adam and {Price-Whelan}, Adrian M. and {Kerzendorf}, Wolfgang E. and {Conley}, Alexander and {Crighton}, Neil and {Barbary}, Kyle and {Muna}, Demitri and {Ferguson}, Henry and {Grollier}, Fr{\'e}d{\'e}ric and {Parikh}, Madhura M. and {Nair}, Prasanth H. and {Unther}, Hans M. and {Deil}, Christoph and {Woillez}, Julien and {Conseil}, Simon and {Kramer}, Roban and {Turner}, James E.~H. and {Singer}, Leo and {Fox}, Ryan and {Weaver}, Benjamin A. and {Zabalza}, Victor and {Edwards}, Zachary I. and {Azalee Bostroem}, K. and {Burke}, D.~J. and {Casey}, Andrew R. and {Crawford}, Steven M. and {Dencheva}, Nadia and {Ely}, Justin and {Jenness}, Tim and {Labrie}, Kathleen and {Lim}, Pey Lian and {Pierfederici}, Francesco and {Pontzen}, Andrew and {Ptak}, Andy and {Refsdal}, Brian and {Servillat}, Mathieu and {Streicher}, Ole},
  journal       = {\aap},
  title         = {{Astropy: A community Python package for astronomy}},
  year          = {2013},
  month         = {Oct},
  pages         = {A33},
  volume        = {558},
  adsurl        = {https://ui.adsabs.harvard.edu/abs/2013A&A...558A..33A},
  archiveprefix = {arXiv},
  doi           = {10.1051/0004-6361/201322068},
  eid           = {A33},
  eprint        = {1307.6212},
  primaryclass  = {astro-ph.IM},
}

@Manual{SpectrumInstruments2015,
  title   = {{Intelligent Reference/TM-4\texttrademark\,GPS Time \& Frequency System - User Manual}},
  address = {{570 E. Arrow Highway, Suite D, San Dimas, CA 91773}},
  year    = {2015},
  key     = {{Spectrum Instruments, Inc.}},
}

@InProceedings{Henden2009,
  author    = {{Henden}, Arne A. and {Welch}, D.~L. and {Terrell}, D. and {Levine}, S.~E.},
  booktitle = {{American Astronomical Society Meeting Abstracts \#214}},
  title     = {{The AAVSO Photometric All-Sky Survey (APASS)}},
  year      = {2009},
  month     = may,
  pages     = {407.02},
  series    = {American Astronomical Society Meeting Abstracts},
  volume    = {214},
  adsurl    = {https://ui.adsabs.harvard.edu/abs/2009AAS...21440702H},
  eid       = {407.02},
}

@Article{Brown2013,
  author        = {{Brown}, T.~M. and {Baliber}, N. and {Bianco}, F.~B. and {Bowman}, M. and {Burleson}, B. and {Conway}, P. and {Crellin}, M. and {Depagne}, {\'E}. and {De Vera}, J. and {Dilday}, B. and {Dragomir}, D. and {Dubberley}, M. and {Eastman}, J.~D. and {Elphick}, M. and {Falarski}, M. and {Foale}, S. and {Ford}, M. and {Fulton}, B.~J. and {Garza}, J. and {Gomez}, E.~L. and {Graham}, M. and {Greene}, R. and {Haldeman}, B. and {Hawkins}, E. and {Haworth}, B. and {Haynes}, R. and {Hidas}, M. and {Hjelstrom}, A.~E. and {Howell}, D.~A. and {Hygelund}, J. and {Lister}, T.~A. and {Lobdill}, R. and {Martinez}, J. and {Mullins}, D.~S. and {Norbury}, M. and {Parrent}, J. and {Paulson}, R. and {Petry}, D.~L. and {Pickles}, A. and {Posner}, V. and {Rosing}, W.~E. and {Ross}, R. and {Sand}, D.~J. and {Saunders}, E.~S. and {Shobbrook}, J. and {Shporer}, A. and {Street}, R.~A. and {Thomas}, D. and {Tsapras}, Y. and {Tufts}, J.~R. and {Valenti}, S. and {Vander Horst}, K. and {Walker}, Z. and {White}, G. and {Willis}, M.},
  journal       = {\pasp},
  title         = {{Las Cumbres Observatory Global Telescope Network}},
  year          = {2013},
  month         = sep,
  number        = {931},
  pages         = {1031},
  volume        = {125},
  adsurl        = {https://ui.adsabs.harvard.edu/abs/2013PASP..125.1031B},
  archiveprefix = {arXiv},
  doi           = {10.1086/673168},
  eprint        = {1305.2437},
  primaryclass  = {astro-ph.IM},
}

@Article{Narita2015,
  author        = {{Narita}, Norio and {Fukui}, Akihiko and {Kusakabe}, Nobuhiko and {Onitsuka}, Masahiro and {Ryu}, Tsuguru and {Yanagisawa}, Kenshi and {Izumiura}, Hideyuki and {Tamura}, Motohide and {Yamamuro}, Tomoyasu},
  journal       = {Journal of Astronomical Telescopes, Instruments, and Systems},
  title         = {{MuSCAT: a multicolor simultaneous camera for studying atmospheres of transiting exoplanets}},
  year          = {2015},
  month         = oct,
  pages         = {045001},
  volume        = {1},
  adsurl        = {https://ui.adsabs.harvard.edu/abs/2015JATIS...1d5001N},
  archiveprefix = {arXiv},
  doi           = {10.1117/1.JATIS.1.4.045001},
  eid           = {045001},
  eprint        = {1509.03154},
  primaryclass  = {astro-ph.IM},
}

@Article{Fukugita1996,
  author   = {{Fukugita}, M. and {Ichikawa}, T. and {Gunn}, J.~E. and {Doi}, M. and {Shimasaku}, K. and {Schneider}, D.~P.},
  journal  = {\aj},
  title    = {{The Sloan Digital Sky Survey Photometric System}},
  year     = {1996},
  month    = apr,
  pages    = {1748},
  volume   = {111},
  adsurl   = {https://ui.adsabs.harvard.edu/abs/1996AJ....111.1748F},
  doi      = {10.1086/117915},
}

@Article{Gulbis2011,
  author        = {{Gulbis}, A.~A.~S. and {Bus}, S.~J. and {Elliot}, J.~L. and {Rayner}, J.~T. and {Stahlberger}, W.~E. and {Rojas}, F.~E. and {Adams}, E.~R. and {Person}, M.~J. and {Chung}, R. and {Tokunaga}, A.~T. and {Zuluaga}, C.~A.},
  journal       = {\pasp},
  title         = {{First Results from the MIT Optical Rapid Imaging System (MORIS) on the IRTF: A Stellar Occultation by Pluto and a Transit by Exoplanet XO-2b}},
  year          = {2011},
  month         = apr,
  number        = {902},
  pages         = {461},
  volume        = {123},
  adsurl        = {https://ui.adsabs.harvard.edu/abs/2011PASP..123..461G},
  archiveprefix = {arXiv},
  doi           = {10.1086/659636},
  eprint        = {1102.5248},
  primaryclass  = {astro-ph.IM},
}

@Article{Lockhart2010,
  author  = {{Lockhart}, Matthew and {Person}, Michael J. and {Elliot}, J.~L. and {Souza}, Steven P.},
  journal = {\pasp},
  title   = {{PICO: Portable Instrument for Capturing Occultations}},
  year    = {2010},
  month   = oct,
  number  = {896},
  pages   = {1207},
  volume  = {122},
  adsurl  = {https://ui.adsabs.harvard.edu/abs/2010PASP..122.1207L},
  doi     = {10.1086/656445},
}

@Article{Gitton1998,
  author   = {{Gitton}, P. and {Noethe}, L.},
  journal  = {The Messenger},
  title    = {{Tuning of the NTT alignment.}},
  year     = {1998},
  month    = jun,
  pages    = {15-18},
  volume   = {92},
  adsurl   = {https://ui.adsabs.harvard.edu/abs/1998Msngr..92...15G},
}

@Article{Noethe2000,
  author   = {{Noethe}, L. and {Guisard}, S.},
  journal  = {\aaps},
  title    = {{Analytical expressions for field astigmatism in decentered two mirror telescopes and application to the collimation of the ESO VLT}},
  year     = {2000},
  month    = may,
  pages    = {157-167},
  volume   = {144},
  adsurl   = {https://ui.adsabs.harvard.edu/abs/2000A&AS..144..157N},
  doi      = {10.1051/aas:2000201},
}

@MastersThesis{Zabel2020,
  author = {{Zabel}, Sven},
  school = {{Universität Stuttgart}},
  title  = {{Optomechanical Design of Shack-Hartmann Test Instrument for the SOFIA Telescope}},
  year   = {2020},
  month  = {jul},
  note   = {IRS-20-S-027},
  type   = {{Master's thesis, IRS-20-S-027}},
}

@Misc{ASCL_Collins2013,
  author        = {{Collins}, Karen and {Kielkopf}, John},
  month         = sep,
  title         = {{AstroImageJ: ImageJ for Astronomy}},
  year          = {2013},
  adsurl        = {https://ui.adsabs.harvard.edu/abs/2013ascl.soft09001C},
  archiveprefix = {ascl},
  eid           = {ascl:1309.001},
  eprint        = {1309.001},
  publisher     = {Astrophysics Source Code Library},
}

@Article{Collins2017,
  author        = {{Collins}, Karen A. and {Kielkopf}, John F. and {Stassun}, Keivan G. and {Hessman}, Frederic V.},
  journal       = {\aj},
  title         = {{AstroImageJ: Image Processing and Photometric Extraction for Ultra-precise Astronomical Light Curves}},
  year          = {2017},
  month         = feb,
  number        = {2},
  pages         = {77},
  volume        = {153},
  adsurl        = {https://ui.adsabs.harvard.edu/abs/2017AJ....153...77C},
  archiveprefix = {arXiv},
  doi           = {10.3847/1538-3881/153/2/77},
  eid           = {77},
  eprint        = {1701.04817},
  primaryclass  = {astro-ph.IM},
}

@Misc{ASCL_Greenfield2013,
  author        = {{Greenfield}, Perry and {Robitaille}, Thomas and {Tollerud}, Erik and {Aldcroft}, Tom and {Barbary}, Kyle and {Barrett}, Paul and {Bray}, Erik and {Crighton}, Neil and {Conley}, Alex and {Conseil}, Simon and {Davis}, Matt and {Deil}, Christoph and {Dencheva}, Nadia and {Droettboom}, Michael and {Ferguson}, Henry and {Ginsburg}, Adam and {Grollier}, Fr{\'e}d{\'e}ric and {Moritz G{\"u}nther}, Hans and {Hanley}, Chris and {Hsu}, J.~C. and {Kerzendorf}, Wolfgang and {Kramer}, Roban and {Lian Lim}, Pey and {Muna}, Demitri and {Nair}, Prasanth and {Price-Whelan}, Adrian and {Shiga}, David and {Singer}, Leo and {Taylor}, James and {Turner}, James and {Woillez}, Julien and {Zabalza}, Victor},
  month         = apr,
  title         = {{Astropy: Community Python library for astronomy}},
  year          = {2013},
  adsurl        = {https://ui.adsabs.harvard.edu/abs/2013ascl.soft04002G},
  archiveprefix = {ascl},
  eid           = {ascl:1304.002},
  eprint        = {1304.002},
  publisher     = {Astrophysics Source Code Library},
}

@Misc{ASCL_Lang2012,
  author        = {{Lang}, Dustin and {Hogg}, David W. and {Mierle}, Keir and {Blanton}, Michael and {Roweis}, Sam},
  month         = aug,
  title         = {{Astrometry.net: Astrometric calibration of images}},
  year          = {2012},
  adsurl        = {https://ui.adsabs.harvard.edu/abs/2012ascl.soft08001L},
  archiveprefix = {ascl},
  eid           = {ascl:1208.001},
  eprint        = {1208.001},
  publisher     = {Astrophysics Source Code Library},
}

@InProceedings{Dunham2004,
  author    = {{Dunham}, Edward W. and {Elliot}, James L. and {Bida}, Thomas A. and {Taylor}, Brian W.},
  booktitle = {{Ground-based Instrumentation for Astronomy}},
  title     = {{HIPO: a high-speed imaging photometer for occultations}},
  year      = {2004},
  editor    = {{Moorwood}, Alan F.~M. and {Iye}, Masanori},
  month     = sep,
  note      = {Society of Photo-Optical Instrumentation Engineers (SPIE) Conference Series},
  pages     = {592-603},
  series    = {\procspie},
  volume    = {5492},
  adsurl    = {https://ui.adsabs.harvard.edu/abs/2004SPIE.5492..592D},
  doi       = {10.1117/12.552152},
}

@Article{Mandel2002,
  author        = {{Mandel}, Kaisey and {Agol}, Eric},
  journal       = {\apjl},
  title         = {{Analytic Light Curves for Planetary Transit Searches}},
  year          = {2002},
  month         = dec,
  number        = {2},
  pages         = {L171-L175},
  volume        = {580},
  adsurl        = {https://ui.adsabs.harvard.edu/abs/2002ApJ...580L.171M},
  archiveprefix = {arXiv},
  doi           = {10.1086/345520},
  eprint        = {astro-ph/0210099},
  primaryclass  = {astro-ph},
}

@InProceedings{Pfueller2012,
  author    = {{Pf{\"u}ller}, Enrico and {Wolf}, J{\"u}rgen and {Hall}, Helen and {R{\"o}ser}, Hans-Peter},
  booktitle = {Ground-based and Airborne Telescopes IV},
  title     = {{Optical characterization of the SOFIA telescope using fast EM-CCD cameras}},
  year      = {2012},
  editor    = {{Stepp}, Larry M. and {Gilmozzi}, Roberto and {Hall}, Helen J.},
  month     = sep,
  pages     = {844413},
  series    = {Society of Photo-Optical Instrumentation Engineers (SPIE) Conference Series},
  volume    = {8444},
  adsurl    = {https://ui.adsabs.harvard.edu/abs/2012SPIE.8444E..13P},
  doi       = {10.1117/12.926367},
  eid       = {844413},
}

@Book{Suiter2009,
  author    = {{Suiter}, Harold Richard},
  publisher = {Willmann-Bell},
  title     = {{Star Testing Astronomical Telescopes: A Manual for Optical Evaluation and Adjustment}},
  year      = {2009},
  edition   = {2nd},
  isbn      = {978-0943396903},
}

@Book{Howell2006,
  author    = {{Howell}, Steve Bruce},
  publisher = {{Cambridge University Press}},
  title     = {{Handbook of CCD Astronomy}},
  year      = {2006},
  edition   = {{2nd}},
  adsurl    = {https://ui.adsabs.harvard.edu/abs/2006hca..book.....H},
  doi       = {10.1017/CBO9780511807909},
}

@Article{Virtanen2020,
  author        = {{Virtanen}, Pauli and {Gommers}, Ralf and {Oliphant}, Travis E. and {Haberland}, Matt and {Reddy}, Tyler and {Cournapeau}, David and {Burovski}, Evgeni and {Peterson}, Pearu and {Weckesser}, Warren and {Bright}, Jonathan and {van der Walt}, St{\'e}fan J. and {Brett}, Matthew and {Wilson}, Joshua and {Millman}, K. Jarrod and {Mayorov}, Nikolay and {Nelson}, Andrew R.~J. and {Jones}, Eric and {Kern}, Robert and {Larson}, Eric and {Carey}, C.~J. and {Polat}, {\.I}lhan and {Feng}, Yu and {Moore}, Eric W. and {VanderPlas}, Jake and {Laxalde}, Denis and {Perktold}, Josef and {Cimrman}, Robert and {Henriksen}, Ian and {Quintero}, E.~A. and {Harris}, Charles R. and {Archibald}, Anne M. and {Ribeiro}, Ant{\^o}nio H. and {Pedregosa}, Fabian and {van Mulbregt}, Paul and {SciPy 1. 0 Contributors}},
  journal       = {Nature Methods},
  title         = {{SciPy 1.0: fundamental algorithms for scientific computing in Python}},
  year          = {2020},
  month         = feb,
  pages         = {261-272},
  volume        = {17},
  adsurl        = {https://ui.adsabs.harvard.edu/abs/2020NatMe..17..261V},
  archiveprefix = {arXiv},
  doi           = {10.1038/s41592-019-0686-2},
  eprint        = {1907.10121},
  primaryclass  = {cs.MS},
}

@Article{Harris2020,
  author        = {{Harris}, Charles R. and {Millman}, K. Jarrod and {van der Walt}, St{\'e}fan J. and {Gommers}, Ralf and {Virtanen}, Pauli and {Cournapeau}, David and {Wieser}, Eric and {Taylor}, Julian and {Berg}, Sebastian and {Smith}, Nathaniel J. and {Kern}, Robert and {Picus}, Matti and {Hoyer}, Stephan and {van Kerkwijk}, Marten H. and {Brett}, Matthew and {Haldane}, Allan and {del R{\'\i}o}, Jaime Fern{\'a}ndez and {Wiebe}, Mark and {Peterson}, Pearu and {G{\'e}rard-Marchant}, Pierre and {Sheppard}, Kevin and {Reddy}, Tyler and {Weckesser}, Warren and {Abbasi}, Hameer and {Gohlke}, Christoph and {Oliphant}, Travis E.},
  journal       = {\nat},
  title         = {{Array programming with NumPy}},
  year          = {2020},
  month         = sep,
  number        = {7825},
  pages         = {357-362},
  volume        = {585},
  adsurl        = {https://ui.adsabs.harvard.edu/abs/2020Natur.585..357H},
  archiveprefix = {arXiv},
  doi           = {10.1038/s41586-020-2649-2},
  eprint        = {2006.10256},
  primaryclass  = {cs.MS},
}

@Article{Hunter2007,
  author  = {{Hunter}, John D.},
  journal = {Computing in Science and Engineering},
  title   = {Matplotlib: A 2D Graphics Environment},
  year    = {2007},
  number  = {3},
  pages   = {90-95},
  volume  = {9},
  doi     = {10.1109/MCSE.2007.55},
}

@Article{AstropyCollaboration2022,
  author        = {{Astropy Collaboration} and {Price-Whelan}, Adrian M. and {Lim}, Pey Lian and {Earl}, Nicholas and {Starkman}, Nathaniel and {Bradley}, Larry and {Shupe}, David L. and {Patil}, Aarya A. and {Corrales}, Lia and {Brasseur}, C.~E. and {N{\"o}the}, Maximilian and {Donath}, Axel and {Tollerud}, Erik and {Morris}, Brett M. and {Ginsburg}, Adam and {Vaher}, Eero and {Weaver}, Benjamin A. and {Tocknell}, James and {Jamieson}, William and {van Kerkwijk}, Marten H. and {Robitaille}, Thomas P. and {Merry}, Bruce and {Bachetti}, Matteo and {G{\"u}nther}, H. Moritz and {Aldcroft}, Thomas L. and {Alvarado-Montes}, Jaime A. and {Archibald}, Anne M. and {B{\'o}di}, Attila and {Bapat}, Shreyas and {Barentsen}, Geert and {Baz{\'a}n}, Juanjo and {Biswas}, Manish and {Boquien}, M{\'e}d{\'e}ric and {Burke}, D.~J. and {Cara}, Daria and {Cara}, Mihai and {Conroy}, Kyle E. and {Conseil}, Simon and {Craig}, Matthew W. and {Cross}, Robert M. and {Cruz}, Kelle L. and {D'Eugenio}, Francesco and {Dencheva}, Nadia and {Devillepoix}, Hadrien A.~R. and {Dietrich}, J{\"o}rg P. and {Eigenbrot}, Arthur Davis and {Erben}, Thomas and {Ferreira}, Leonardo and {Foreman-Mackey}, Daniel and {Fox}, Ryan and {Freij}, Nabil and {Garg}, Suyog and {Geda}, Robel and {Glattly}, Lauren and {Gondhalekar}, Yash and {Gordon}, Karl D. and {Grant}, David and {Greenfield}, Perry and {Groener}, Austen M. and {Guest}, Steve and {Gurovich}, Sebastian and {Handberg}, Rasmus and {Hart}, Akeem and {Hatfield-Dodds}, Zac and {Homeier}, Derek and {Hosseinzadeh}, Griffin and {Jenness}, Tim and {Jones}, Craig K. and {Joseph}, Prajwel and {Kalmbach}, J. Bryce and {Karamehmetoglu}, Emir and {Ka{\l}uszy{\'n}ski}, Miko{\l}aj and {Kelley}, Michael S.~P. and {Kern}, Nicholas and {Kerzendorf}, Wolfgang E. and {Koch}, Eric W. and {Kulumani}, Shankar and {Lee}, Antony and {Ly}, Chun and {Ma}, Zhiyuan and {MacBride}, Conor and {Maljaars}, Jakob M. and {Muna}, Demitri and {Murphy}, N.~A. and {Norman}, Henrik and {O'Steen}, Richard and {Oman}, Kyle A. and {Pacifici}, Camilla and {Pascual}, Sergio and {Pascual-Granado}, J. and {Patil}, Rohit R. and {Perren}, Gabriel I. and {Pickering}, Timothy E. and {Rastogi}, Tanuj and {Roulston}, Benjamin R. and {Ryan}, Daniel F. and {Rykoff}, Eli S. and {Sabater}, Jose and {Sakurikar}, Parikshit and {Salgado}, Jes{\'u}s and {Sanghi}, Aniket and {Saunders}, Nicholas and {Savchenko}, Volodymyr and {Schwardt}, Ludwig and {Seifert-Eckert}, Michael and {Shih}, Albert Y. and {Jain}, Anany Shrey and {Shukla}, Gyanendra and {Sick}, Jonathan and {Simpson}, Chris and {Singanamalla}, Sudheesh and {Singer}, Leo P. and {Singhal}, Jaladh and {Sinha}, Manodeep and {Sip{\H{o}}cz}, Brigitta M. and {Spitler}, Lee R. and {Stansby}, David and {Streicher}, Ole and {{\v{S}}umak}, Jani and {Swinbank}, John D. and {Taranu}, Dan S. and {Tewary}, Nikita and {Tremblay}, Grant R. and {de Val-Borro}, Miguel and {Van Kooten}, Samuel J. and {Vasovi{\'c}}, Zlatan and {Verma}, Shresth and {de Miranda Cardoso}, Jos{\'e} Vin{\'\i}cius and {Williams}, Peter K.~G. and {Wilson}, Tom J. and {Winkel}, Benjamin and {Wood-Vasey}, W.~M. and {Xue}, Rui and {Yoachim}, Peter and {Zhang}, Chen and {Zonca}, Andrea and {Astropy Project Contributors}},
  journal       = {\apj},
  title         = {{The Astropy Project: Sustaining and Growing a Community-oriented Open-source Project and the Latest Major Release (v5.0) of the Core Package}},
  year          = {2022},
  month         = aug,
  number        = {2},
  pages         = {167},
  volume        = {935},
  adsurl        = {https://ui.adsabs.harvard.edu/abs/2022ApJ...935..167A},
  archiveprefix = {arXiv},
  doi           = {10.3847/1538-4357/ac7c74},
  eid           = {167},
  eprint        = {2206.14220},
  primaryclass  = {astro-ph.IM},
}

@Misc{Bradley2025,
  author    = {Larry {Bradley} and Brigitta {Sipőcz} and Thomas {Robitaille} and Erik {Tollerud} and Zé {Vinícius} and Christoph {Deil} and Kyle {Barbary} and Tom J {Wilson} and Ivo {Busko} and Axel {Donath} and Hans Moritz {Günther} and Mihai {Cara} and P. L. {Lim} and Sebastian {Meßlinger} and Zach {Burnett} and Simon {Conseil} and Michael {Droettboom} and Azalee {Bostroem} and E. M. {Bray} and Lars Andersen {Bratholm} and William {Jamieson} and Adam {Ginsburg} and Geert {Barentsen} and Matt {Craig} and Sergio {Pascual} and Shivangee {Rathi} and Marshall {Perrin} and Brett M. {Morris}},
  month     = feb,
  title     = {astropy/photutils: 2.2.0},
  year      = {2025},
  doi       = {10.5281/zenodo.14889440},
  publisher = {Zenodo},
  swhid     = {swh:1:dir:11159107f27a28985192ed1118b1f2055709d093 ;origin=https://doi.org/10.5281/zenodo.596036;visi t=swh:1:snp:ae8c4a55d349d43e53cfe9ce92e678fcfe840f 3b;anchor=swh:1:rel:0117f67e8888adcdfc85308287dd9c 854b466389;path=astropy-photutils-ffb96c5},
}

@Article{Ginsburg2019,
  author        = {{Ginsburg}, A. and {Sip{\H o}cz}, B.~M. and {Brasseur}, C.~E. and {Cowperthwaite}, P.~S. and {Craig}, M.~W. and {Deil}, C. and {Guillochon}, J. and {Guzman}, G. and {Liedtke}, S. and {Lian Lim}, P. and {Lockhart}, K.~E. and {Mommert}, M. and {Morris}, B.~M. and {Norman}, H. and {Parikh}, M. and {Persson}, M.~V. and {Robitaille}, T.~P. and {Segovia}, J.-C. and {Singer}, L.~P. and {Tollerud}, E.~J. and {de Val-Borro}, M. and {Valtchanov}, I. and {Woillez}, J. and {The Astroquery collaboration} and {a subset of the astropy collaboration}},
  journal       = {\aj},
  title         = {{astroquery: An Astronomical Web-querying Package in Python}},
  year          = {2019},
  month         = mar,
  pages         = {98},
  volume        = {157},
  adsurl        = {https://adsabs.harvard.edu/abs/2019AJ....157...98G},
  archiveprefix = {arXiv},
  doi           = {10.3847/1538-3881/aafc33},
  eid           = {98},
  eprint        = {1901.04520},
  primaryclass  = {astro-ph.IM},
}

@Article{Pedregosa2011,
  author        = {{Pedregosa}, Fabian and {Varoquaux}, Ga{\"e}l and {Gramfort}, Alexandre and {Michel}, Vincent and {Thirion}, Bertrand and {Grisel}, Olivier and {Blondel}, Mathieu and {M{\"u}ller}, Andreas and {Nothman}, Joel and {Louppe}, Gilles and {Prettenhofer}, Peter and {Weiss}, Ron and {Dubourg}, Vincent and {Vanderplas}, Jake and {Passos}, Alexandre and {Cournapeau}, David and {Brucher}, Matthieu and {Perrot}, Matthieu and {Duchesnay}, {\'E}douard},
  journal       = {Journal of Machine Learning Research},
  title         = {{Scikit-learn: Machine Learning in Python}},
  year          = {2011},
  month         = oct,
  pages         = {2825-2830},
  volume        = {12},
  adsurl        = {https://ui.adsabs.harvard.edu/abs/2011JMLR...12.2825P},
  archiveprefix = {arXiv},
  doi           = {10.48550/arXiv.1201.0490},
  eprint        = {1201.0490},
  primaryclass  = {cs.LG},
}

@InProceedings{McKinney2010,
  author    = {{McKinney}, Wes},
  booktitle = {Proceedings of the 9th Python in Science Conference},
  title     = {Data Structures for Statistical Computing in Python},
  year      = {2010},
  pages     = {56--61},
  publisher = {SciPy},
  doi       = {10.25080/majora-92bf1922-00a},
  issn      = {2575-9752},
}

@Article{Martinez2010a,
  author  = {{Martinez}, P. and {Kolb}, J. and {Sarazin}, M. and {Tokovinin}, A.},
  journal = {The Messenger},
  title   = {{On the Difference between Seeing and Image Quality: When the Turbulence Outer Scale Enters the Game}},
  year    = {2010},
  month   = sep,
  pages   = {5-8},
  volume  = {141},
  adsurl  = {https://ui.adsabs.harvard.edu/abs/2010Msngr.141....5M},
}

@InBook{Quirrenbach2006,
  author    = {{Quirrenbach}, Andreas},
  chapter   = {{The Effects of Atmospheric Turbulenceon Astronomical Observations}},
  pages     = {130--142},
  publisher = {{Springer Berlin Heidelberg}},
  title     = {{Detection and Characterization of Extrasolar Planets}},
  year      = {2006},
  isbn      = {978-3-540-31470-7},
  series    = {Saas-Fee Advanced Course, Vol. 31},
  booktitle = {Extrasolar Planets: Swiss Society for Astrophysics and Astronomy},
  doi       = {10.1007/978-3-540-31470-7_1},
  url       = {https://doi.org/10.1007/978-3-540-31470-7_1},
}

@Article{Martinez2010b,
  author        = {{Martinez}, P. and {Kolb}, J. and {Tokovinin}, A. and {Sarazin}, M.},
  journal       = {\aap},
  title         = {{Atmospheric image blur with finite outer scale or partial adaptive correction}},
  year          = {2010},
  month         = jun,
  pages         = {A90},
  volume        = {516},
  adsurl        = {https://ui.adsabs.harvard.edu/abs/2010A&A...516A..90M},
  archiveprefix = {arXiv},
  doi           = {10.1051/0004-6361/201014413},
  eid           = {A90},
  eprint        = {1003.4593},
  primaryclass  = {astro-ph.IM},
}

@Article{Ofek2020,
  author        = {{Ofek}, E.~O. and {Ben-Ami}, S.},
  journal       = {\pasp},
  title         = {{Seeing-limited Imaging Sky Surveys{\textemdash}Small versus Large Telescopes}},
  year          = {2020},
  month         = dec,
  number        = {1018},
  pages         = {125004},
  volume        = {132},
  adsurl        = {https://ui.adsabs.harvard.edu/abs/2020PASP..132l5004O},
  archiveprefix = {arXiv},
  doi           = {10.1088/1538-3873/abc14c},
  eid           = {125004},
  eprint        = {2011.04674},
  primaryclass  = {astro-ph.IM},
}

@Book{Yoder2006,
  author    = {{Yoder}, Jr., Paul R.},
  publisher = {{CRC Press}},
  title     = {{Opto-Mechanical Systems Design}},
  year      = {2006},
  edition   = {3rd},
  doi       = {10.1201/9781420027235},
}

@Article{Trujillo2001,
  author        = {{Trujillo}, I. and {Aguerri}, J.~A.~L. and {Cepa}, J. and {Guti{\'e}rrez}, C.~M.},
  journal       = {\mnras},
  title         = {{The effects of seeing on S{\'e}rsic profiles - II. The Moffat PSF}},
  year          = {2001},
  month         = dec,
  number        = {3},
  pages         = {977-985},
  volume        = {328},
  adsurl        = {https://ui.adsabs.harvard.edu/abs/2001MNRAS.328..977T},
  archiveprefix = {arXiv},
  doi           = {10.1046/j.1365-8711.2001.04937.x},
  eprint        = {astro-ph/0109067},
  primaryclass  = {astro-ph},
}

@Article{Moffat1969,
  author  = {{Moffat}, A.~F.~J.},
  journal = {\aap},
  title   = {{A Theoretical Investigation of Focal Stellar Images in the Photographic Emulsion and Application to Photographic Photometry}},
  year    = {1969},
  month   = dec,
  pages   = {455},
  volume  = {3},
  adsurl  = {https://ui.adsabs.harvard.edu/abs/1969A&A.....3..455M},
}

@Article{Mahajan1981,
  author   = {{Mahajan}, V.~N.},
  journal  = {Journal of the Optical Society of America (1917-1983)},
  title    = {{Zernike annular polynomials for imaging systems with annular pupils.}},
  year     = {1981},
  month    = jan,
  pages    = {75-85},
  volume   = {71},
  adsurl   = {https://ui.adsabs.harvard.edu/abs/1981JOSA...71...75M},
  doi      = {10.1364/JOSA.71.000075},
}

@InProceedings{Erickson2000,
  author    = {{Erickson}, Edwin F. and {Dunham}, Edward W.},
  booktitle = {Airborne Telescope Systems},
  title     = {{Image stability requirement for the SOFIA telescope}},
  year      = {2000},
  editor    = {{Melugin}, Ramsey K. and {R{\"o}ser}, Hans-Peter},
  month     = jun,
  pages     = {2-13},
  series    = {Society of Photo-Optical Instrumentation Engineers (SPIE) Conference Series},
  volume    = {4014},
  adsurl    = {https://ui.adsabs.harvard.edu/abs/2000SPIE.4014....2E},
  doi       = {10.1117/12.389101},
}

@Conference{Schindler2021,
  author    = {{Schindler}, Karsten},
  booktitle = {{The Future of Airborne Infrared/Submm Astronomy: Instrument Solutions}},
  title     = {{A new near-infrared channel for SOFIA's Focal Plane Imager}},
  year      = {2021},
  month     = nov,
  pages     = {3-08},
}

@Misc{Person2018,
  author    = {{Person}, Michael J.},
  title     = {{Near-infrared Extension to the FPI+ Camera Focusing on Occultation Science}},
  year      = {2018},
  publisher = {NASA Proposal Number 18-S4THG18-0002},
}

@InProceedings{Narita2024,
  author    = {{Narita}, Norio and {Fukui}, Akihiko and {MuSCAT team}},
  booktitle = {AASTCS10, Extreme Solar Systems V},
  title     = {{Development of the MuSCAT series and their latest scientific results}},
  year      = {2024},
  month     = apr,
  pages     = {628.13},
  series    = {AAS/Division for Extreme Solar Systems Abstracts},
  volume    = {56},
  adsurl    = {https://ui.adsabs.harvard.edu/abs/2024ESS.....562813N},
  eid       = {628.13},
}

@Article{Hainaut2020,
  author        = {{Hainaut}, Olivier R. and {Williams}, Andrew P.},
  journal       = {\aap},
  title         = {{Impact of satellite constellations on astronomical observations with ESO telescopes in the visible and infrared domains}},
  year          = {2020},
  month         = apr,
  pages         = {A121},
  volume        = {636},
  adsurl        = {https://ui.adsabs.harvard.edu/abs/2020A&A...636A.121H},
  archiveprefix = {arXiv},
  doi           = {10.1051/0004-6361/202037501},
  eid           = {A121},
  eprint        = {2003.01992},
  primaryclass  = {astro-ph.IM},
}

@Article{Person2021,
  author   = {{Person}, Michael J. and {Bosh}, Amanda S. and {Zuluaga}, Carlos A. and {Sickafoose}, Amanda A. and {Levine}, Stephen E. and {Pasachoff}, Jay M. and {Babcock}, Bryce A. and {Dunham}, Edward W. and {McLean}, Ian S. and {Wolf}, J{\"u}rgen and {Abe}, Fumio and {Becklin}, E.~E. and {Bida}, Thomas A. and {Bright}, Len P. and {Brothers}, Tim and {Christie}, Grant and {Durst}, Rebecca F. and {Gilmore}, Alan C. and {Hamilton}, Ryan T. and {Harris}, Hugh C. and {Johnson}, Chris and {Kilmartin}, Pamela M. and {Kosiarek}, Molly and {Leppik}, Karina and {Logsdon}, Sarah E. and {Lucas}, Robert and {Mathers}, Shevill and {Morley}, C.~J.~K. and {Nelson}, Peter and {Ngan}, Haydn and {Pf{\"u}ller}, Enrico and {Natusch}, Tim and {Sallum}, Stephanie and {Savage}, Maureen L. and {Seeger}, Christina H. and {Siu}, Ho Chit and {Stockdale}, Chris and {Suzuki}, Daisuke and {Thanathibodee}, Thanawuth and {Tilleman}, Trudy and {Tristram}, Paul J. and {Vacca}, William D. and {Van Cleve}, Jeffrey and {Varughese}, Carolle and {Weisenbach}, Luke W. and {Widen}, Elizabeth and {Wiedemann}, Manuel},
  journal  = {\icarus},
  title    = {{Haze in Pluto's atmosphere: Results from SOFIA and ground-based observations of the 2015 June 29 Pluto occultation}},
  year     = {2021},
  month    = mar,
  pages    = {113572},
  volume   = {356},
  adsurl   = {https://ui.adsabs.harvard.edu/abs/2021Icar..35613572P},
  doi      = {10.1016/j.icarus.2019.113572},
  eid      = {113572},
}

@Article{FernandezValenzuela2026,
  author  = {{Fern\'andez-Valenzuela}, Estela},
  journal = {submitted to ApJL},
  title   = {{Weywot, an unusually low albedo satellite in the trans-Neptunian region}},
  year    = {2026},
}

@Article{Knieling2024,
  author   = {{Knieling}, Bastian and {Schindler}, Karsten and {Sickafoose}, Amanda A. and {Person}, Michael J. and {Levine}, Stephen E. and {Krabbe}, Alfred},
  journal  = {\psj},
  title    = {{Stellar Occultations in the Era of Data Mining and Modern Regression Models: Using Gaussian Processes to Analyze Light Curves and Improve Predictions}},
  year     = {2024},
  month    = apr,
  number   = {4},
  pages    = {104},
  volume   = {5},
  adsurl   = {https://ui.adsabs.harvard.edu/abs/2024PSJ.....5..104K},
  doi      = {10.3847/PSJ/ad3819},
  eid      = {104},
}

@InProceedings{Sickafoose2019a,
  author    = {{Sickafoose}, A.~A. and {Bosh}, A.~S. and {Levine}, S.~E. and {Person}, M.~J. and {Schindler}, K. and {Zuluaga}, C.~A.},
  booktitle = {Pluto System After New Horizons},
  title     = {{Stellar Occultations by Pluto: 2017-2018}},
  year      = {2019},
  editor    = {{LPI Editorial Board}},
  month     = jul,
  pages     = {7026},
  series    = {LPI Contributions},
  volume    = {2133},
  adsurl    = {https://ui.adsabs.harvard.edu/abs/2019LPICo2133.7026S},
  eid       = {7026},
}

@InProceedings{Sickafoose2023,
  author    = {{Sickafoose}, Amanda and {Person}, Michael and {Zuluaga}, Carlos and {Bosh}, Amanda and {Levine}, Stephen and {Brothers}, Tim and {Knieling}, Bastian and {Lister}, Tim and {Osip}, David and {Rojo}, Patricio and {Schindler}, Karsten and {Brimacombe}, Joseph and {Carruthers}, Timothy and {Colclasure}, Abigail and {Janse van Rensburg}, Petro and {Genade}, Anja and {Potter}, Stephen},
  booktitle = {55th Annual Meeting of the Division for Planetary Sciences},
  title     = {{Pluto's Atmosphere Persists}},
  year      = {2023},
  month     = oct,
  pages     = {308.02},
  series    = {AAS/Division for Planetary Sciences Meeting Abstracts},
  volume    = {55},
  adsurl    = {https://ui.adsabs.harvard.edu/abs/2023DPS....5530802S},
  eid       = {308.02},
}

@Book{Eickhoff2016,
  editor    = {{Eickhoff}, Jens},
  publisher = {{Springer Cham}},
  title     = {{The FLP Microsatellite Platform -- Flight Operations Manual}},
  year      = {2016},
  series    = {{Springer Aerospace Technology}},
  date      = {2016},
  doi       = {10.1007/978-3-319-23503-5},
  url       = {https://app.dimensions.ai/details/publication/pub.1028832577},
}

@Article{Adams2024,
  author        = {{Adams}, Elisabeth R. and {Jackson}, Brian and {Sickafoose}, Amanda A. and {Morgenthaler}, Jeffrey P. and {Worters}, Hannah and {Stubbers}, Hailey and {Carlson}, Dallon and {Bhure}, Sakhee and {Dekeyser}, Stijn and {Huang}, Chelsea X. and {Weinberg}, Nevin N.},
  journal       = {\psj},
  title         = {{Doomed Worlds. I. No New Evidence for Orbital Decay in a Long-term Survey of 43 Ultrahot Jupiters}},
  year          = {2024},
  month         = jul,
  number        = {7},
  pages         = {163},
  volume        = {5},
  adsurl        = {https://ui.adsabs.harvard.edu/abs/2024PSJ.....5..163A},
  archiveprefix = {arXiv},
  doi           = {10.3847/PSJ/ad3e80},
  eid           = {163},
  eprint        = {2404.07339},
  primaryclass  = {astro-ph.EP},
}

@Article{Harvey1995,
  author    = {{Harvey}, James E. and {Ftaclas}, Christ},
  journal   = {Appl. Opt.},
  title     = {Diffraction effects of telescope secondary mirror spiders on various image-quality criteria},
  year      = {1995},
  month     = {Oct},
  number    = {28},
  pages     = {6337--6349},
  volume    = {34},
  doi       = {10.1364/AO.34.006337},
  publisher = {Optica Publishing Group},
  url       = {https://opg.optica.org/ao/abstract.cfm?URI=ao-34-28-6337},
}

@Article{Borlaff2025,
  author   = {{Borlaff}, Alejandro S. and {Marcum}, Pamela M. and {Howell}, Steve B.},
  journal  = {\nat},
  title    = {{Satellite megaconstellations will threaten space-based astronomy}},
  year     = {2025},
  month    = dec,
  number   = {8092},
  pages    = {51-57},
  volume   = {648},
  adsurl   = {https://ui.adsabs.harvard.edu/abs/2025Natur.648...51B},
  doi      = {10.1038/s41586-025-09759-5},
}

@Article{Rothmeier2025,
  author        = {{Rothmeier}, Marvin and {Adams}, Elisabeth R. and {Schindler}, Karsten and {Beck}, Andr{\'e} and {Jackson}, Brian and {Morgenthaler}, Jeffrey P. and {Sickafoose}, Amanda A. and {Barker}, Malia and {Mancini}, Luigi and {Southworth}, John and {Evans}, Daniel and {Krabbe}, Alfred},
  journal       = {\psj},
  title         = {{Doomed Worlds. II. Reassessing Suggestions of Orbital Decay for TrES-5 b}},
  year          = {2025},
  month         = dec,
  number        = {12},
  pages         = {292},
  volume        = {6},
  adsurl        = {https://ui.adsabs.harvard.edu/abs/2025PSJ.....6..292R},
  archiveprefix = {arXiv},
  doi           = {10.3847/PSJ/ae1b9c},
  eid           = {292},
  eprint        = {2512.13937},
  primaryclass  = {astro-ph.EP},
}

@Misc{pandasdevelopmentteam2024,
  month     = sep,
  title     = {pandas-dev/pandas: Pandas 2.2.3},
  year      = {2024},
  doi       = {10.5281/zenodo.13819579},
  key       = {The pandas Development Team},
  publisher = {Zenodo},
}

@InCollection{Vukobratovich2017,
  author    = {{Vukobratovich}, Daniel},
  booktitle = {{Handbook of Optomechanical Engineering}},
  publisher = {CRC Press},
  title     = {{Lightweight Mirror Design}},
  year      = {2017},
  address   = {Boca Raton, FL},
  chapter   = {6},
  edition   = {2nd},
  editor    = {Ahmad, Anees},
  isbn      = {9781498761482},
  pages     = {173--205},
  doi       = {10.4324/9781315153247},
}

@Misc{OharaCorporation,
  title = {{CLEARCERAM-Z}: Ultra-Low Expansion Glass-Ceramics},
  year  = {2005},
  key   = {{Ohara Corporation}},
  url   = {https://www.ohara-gmbh.com/fileadmin/user_upload/export-data/pdf/Ohara_Clearceram-Z.pdf},
}

@InProceedings{Abdulkadyrov2012,
  author    = {{Abdulkadyrov}, Magomed A. and {Ignatov}, Alexandr N. and {Patrikeev}, Alexey P. and {Semenov}, Aleksandr P. and {Sharov}, Yury A.},
  booktitle = {Modern Technologies in Space- and Ground-based Telescopes and Instrumentation II},
  title     = {{Astrositall application in astronomical and space optics production}},
  year      = {2012},
  editor    = {{Navarro}, Ram{\'o}n and {Cunningham}, Colin R. and {Prieto}, Eric},
  month     = sep,
  pages     = {84502L},
  series    = {Society of Photo-Optical Instrumentation Engineers (SPIE) Conference Series},
  volume    = {8450},
  adsurl    = {https://ui.adsabs.harvard.edu/abs/2012SPIE.8450E..2LA},
  doi       = {10.1117/12.924648},
  eid       = {84502L},
}

@Book{Meeus1998,
  author    = {{Meeus}, Jean},
  publisher = {Willmann-Bell},
  title     = {{Astronomical Algorithms}},
  year      = {1998},
  address   = {Richmond, VA},
  edition   = {2nd},
  isbn      = {978-0-943396-61-3},
  adsurl    = {https://ui.adsabs.harvard.edu/abs/1998aalg.book.....M},
}

@Misc{ASCOM_Platform,
  title         = {{ASCOM Platform}},
  year          = {2024},
  archiveprefix = {github},
  key           = {{ASCOM Initiative}},
  publisher     = {GitHub},
  url           = {https://github.com/ASCOMInitiative/ASCOMPlatform},
}

@Misc{APCC,
  author = {{Gralak}, Ray and {Astro-Physics, Inc.}},
  title  = {{Astro-Physics Command Center (APCC) Pro v1.9}},
  year   = {2024},
  url    = {https://www.astro-physics.com/apcc},
}

@Misc{SkyTrack,
  author = {Boshart, Brent},
  title  = {{SkyTrack}},
  year   = {2022},
  url    = {https://www.heavenscape.com},
}

@Misc{PhD2,
  author    = {Galasso, Andy and Waddington, Bruce and Stark, Craig and {Open PHD Guiding contributors}},
  title     = {{PHD2}},
  year      = {2024},
  publisher = {GitHub},
  source    = {\url{https://openphdguiding.org}},
  url       = {https://github.com/OpenPHDGuiding/phd2},
}

@Misc{ansvr,
  author    = {{Galasso}, Andy},
  title     = {{ANSVR: Astrometry.net Local Plate Solver for Windows}},
  year      = {2024},
  publisher = {Freeware},
  url       = {https://adgsoftware.com/ansvr/},
}

@Misc{PinPoint,
  author = {{Denny}, Robert B.},
  title  = {{PinPoint Astrometric Engine v7.0}},
  year   = {2021},
  url    = {http://pinpoint.dc3.com/},
}

@Article{Groth1986,
  author   = {{Groth}, E.~J.},
  journal  = {\aj},
  title    = {{A pattern-matching algorithm for two-dimensional coordinate lists}},
  year     = {1986},
  month    = may,
  pages    = {1244-1248},
  volume   = {91},
  adsurl   = {https://ui.adsabs.harvard.edu/abs/1986AJ.....91.1244G},
  doi      = {10.1086/114099},
}

@Misc{Cygwin,
  title = {{Cygwin: A POSIX-compatible environment for Windows}},
  year  = {2024},
  key   = {{The Cygwin Project}},
  url   = {https://www.cygwin.com/},
}

@Book{Scheiner1897,
  author    = {{Scheiner}, Julius},
  publisher = {{Wilhelm Engelmann}},
  title     = {{Die Photographie der Gestirne}},
  year      = {1897},
  address   = {Leipzig},
  adsurl    = {https://ui.adsabs.harvard.edu/abs/1897dpdg.book.....S},
  doi       = {10.3931/e-rara-24897},
  url       = {https://archive.org/details/diephotographied00sche},
}

@Article{Scheiner1892,
  author    = {{Scheiner}, Julius},
  journal   = {{Bulletin du Comité International Permanent pour l'Exécution Photographique de la Carte du Ciel}},
  title     = {{Sur une méthode très simple permettant d'orienter un instrument à monture parallactique plus exactement qu'on ne peut le faire en général par des lectures des cercles}},
  year      = {1892},
  note      = {Translated by D. Klumpke},
  number    = {6},
  pages     = {385--388},
  volume    = {1},
  address   = {Paris},
  publisher = {{Gauthier-Villars}},
  url       = {https://gallica.bnf.fr/ark:/12148/bpt6k6570617p/f399},
}

@Misc{Garrett2021,
  author    = {John D. {Garrett}},
  month     = sep,
  title     = {{garrettj403/SciencePlots}},
  year      = {2021},
  doi       = {10.5281/zenodo.4106649},
  publisher = {Zenodo},
  url       = {http://doi.org/10.5281/zenodo.4106649},
  version   = {1.0.9},
}

@Misc{SCHOTT2023,
  month        = dec,
  title        = {{Optical Glass Datasheet SCHOTT N-BK7}},
  year         = {2023},
  address      = {Mainz, Germany},
  key          = {{SCHOTT AG}},
  organization = {SCHOTT AG},
  url          = {https://media.schott.com/api/public/content/41e799d0bf874807a0bb8e702fbb75b5?v=54856406},
  urldate      = {2025-02-10},
}

@Misc{AtlasSteels2021,
  month        = apr,
  title        = {{Stainless Steel 416 Grade Data Sheet}},
  year         = {2021},
  key          = {{Atlas Steels}},
  organization = {Atlas Steels},
  url          = {https://atlassteels.com.au/wp-content/uploads/2021/06/Stainless-Steel-416-Grade-Data-Sheet-28-04-21.pdf},
  urldate      = {2025-02-10},
}

@Article{Bonnarel2000,
  author   = {{Bonnarel}, F. and {Fernique}, P. and {Bienaym{\'e}}, O. and {Egret}, D. and {Genova}, F. and {Louys}, M. and {Ochsenbein}, F. and {Wenger}, M. and {Bartlett}, J.~G.},
  journal  = {\aaps},
  title    = {{The ALADIN interactive sky atlas. A reference tool for identification of astronomical sources}},
  year     = {2000},
  month    = apr,
  pages    = {33-40},
  volume   = {143},
  adsurl   = {https://ui.adsabs.harvard.edu/abs/2000A&AS..143...33B},
  doi      = {10.1051/aas:2000331},
}
\bibliographystyle{aasjournalv7}

\end{document}